\documentclass[12pt]{article}
\usepackage[slantedGreek]{mathptmx}
\usepackage{fullpage}
\usepackage{graphicx}        
\usepackage{amssymb, amsmath,bm}
\usepackage{stmaryrd}
\usepackage{enumitem}
\usepackage{comment}
\usepackage{caption, rotating}
\usepackage[small]{subfigure}
\usepackage[small]{subfigure}
\usepackage{amsfonts}
\usepackage{pdflscape}
\renewcommand{\d}{\text{d}}

\numberwithin{equation}{section}
\usepackage[numbers,square,comma,sort&compress]{natbib}
\graphicspath{{./figures/}} 

\title{Analysis of shear localization in viscoplastic solids with pressure-sensitive structural transformations} 

\author{J.D. Clayton$^{1}$\\ \\
$^1$Terminal Effects Division, Army Research Directorate
\\
DEVCOM ARL, Aberdeen, MD 21005, USA 
}

\date{}

\begin{document}
\graphicspath{{figures/}}
\maketitle

\begin{abstract}
Localization, in the form of adiabatic shear, is analyzed in viscoplastic solids
that may undergo structural transformation driven by pressure, shear stress, temperature,
and magnetic field. As pertinent to polycrystalline metals, transformations may include solid-solid phase transitions, twinning, and dynamic recrystallization. A finite-strain constitutive framework for isotropic metals
 is used to solve a boundary value problem involving simple shearing with superposed hydrostatic pressure and constant external magnetic field.
Three-dimensional theory is reduced to a formulation simple enough to facilitate 
 approximate analytical solutions yet sophisticated enough to maintain the salient physics. 
 Ranges of constitutive parameters (e.g., strain hardening, strain-rate sensitivity, thermal softening, and strain-driven structure transformation limits influenced by pressure and magnetic field) are obtained for which localization to infinite shear strain is possible. Motivated by experimental and
 theoretical studies suggesting a non-negligible role of shear on
 phase transformations in iron (Fe), the model is used to understand influences of pressure and phase transitions on applied strains for which localization should occur in pure Fe and a high-strength steel.
 Results show, among other trends for these two materials, that shear localization in conjunction with phase transformation is promoted when the transformed phase is softer than the parent phase.
 Localization that would occur in the isolated parent phase can be mitigated if the strain hardening or thermal softening tendencies of the transformed phase are sufficiently increased or reduced, respectively.
 \end{abstract}
\noindent \textbf{Key words}: shear band; strain localization; plasticity; phase transition; ferromagnetism; iron; steel \\

\noindent


\section{Introduction \label{sec1}}
Adiabatic shear localization is an important phenomenon in
settings that witness high strain rates and large shear deformations,
including impact, fragmentation, and industrial operations such as machining.
A solid undergoes large and rapid shear deformation in a relatively thin
band compared to surrounding material, often accompanied by substantial temperature
rise due to dissipative processes such as plasticity. The current work focuses on relatively ductile, polycrystalline metallic materials
with baseline constitutive behaviors of strain hardening, strain-rate sensitivity (at least mild),
and thermal softening.
Localization is promoted by strain softening, thermal softening, and defects in the material,
and is impeded by strain-rate hardening, inertia, and heat conduction.
Regarding the latter, the process is truly adiabatic only in the theoretical limit of null thermal conductivity;
conduction provides regularization and can affect band morphology and width \cite{wright2002}.

The present work is directed toward metals that may undergo other structural transitions in conjunction
with dislocation-mediated plasticity, namely solid-solid phase transitions, deformation twinning, and dynamic recrystallization (DRX).  Ferrous metals (e.g., Fe and steel) provide impetus, but the analysis can be applied to other solids sharing similar mechanisms.
The pressure-temperature phase diagram of pure Fe \cite{curran1971,li2017} includes three phases: $\alpha$ (body centered cubic or BCC) at low temperature and pressure, $\epsilon$ (hexagonal close packed or HCP) at high pressure, and $\gamma$ (face centered cubic or FCC) at high temperature. 
Martensitic and bainitic steels such as a nickel-chromium (Ni-Cr) steel \cite{hauver1976,hauver1979} modeled later and 4340 steel \cite{franz1979} show $\alpha \leftrightarrow \epsilon$ transformations at pressures on the order of 13 GPa,
noting that the starting phase could be BCC or body centered tetragonal (BCT) depending on composition and processing.
Although traditionally omitted in models for Fe \cite{boettger1997,duvall1977,claytonCMT2022,lloyd2022}, shear stress has been posited to
serve as an additional driving force based on experimental \cite{barge1990,ma2006,rittel2006}
and theoretical \cite{caspersen2004,lew2006} work, the latter merging density functional theory (DFT) and continuum laminate theory.  A continuum model \cite{sadjadpour2015} was developed for shear-driven $\alpha \leftrightarrow \epsilon$ transformations in pure Fe, but it omits phase mass density changes and corresponding pressure-driven transitions.

If the superposed pressure is high enough, $\alpha \rightarrow \epsilon$ transformation should occur in conjunction with shear at low temperatures, and if the temperature rise is high enough in a shear band, $\alpha \rightarrow \gamma$ transformation may be possible \cite{syn2005,jo2020}. The high-pressure and high-temperature phases are unstable under ambient conditions, and transformations are typically reversible, though retained austenite is not impossible. Therefore, material recovered after experiments (e.g., via shock recovery techniques \cite{williams2019,claytonENC2024}) may not contain discernible fractions of these phases \cite{rittel2006,boakye2013}.  However, recovered microstructures often contain other features (e.g., tertiary phase inclusions, dislocation structures, and twins) as well as mechanical behaviors that suggest transformations did take place \cite{cho1990,rittel2006,syn2005}.  For example, twinning is infrequent in $\alpha$-Fe but prevalent in $\epsilon$-Fe under high-pressure shear \cite{gandhi2022}. Rotational DRX has been observed in Fe, steels, and other alloys undergoing shear localization \cite{rittel2008,lieou2018,lieou2019}. Rotational DRX has been viewed as a softening mechanism important in shear band initiation, accompanying or even preceding thermal softening \cite{rittel2008,lieou2018,lieou2019,magagnosc2021}.

Shear bands do not form in all high-rate experiments on Fe and steels:
boundary conditions, processing, and defects all affect shear band susceptibility \cite{fellows2001,fellows2001b,xue2005,cerreta2013}.
Magnetic fields influence transformation behaviors in ferrous metals, noting $\alpha$-Fe is ferromagnetic but other phases are weakly or negligibly magnetic \cite{curran1971,murdoch2021,claytonCMT2022,claytonZAMM2024}.
It has been suggested \cite{moss1980,moss1981} that softening from a magnetic transition affects localization in bainitic and martensitic steels.
Somewhat more recent work \cite{egor2018} proposed and modeled how magnetic fields might affect shear bands.

Analytical and numerical studies to understand and predict instability and localization of shear
in viscoplastic solids have been categorized in a monograph \cite{wright2002} and a more recent review \cite{yan2021}. Loss of stability is considered a necessary, but not sufficient, condition for localization \cite{shawki1988,yan2021}.
Even if the material response is unstable, localization can be inhibited by strain-rate sensitivity, conduction, and inertia \cite{fressengeas1987,wright2002}.
Perhaps the simplest analytical treatment predicts instability when the shear stress-strain curve attains a local maximum (i.e., zero slope at the onset of softening) \cite{staker1981,voyiadjis2012}. This treatment only considers those properties governing the homogeneous stress-strain response, ignoring defects, conduction, and inertia. Linear perturbation analyses \cite{clifton1980,bai1982,anand1987,shawki1994} provide more insight, supplying conditions and material property combinations under which perturbations from the homogeneous solution satisfy the governing equations, including inertia and conduction in some cases. However, such analyses do not account for the magnitude of the perturbation relative to the response of surrounding material. In this regard, a relative linear stability analysis \cite{fressengeas1987} suggests more stringent and realistic sufficiency conditions for non-localization of shear strain. 

No aforementioned perturbation methods based on linear stability can quantify the critical applied (i.e., average) strain, beyond the point of instability, at which localization into a fully formed shear band of very high strain, and most often temperature, occurs \cite{molinari1988}.
Analytical \cite{semiatin1984,molinari1986,molinari1987,wright1990,grady1991,wright1994} or numerical \cite{wright1987,raftenberg2001,schoenfeld2003,batra2004,fermencoker2004,fermencoker2005,langer2017,le2018,lieou2018,lieou2019,jin2019} treatments accounting for the unstable nonlinear response are required.
Analytical solutions are usually restricted to relatively simple constitutive frameworks and one-dimensional (1-D) problems (i.e., simple shear), though two-dimensional (2-D) analyses exist \cite{anand1987,chen1999}.
Elasticity, inertia, and conduction are frequently omitted to permit tractable solutions \cite{molinari1986,molinari1987,molinari1988,shawki1994}.
Numerical studies enable more sophisticated constitutive theory and more complex initial-boundary conditions, but challenges arise for resolving band geometry, typically small relative to the discretization (i.e., mesh size). 
Phase-field methods offer another avenue toward regularized numerical solutions to problems of shear localization \cite{claytonPM2015,mcauliffe2015,xu2020,zeng2022}, deformation twinning \cite{claytonPHYSD2011}, and phase changes in ferrous metals under magnetic fields \cite{zeng2022b} or high pressure \cite{yao2024}. Extraction of localization susceptibility is less straightforward as model and problem complexity increase.

The current work advances the analytical approach of Molinari and Clifton \cite{molinari1986,molinari1987,molinari1988} to consider structural transitions as may occur under large shear, high temperature, or high pressure.  Transitions can induce a transformation shear, tangential to the band, and a transformation volume change from expansion or contraction normal to the band. The ubiquitous 1-D simple shear problem is augmented with pressure loading and compressibility due to a finite bulk modulus and density changes commensurate with phase transitions. The nonlinear analysis provides ranges of material properties and external fields for which full localization is admitted and, at least implicitly, the applied shear strain required. Effects of an external magnetic field on transformations pertinent to ferrous metals are incorporated. This field must be of low enough intensity that stresses of electromagnetic origin do not appreciably affect flow behavior.
No prior analytical solutions accounting for simultaneous phase transitions and shear bands in metals seem to exist despite frequent mentioning of their likely coexistence \cite{yan2021,moss1980,moss1981,molinari1988}.
Pressure-sensitive plastic flow and pressure-sensitive ductile failure have been incorporated in a few shear band studies for metals \cite{anand1987,raftenberg2001,voyiadjis2012}; such phenomena were also considered in a bifurcation analysis for geomaterials \cite{rudnicki1975}.

This paper is outlined as follows.
In Sec.~\ref{sec2}, a general 3-D constitutive framework and balance laws are summarized; these are
based on prior theoretical modeling of viscoplasticity and phase transitions in ferrous metals \cite{claytonCMT2022,claytonZAMM2024}.
In Sec.~\ref{sec3}, a boundary value problem combining simple shear with superposed pressure is posited; constitutive and governing equations are suitably reduced to address this problem.
In Sec.~\ref{sec4}, homogeneous solutions and localization conditions (i.e., inhomogeneous fields) for shear bands are derived, 
building on prior analytical work of Molinari and Clifton \cite{molinari1986,molinari1987,molinari1988} in rigid-viscoplastic, non-conducting solids to newly include effects of compressibility and structural transitions.
In Sec.~\ref{sec5}, a reduced-order continuum model for $\alpha \rightarrow \epsilon$ transformations in Fe and a low-carbon, high-strength Ni-Cr steel is formulated that includes pressure and shear effects, parameterized by results in Refs.~\cite{caspersen2004,lew2006,rittel2006,sadjadpour2015,hauver1976,hauver1979}. The $\alpha \rightarrow \gamma$ transformation is modeled in a similar context, whereby thermal effects dominate over pressure. The transformation theory is incorporated in the localization analysis, offering new insight into coupling among shear strength, pressure, temperature, phase transitions, and adiabatic shear bands in Fe and initially BCC steel. Lastly, conclusions are summarized in Sec.~\ref{sec6}.

\section{General 3-D constitutive framework and governing equations \label{sec2}}

\subsection{Material model}
Let ${\bf x} $ = $ {\bm \varphi } ({\bf X},t)$ 
be the spatial position at time $t$ of a material particle whose location in a fixed reference configuration is ${\bf X}$. Denote by ${\bf F}$, ${\bf F}^E$, ${\bf F}^\xi$, and ${\bf F}^P$ the total,
thermoelastic, transformation, and plastic deformation gradients;
none of the latter three necessarily obeys compatibility conditions for the gradient of a vector field \cite{claytonDGKC2014}. With $\nabla_0(\cdot)$ the gradient with respect to ${\bf X}$ and $J$ the Jacobian determinant, in coincident Cartesian coordinate systems on each configuration \cite{claytonNCM2011},
\begin{align}
\label{eq:defgrad}
{\bf F} = \nabla_0 {\bm \varphi} = {\bf F}^E {\bf F}^\xi {\bf F}^P, 
\qquad
J = \det {\bf F} = \det {\bf F}^E \det {\bf F}^\xi \det{\bf F}^P = J^E J^\xi J^P > 0.
\end{align}
Denote by $\xi ({\bf X},t)$ and $\chi({\bf X},t)$ dimensionless scalar internal state variables associated
with structural transformation and plastic deformation processes, respectively.
The former could be the volume or mass fraction of a crystallographic phase \cite{claytonCMT2022,levitas1998,boettger1997}, the twinned volume fraction \cite{kalidindi1998,claytonRSPA2009}, or a local measure of DRX such as normalized grain boundary density \cite{lieou2018}.  The latter is often a measure of dislocation density \cite{claytonJMPS2005,claytonNCM2011}.
Transformation deformation is a state function of $\xi$. Plastic flow is isochoric; typically small volume changes from dislocations \cite{wright2002,holder1969,horie1980,claytonJEMT2009,claytonQJMAM2014} are omitted per the usual assumptions in metal plasticity \cite{asaro1983,nematnasser2004}:
\begin{align}
\label{eq:FxiJp}
{\bf F}^\xi = {\bf F}^\xi(\xi), \qquad J^P = \det {\bf F}^P = 1, \qquad {\bf C}^E = ({\bf F}^E)^{\rm T}{\bf F}^E. 
\end{align}
Symmetric elastic deformation is ${\bf C}^E$. Absolute temperature is $\theta$. Entropy, magnetization, and free energy per unit reference volume are $\eta$, ${\bf M}_0$, and $\Psi$.  The latter is of form similar to Refs.~\cite{claytonCMT2022,claytonZAMM2024}:
\begin{align}
\label{eq:Psi}
\Psi = \Psi({\bf F}^E, {\bf M}_0, \theta, \xi, \chi) = \psi + J {\bf B} \cdot {\bf M},
\end{align}
with $\psi$ Helmholtz free energy, ${\bf B}$ magnetic induction, ${\bf M} = J^{-1} {\bf M}_0$, and in MKS units \cite{claytonNCM2011,maugin1988},
\begin{align}
\label{eq:BHM}
{\bf B} = \hat{\mu}_0 ( {\bf H} + {\bf M}).
\end{align}
Vacuum permeability is $\hat{\mu}_0$. Magnetic field ${\bf H}$ is equal among coexisting phases at ${\bf X}$ \cite{daniel2008}, whereas spatial magnetization ${\bf M}$ is a local volume average.

Helmholtz free energy per unit volume is the sum of four terms. Thermoelastic strain energy is $W$, thermal energy is $Q$, microstructure energy is $R$, and isotropic magnetostatic energy is $\Phi$:
\begin{align}
\label{eq:psi}
\psi({\bf C}^E,|{\bf H}|,\theta,\xi,\chi) = W({\bf C}^E,\theta) + Q(\theta,\xi) + R(\chi,\xi) + \Phi(|{\bf H}|, \xi).
\end{align}
Exercising proven, simplified versions of prior theory \cite{claytonCMT2022,claytonZAMM2024},
elastic constants, thermal expansion, and specific heat are equal among coexisting phases. Magnetostriction energy is omitted: strains associated with magnetostriction are on the order of $10^{-5}$ in ferrous metals of interest \cite{daniel2008,claytonCMT2022}, small compared to the yield strain.
Couplings with temperature in $R$ and $\Phi$ (e.g., $\theta$-dependence of the shear modulus and saturation magnetization) are secondary effects likewise omitted for simplicity.

Let $\theta_0$ be a reference temperature and $\Delta \theta = \theta - \theta_0$. The isothermal bulk modulus is $B_0$ with pressure derivative $B'_0$, volumetric thermal expansion is $A_0$, and shear modulus is $\mu$. Strain energy merges a logarithmic equation of state \cite{claytonIJES2014,claytonEML2015} with
a polyconvex deviatoric part from $\tilde{\bf C}^E$ \cite{balzani2006,claytonSYMM2023}:
\begin{align}
\label{eq:W}
& W = {\textstyle{\frac{1}{2}}} B_0 (\ln J^E)^2 [ 1 -   {\textstyle{\frac{1}{3}}} (B'_0 - 2) \ln J^E ]
- A_0 B_0 \Delta \theta  \ln J^E 
+  {\textstyle{\frac{1}{2}}} \mu ( {\rm tr} \, \tilde{\bf C}^E - 3), \\
\label{eq:CEtilde}
& \tilde{\bf C}^E = (J^E)^{-2/3} {\bf C}^E.
\end{align}
Lattice pressure $p^E$ and lattice deviatoric Cauchy stress $\tilde{ \bm \sigma}^E$ are
\begin{align}
\label{eq:lattp}
& p^E = - \frac{1}{J} \frac{\partial \psi}{\partial \ln J^E} = 
- \frac{B_0}{J} \{   \ln J^E  [ 1 - {\textstyle{\frac{1}{2}}}  (B'_0 - 2) \ln J^E] - A_0  \Delta \theta \},
\\
\label{eq:lattsig}
& \tilde{ \bm \sigma}^E = \frac{2}{J} {\bf F}^E \frac{ \partial  \psi }{\partial  \tilde{\bf C}^E}: \frac{\partial  \tilde{\bf C}^E}{\partial {\bf C}^E} ({\bf F}^E)^{\rm T} = \frac{\mu}{J} \tilde{\bf B}^E, \quad
\tilde{\bf B}^E = (J^E)^{-2/3} {\bf F}^E ({\bf F}^E)^{\rm T} - \frac{1}{3} {\rm tr} \, [ (J^E)^{-2/3} {\bf F}^E ({\bf F}^E)^{\rm T} ] {\bf 1}.
\end{align}

Let $c_V$ be specific heat per unit volume, 
$\lambda_T$ a latent heat \cite{turt2005,sadjadpour2015,claytonJDBM2021} with $\theta_T$ a corresponding transformation temperature, and $\psi_0$ a phase energy difference, all constants. Thermal energy is
\begin{align}
\label{eq:Q}
Q = - c_V [ \theta \ln (\theta / \theta_0) - \Delta \theta] - \xi [(\lambda_T / \theta_T) (\theta - \theta_T) - \psi_0].
\end{align}
The term proportional to $\xi$ models phase transitions, $\xi$ and $1-\xi$ then being volume fractions of  transformed and parent phases. This term can be omitted when $\xi$ measures twins or DRX, whereby effects of $\psi_0$ are embedded in $R$. Microstructure energy $R$ is deferred to specific forms for metals of interest in Secs.~\ref{sec3}.2 and \ref{sec5}.1.

Regarding $\Phi$, attention is restricted here to discrete, saturated ferromagnetic states or non-magnetic states.  For ferromagnetic phases of isotropic ferrous metals (e.g., $\alpha$-Fe), net magnetization vanishes in the absence of an aligning field of sufficient magnitude $H_S$ since domains are randomly oriented at a point ${\bf X}$ \cite{landau1960,maugin1988,claytonCMT2022}.  For fields approximately exceeding $H_S$, magnetization is saturated at constant magnitude $| {\bm M}_0| = M_S$. Magnetization and magnetic field vectors are collinear for isotropy \cite{landau1960,claytonZAMM2024}. Magnetostatic energy and average magnetization magnitude $M_0 = | {\bf M}_0 |$ are 
\begin{align}
\label{eq:Mfield}
\Phi = - \hat{\mu}_0 | {\bf H} | M_0, \quad
M_0 = (1 - \xi) M_0^{(0)} + \xi M_0^{(1)}, \quad
M_0^{(\cdot)} = \begin{cases}
& M_S^{(\cdot)} \quad ( |{\bf H}| \gtrsim H_S^{(\cdot)} ), \\
& 0 \qquad \,  (|{\bf H}| \lesssim H_S^{(\cdot)} ).
\end{cases}
\end{align}
This expression, which agrees with other works \cite{curran1971,james1990,daniel2008}, assumes $\xi$ is the transformed fraction; superscripts $(\cdot)^{(0)}$ and
$(\cdot)^{(1)}$ label the parent and transformed phase.
The present coarse-grained theory, like those of Refs.~\cite{daniel2008,james1990} and Ch.~3 of Ref.~\cite{maugin1988} omits explicit exchange effects, magnetization gradients in the free energy, and a separate angular momentum balance for electronic spin. Such microscopic phenomena have been addressed in other sophisticated theories \cite{maugin1972a,claytonMMS2022}.

\subsection{Balance laws}
Attention is restricted to quasi-magnetostatics \cite{maugin1988,claytonCMT2022}, further in the theoretical limit of negligible electric current for the problem of interest.  For isotropy, ${\bf B} || {\bf M}$ so Cauchy stress ${\bm \sigma}$ is symmetric \cite{maugin1988,claytonCMT2022}. Let $\rho$ and $\rho_0$ be spatial and referential mass densities and ${\bm \upsilon} = \dot{\bf x}$ be particle velocity.
Continuum balances of mass and momentum and the two pertinent Maxwell's equations are
\begin{align}
\label{eq:balances}
\rho_0 = \rho J, \quad \nabla \cdot {\bm \sigma} + {\bf b} = \rho \dot{\bm \upsilon}, \quad {\bm \sigma} = {\bm \sigma}^{\rm T}; \qquad
\nabla \cdot {\bf B} = 0, \quad \nabla \times {\bf H} = {\bf 0}.
\end{align}
The gradient with respect to ${\bf x}$ is $\nabla(\cdot)$, and the body force per unit spatial volume is ${\bf b}$. Electromagnetic fields contribute terms of order $ \hat{\mu}_0^{-1} | {\bf B}|^2$ to the total stress tensor 
\cite{maugin1988,claytonCMT2022}. The yield stress of ferrous metals of interest is on the order of 0.5 GPa.
Attention is restricted to fields for which $ \hat{\mu}_0^{-1} | {\bf B}|^2$ is significantly smaller than this,
noting $\hat{\mu}_0 M_S \approx 2$ T in Fe. An upper limit is then $ \hat{\mu}_0 | {\bf H} | \lesssim 5$ T, by which
electromagnetic stress and body force terms are safely omitted, as in Refs.~\cite{curran1971,barker1974}. Cauchy stress becomes the sum of \eqref{eq:lattp} and \eqref{eq:lattsig}, with $p$ the Cauchy pressure:
\begin{align}
\label{eq:Cauchy}
{\bm \sigma} \approx -p^E {\bf 1} + \tilde{\bm \sigma}^E, \qquad p = - {\textstyle{\frac{1}{3}}} {\rm tr} \, {\bm \sigma} \approx p^E.
\end{align}
The first Piola-Kirchhoff stress ${\bf P}$, elastic second Piola-Kirchhoff stress ${\bf S}$, Mandel stress $\bar{\bf S}$, and Gibbs free energy per unit reference volume ${\mathbb G}$ are \cite{turt2005,claytonCMT2022,claytonZAMM2024}
\begin{align}
\label{eq:PK1}
{\bf P} = J {\bm \sigma} {\bf F}^{- \rm{T}}, \quad
{\bf S} = J ({\bf F}^E)^{-1} {\bm \sigma} ({\bf F}^E)^{- \rm{T}} = 2 \, {\partial \psi}/{\partial {\bf C}^E}, \quad
\bar{ \bf S} = {\bf C}^E {\bf S}, \quad
\mathbb{G} = \psi - {\bf P}:{\bf F}.
\end{align}

Assume isotropic thermal conductivity $\kappa = \rm{constant}$ with Fourier conduction 
${\bf q} = - \kappa \nabla_0 \theta$.  Invoking the model of \eqref{eq:defgrad}--\eqref{eq:Mfield},
the local balance of energy in the form of a temperature rate equation, in the absence of point heat sources, and the entropy inequality in reduced form are derived as follows \cite{claytonCMT2022,claytonZAMM2024}:
\begin{align}
\label{eq:1stlaw}
& c_V \dot{\theta} = [( {\bf F}^\xi)^{\rm T} \bar{\bf S} ({\bf F}^\xi)^{\rm -T}]:{\bf L}^P - \left[ \Delta^* \mathbb{G} + \frac{ \lambda_T }{\theta_T} \theta \right] \dot{\xi} -  \frac{\theta A_0 B_0}{J^E} \dot{J}^E -
\frac {\partial R }{ \partial \chi} \dot{\chi} + \kappa \nabla_0^2 \theta,
\\
\label{eq:2ndlaw}
&  [( {\bf F}^\xi)^{\rm T} \bar{\bf S} ({\bf F}^\xi)^{\rm -T}]:{\bf L}^P -  [\Delta^* \mathbb{G}] \dot{\xi}
- \frac {\partial R }{ \partial \chi} \dot{\chi} + \frac{ \kappa}{\theta}  | \nabla_0 \theta |^2 \geq 0;
\qquad {\bf L}^P = \dot{\bf F}^P {\bf F}^{P -1},
\end{align}
with ${\bf L}^P$ the plastic velocity gradient and $\Delta^* \mathbb{G}$ the Gibbs driving force for structural transitions:
\begin{align}
\label{eq:Gibbs}
\Delta^* \mathbb{G} = \partial \psi / \partial \xi - {\bf P}: (\partial {\bf F}/\partial \xi)
= \partial (Q + R + \Phi)/\partial \xi - [\bar{\bf S} ( {\bf F}^\xi)^{- {\rm T}}]: (\partial {\bf F}^\xi / \partial \xi).
\end{align}
Derivatives of $Q$, $R$, and $\Phi$ furnish thermochemical, microstructure, and magnetic driving forces, while the rightmost term is a mechanical driving force. Entropy obeys $\eta = - \partial \psi / \partial \theta$ as usual.

\section{Boundary value problem and model reduction \label{sec3}}

\subsection{Pressurized simple shear}

\begin{figure}
\centering
\includegraphics[width=0.7\textwidth]{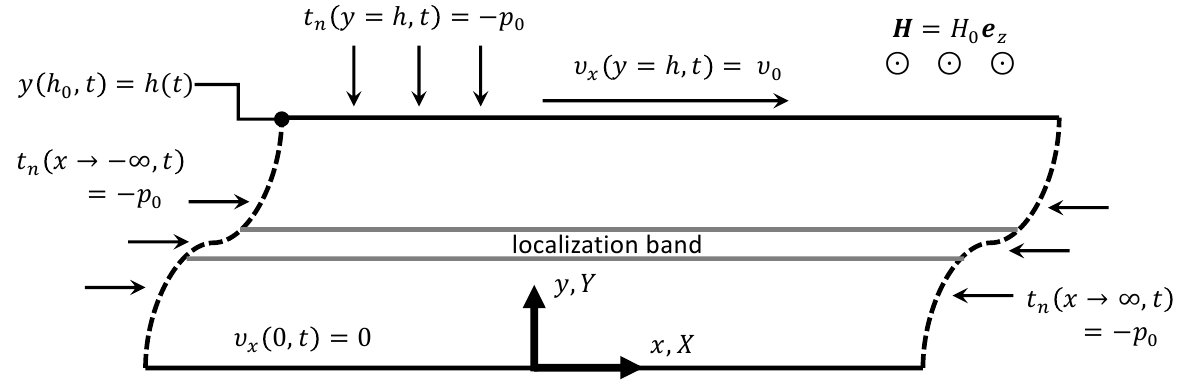}
\caption{Boundary value problem: tangential velocity at $y=h$ is $\upsilon_0$, applied normal traction $t_n$ on all boundaries is $p_0$, including $t_n (z \rightarrow \pm \infty,t) = -p_0$ (not shown). Remote magnetic field is of constant magnitude $H_0$, aligned tangential to plane $y = h$ and orthogonal to $\upsilon_0$. Slab is infinite in $X,x,Z,z$ directions; initial (current) height of slab is $h_0$ ($h$). 
Material is thermally insulated: $\partial \theta / \partial y = 0$ at $y = 0$ and at $y = h$ for $t \geq 0^+$.
Initial state variable $\chi(y,t = 0)$ and temperature $\theta(y,t=0)$ can vary modestly within $y \in (0,h)$.
\label{fig1}}
\end{figure}

Geometry and boundary conditions for the problem analyzed henceforth are shown in Fig.~\ref{fig1}.
Cartesian reference and spatial coordinates are $(X,Y,Z)$ and $(x,y,z)$.
A slab of material of initial height $h_0$ and current height $h(t)$ is infinitely extended in $X,x$- and $Z,z$- directions. Mixed mechanical boundary conditions referred to the spatial frame are used. A constant velocity of $\upsilon_x = \upsilon_0$ is applied at $y = h$, and $\upsilon_x = 0$ at $y=0$. The normal traction on all boundaries (finite and infinite) for $t \geq 0^+$ is a constant pressure field $t_n = - p_0$.  
Tangential traction due to $\sigma_{xy}$ for any cross section $x = \text{constant}$ exists but is not shown in Fig.~\ref{fig1}. Taking $(X,Y,Z)$ to be coordinates \textit{minus} application of pressure (e.g., corresponding to a stress-free reference state), initial height $h_0$ will differ from the $Y$ value of the planar boundary of the unstressed material due to compressibility: $h_0$ depends on $p_0$.
Initial pressure $p_0$ is applied slowly enough to be considered isothermal in an average sense and quasi-static. Subsequently for $t \geq 0^+$, boundaries $y=0$ and $y=h$ are thermally insulated.
Remote magnetic field is of constant magnitude $H_0 \geq 0 $ for $t \geq 0^+$ and is aligned parallel to $z,Z$.  

Under shearing for $t > 0^+$, $h$ may displace from $h_0$ to maintain constant $p_0$ at $y= h$.  This is  expected if the material undergoes a volume-changing structural transformation under shear.
Velocity $\upsilon_0$ is time-independent for $t \geq 0^+$: the necessarily inertial accelerative ramp-up from a resting state to, say, a uniform initial velocity gradient $(\partial \upsilon_x / \partial y)|_{t = 0^+}$, is not analyzed.
When $p_0 = 0$ and $H_0 = 0$, the boundary value problem in Fig.~\ref{fig1} and aforementioned assumptions degenerate to those standard for shear band analysis \cite{wright2002,molinari1986,molinari1987,molinari1988,fressengeas1987,wright1994,shawki1994,langer2017,le2018}.  For $t \geq 0^+$, the problem with infinite $x,z$ boundaries is 1-D in the sense that spatial fields depend at most only on $(y,t)$, viz., 
\begin{align}
\label{eq:1D}
{\bm \upsilon} = {\bm \upsilon}(y,t), \quad {\bf F} = {\bf F}(y,t), \quad {\bm \sigma } = {\bm \sigma}(y,t), \quad \theta = \theta(y,t), \quad {\bf H} = {\bf H}(y,t).\
\end{align}

The orientation of any single localization band, within which ${\bf F}$ differs radically from most of the surrounding material, is restricted to that of Fig.~\ref{fig1}, but multiple parallel bands are not excluded. A 2-D perturbation analysis \cite{anand1987} suggested two band orientations are equally favorable, and that pressure dependency of flow stress affects the angle between these orientations. Such more complex geometries, due to allowance of linear pressure perturbations in 2-D, are excluded in the present 1-D setting, as elsewhere \cite{wright2002,wright1994,molinari1986,molinari1987}.
Properties and initial temperature need not be uniform in $y$, but $x$ and $z$ variations are excluded. Variations in initial strength (e.g., from $\chi$ perturbations) and temperature \cite{molinari1988} can trigger localization at critical $y$ locations. 
The former are a pragmatic substitute to $z$-thickness changes from manufacturing flaws in torsional Kolsky bar specimens \cite{molinari1986,molinari1987}. 

The reference configuration spanned by $(X,Y,Z)$ is defined more precisely as follows.
This configuration is a model construction useful for analysis and need not be attained in experimental practice. Pressure $p_0$ and field $H_0$ are removed instantaneously at fixed $\theta$ and $\xi$ from an unsheared state, before $\upsilon_x = \upsilon_0$ is applied. In this undeformed and unloaded state, by construction, ${\bf F} = {\bf F}^E = {\bf F}^\xi = {\bf F}^P = {\bf 1}$, ${\bm \sigma} = {\bf 0}$,
and $\xi = \xi_0 = \text{constant}$.  Almost everywhere, $\theta \approx \theta_0 = \text{constant}$ and 
$\chi \approx \chi_0 = \text{constant}$.  
Initial perturbations in $\theta$ and $\chi$ along $Y$ are allowed, with magnitudes small enough that initial perturbations in $J$, $p$, and $\xi$ can be omitted (e.g., $\theta$ variations not exceeding 10 to 100 K given $A_0$).
The uniform value $\xi_0$ corresponds to an equilibrium state at $p = p_0$, $\theta = \theta_0$, and $\chi = \chi_0$. For the case of phase transitions, $\xi_0$ is the volume fraction of the second phase
and $1-\xi_0$ the volume fraction of the parent.  Referential mass density is $\rho_0 = (1-\xi_0) \rho_0^{(0)} + \xi_0 \rho_0^{(1)} = \text{constant}$.

Nonzero components of ${\bf F}(y(Y,t),t)$ are $F_{xX}$, $F_{yY}$, $F_{zZ}$ and $F_{xY}$.
For a generic differentiable function $f = f(y(Y,t),t)$, it follows that $\partial f / \partial Y =  (\partial f / \partial y) F_{yY} $,
and $\partial f / \partial X = \partial f / \partial Z = 0$.
From the second of \eqref{eq:1D}, compatibility conditions $\nabla_0 \times {\bf F} = {\bf 0}$ \cite{claytonDGKC2014} require
\begin{align}
\label{eq:compat1D}
\partial F_{xX} / \partial Y = \partial F_{xY} / \partial X = 0, \quad
\partial F_{zZ} / \partial Y = \partial F_{zY} / \partial Z = 0.
\end{align}
Nonzero magnetic components are $H_z$, $M_z$, and $B_z = \hat{\mu}_0 (H_z + M_z)$.
Jump conditions \cite{maugin1988,claytonCMT2022} require $H_z$ continuity
across $y=0$ and $y=h$.  The last of  \eqref{eq:balances} and \eqref{eq:1D} lead to 
$\partial H_z / \partial y = \partial H_y / \partial z = 0$.
The trivial solution to Maxwell's equations for $t \geq 0^+$ follows as
\begin{align}
\label{eq:maxsoln}
H_z(y,t) = H_0 = \text{constant} \quad \Rightarrow \quad B_z(y,t) = \hat{\mu}_0 [H_0 + M_z(y,t)].
\end{align}
Strictly, if $M_z$ varies with $t$ as in a ferromagnetic transition, $B_z$ will vary with $t$, inducing a transient electric field and current. In the limits of quasi-magnetostatics or null electric conductivity, electric current is omitted in \eqref{eq:balances}, \eqref{eq:1stlaw}, and \eqref{eq:2ndlaw}.

For the isotropic constitutive model, nonzero stress components are $\sigma_{xx}$, $\sigma_{yy}$, $\sigma_{zz}$, and $\sigma_{xy} = \sigma_{yx}$, and heat flux is limited to $q_Y$. Taking ${\bf b} = {\bf 0}$, the second of \eqref{eq:balances} and heat conduction energy rate in \eqref{eq:1stlaw}, with \eqref{eq:1D}, are, respectively,
\begin{align}
\label{eq:mombal1D}
\frac{\partial \sigma_{xy}}{ \partial y} = \rho \dot{\upsilon}_x,
\quad  
\frac{\partial \sigma_{yy}}{ \partial y }= \rho \dot{\upsilon}_y;
\qquad
q_Y = -  \kappa F_{yY} \frac{\partial \theta} {\partial y},
\quad
\kappa \nabla_0^2 {\theta} = \kappa  F_{yY} \frac{\partial}{\partial y} \left( F_{yY} \frac {\partial \theta}{\partial y} \right).
\end{align}
For the more restrictive settings of null inertia (finite mass $\rho > 0$ with $\dot{\bm \upsilon} \rightarrow {\bf 0}$) and locally adiabatic conditions (non-conductor $\kappa \rightarrow 0$ with  $|\partial \theta /
\partial y| >0 $ admitted), \eqref{eq:mombal1D} degenerates to, respectively,
\begin{align}
\label{eq:statics}
\frac{\partial \sigma_{xy}}{\partial y} = 0, \, \, \frac{\partial \sigma_{yy}}{\partial y} = 0
\, \, \Rightarrow \, \, \sigma_{xy} = \sigma_{xy}(t), \, \, \sigma_{yy} = \sigma_{yy}(t);
\quad
\kappa = 0 \, \Rightarrow \, q_Y = 0, \quad [\theta = \theta(y,t)].
\end{align}

Analysis of the problem proceeds as follows. Particular forms of ${\bm \varphi}$, ${\bf F}^E$, ${\bf F}^\xi$, and ${\bf F}^P$ are postulated a priori, based on the geometry and physics of the problem in Fig.~\ref{fig1}.
Then, given the full 3-D constitutive theory of Sec.~\ref{sec2} and these defined kinematics, assumptions (e.g., linearization or omission of higher-order terms) needed for satisfaction of \eqref{eq:mombal1D} or \eqref{eq:statics} are deduced.
A reduced-order version of the 3-D constitutive framework, including slip and transition kinetics motivated from prior work \cite{claytonCMT2022,claytonZAMM2024} and consistent with these assumptions, but simple enough to accommodate implicit analytical solutions, is posited in Sec.~\ref{sec3}.2. The approach yields a closed set of governing equations in Sec.~\ref{sec3}.3.

Kinematic variables in \eqref{eq:defgrad} are postulated as follows. Recall that the reference configuration is unstressed: ${\bf F}^E$ includes the isothermal elastic volume change resulting from initial pressurization by $p_0$. Motion ${\bf x} = {\bm \varphi}({\bf X},t)$ and all deformation gradient terms in $3 \times 3$ matrix form are
\begin{align}
\label{eq:varphi}
& x = x(X,Y,t), \qquad y = y(Y,t), \qquad z = z(Z,t); \qquad \gamma = J^\xi \gamma^E + \gamma^\xi + \gamma^P;
\\ 
\label{eq:FE}
& [{\bf F}^E(Y,t)] = (J^{E}(t))^{1/3} 
\begin{bmatrix}
& 1 &  \gamma^E(Y,t) & 0 \\
& 0 & 1 & 0 \\
& 0 & 0 & 1
\end{bmatrix},
\qquad
 [{\bf F}^\xi (Y,t)] = 
\begin{bmatrix}
& 1 &  \gamma^\xi(Y,t) & 0 \\
& 0 & J^\xi(Y,t) & 0 \\
& 0 & 0 & 1
\end{bmatrix},
\\
\label{eq:FP}
& [{\bf F}^P (Y,t)] = 
\begin{bmatrix}
& 1 &  \gamma^P(Y,t) & 0 \\
& 0 & 1 & 0 \\
& 0 & 0 & 1
\end{bmatrix},
\qquad
 [{\bf F} (Y,t)] = 
\begin{bmatrix}
& 1 &  \gamma(Y,t) & 0 \\
& 0 & J^\xi(Y,t) & 0 \\
& 0 & 0 & 1
\end{bmatrix}
 (J^{E}(t))^{1/3} .
\end{align}
Thermoelastic deformation consists of spatially homogeneous volume change $J^E$ and simple
 shear $\gamma^E$.  
 Total shear \textit{after} thermoelastic volume change in the last of \eqref{eq:FP} is $\gamma$. Plastic deformation is the simple shear $\gamma^P$.
Structural transitions include simple shear $\gamma^\xi$ and strain normal to the band,
with volume change $J^\xi$, each interpolated via $\xi(Y,t)$ \cite{claytonCMT2022}:
\begin{align}
\label{eq:Fxi}
\gamma^\xi = \gamma^\xi_0 (\xi - \xi_0), \qquad J^\xi = 1 + \delta^\xi_0 (\xi -\xi_0); 
\qquad \delta^\xi_0 = \rho^{(0)}/ \rho^{(1)} - 1.
\end{align}
Both $\gamma^\xi_0$ and $\delta^\xi_0$ are material constants; either or both can be zero, positive, or negative.
Note ${\bf F}$ of \eqref{eq:FP} obeys \eqref{eq:compat1D}.
From ${\bf F}^E$ in \eqref{eq:FE}, the Cauchy stress of \eqref{eq:lattp}, \eqref{eq:lattsig}, and \eqref{eq:Cauchy} is
\begin{align}
\label{eq:Cauchymatrix}
& [ {\bm \sigma} ] = 
\frac{1}{J^E J^\xi}
\begin{bmatrix}
& -\tilde{p} + \frac{2}{3} \mu (\gamma^E)^2 & \mu \gamma^E & 0 \\
&  \mu \gamma^E  &  -\tilde{p} - \frac{1}{3} \mu (\gamma^E)^2 & 0 \\
& 0 & 0 & -\tilde{p} - \frac{1}{3} \mu (\gamma^E)^2 
\end{bmatrix},
\\
\label{eq:barp}
& \tilde{p}(J^E,\theta) = - {B_0} \{   \ln J^E  [ 1 - {\textstyle{\frac{1}{2}}}  (B'_0 - 2) \ln J^E] - A_0  (\theta - \theta_0) \}.
\end{align}

Some approximations are now introduced. Elastic shear is assumed small such that terms of $O((\gamma^E)^2)$ are negligible. 
Thermal expansion in $\tilde{p}$ is omitted (e.g., $A_0 \rightarrow 0$), typical in analysis of shear bands \cite{molinari1986,molinari1987,wright2002,langer2017,le2018}.
Lastly, terms of $O(\delta^\xi_0)$ are omitted in the prefactor of \eqref{eq:Cauchymatrix}. For example of the consequence of the last approximation, for the $\alpha \rightarrow \epsilon$ transition 
in Fe, $\delta^\xi_0 = -0.0512$ \cite{boettger1997,claytonZAMM2024}, so underestimation of stress in the $\epsilon$ phase would be $ \approx 5$\%. From the boundary conditions of Fig.~\ref{fig1}, $\sigma_{xx} \approx \sigma_{yy} \approx \sigma_{zz} \approx -p_0 = \text{constant}$, so
\begin{align}
\label{eq:stressred}
& [ {\bm \sigma} ] \approx
\begin{bmatrix}
& -{p}_0 & \mu \gamma^E/J^E & 0 \\
&  \mu \gamma^E/J^E  &  -{p}_0 & 0 \\
& 0 & 0 & -{p}_0 
\end{bmatrix}, \quad
J^E \approx \underset{\hat{J}^E > 0}{\arg} \{ p_0 = - {B_0}   \frac{ \ln \hat{J}^E}{\hat{J}^E}  [ 1 - {\textstyle{\frac{1}{2}}}  (B'_0 - 2) \ln \hat{J}^E] \}.
\end{align}
Since $p_0 = \text{constant}$, $J^E(p_0) = \text{constant}$ for $t \geq 0^+$.
Since $\sigma_{yy} \approx -p_0$, the second of \eqref{eq:statics} is satisfied.
The first of \eqref{eq:mombal1D} becomes $ \partial \tau / \partial y \approx \partial ( \mu \gamma^E) / \partial y = J^E \rho \dot{\upsilon}_x$.
When $\tau (t)= J^E \sigma_{xy}(t)$ is independent of $y$, then inertial force is negligible.
On the other hand, if higher-order terms are retained in \eqref{eq:Cauchymatrix}, then
inertial forces, or departures from the kinematic ansatz of \eqref{eq:varphi}--\eqref{eq:FP}, would be expected.

From \eqref{eq:FE}--\eqref{eq:stressred}, dissipation from plastic and
structure deformations in \eqref{eq:1stlaw} and \eqref{eq:2ndlaw} is
\begin{align}
\label{eq:pdiss}
& [( {\bf F}^\xi)^{\rm T} \bar{\bf S} ({\bf F}^\xi)^{\rm -T}]:{\bf L}^P = \tau \dot{\gamma}^P ; \qquad
\tau = J^E \sigma_{xy}  = \mu \gamma^E / J^\xi \approx \mu \gamma^E, \qquad \bar{p}_0 = J^E p_0 ;
\\
\label{eq:xidiss}
&  [\bar{\bf S} ( {\bf F}^\xi)^{- {\rm T}}]: (\partial {\bf F}^\xi / \partial \xi) \dot{\xi} =
[\tau \gamma_0^\xi - J^E \{p -  {\textstyle{\frac{2}{3}}} J^{-1} \mu (\gamma^E)^2 \} \delta^\xi_0 ] \dot{\xi}
\approx [\tau \gamma^\xi_0 - \bar{p}_0 \delta^\xi_0] \dot{\xi}.
\end{align}
The mechanical driving force for both shears $\gamma^P$ and $\gamma^\xi$ is $\tau$; that for volume change $J^\xi$ is $\bar{p}_0$.
Taking the parent phase to be ferromagnetic (e.g., $\alpha$-Fe) and the product essentially nonmagnetic (e.g., $\epsilon$- or $\gamma$-Fe), $\Phi = - \hat{\mu}_0 H_0 M_0$ with $M_0 = (1-\xi) M_S$ for $H_0 \gtrsim H_S$ in \eqref{eq:Mfield}. Gibbs driving force \eqref{eq:Gibbs} becomes
\begin{align}
\label{eq:Gibbs1D}
-\Delta^* \mathbb{G} (\gamma^E,\theta,\xi,\chi; p_0, H_0) \approx \tau \gamma^\xi_0 - \bar{p}_0 \delta^\xi_0 
+ (\lambda_T / \theta_T) (\theta - \theta_T) - \psi_0
- \partial R / \partial \xi
- \hat{\mu}_0 H_0 M_S,
\end{align}
where the rightmost term vanishes for $H_0 \lesssim H_S$.  The energy balance in \eqref{eq:1stlaw} is now,
with generally transient Taylor-Quinney factor $\beta(y(Y,t),t)$ \cite{claytonJMPS2005,claytonCOMPB2009,lieou2021}, $\theta = \theta(y(Y,t),t)$ and $\dot{J}^E(p_0) = 0$, 
\begin{align}
\label{eq:1stbeta}
c_V \dot{\theta} = \beta \tau \dot{\gamma}^P - [ \Delta^* \mathbb{G} + ( \lambda_T /{\theta_T}) \theta  ]\dot{\xi} + \kappa(\partial^2 \theta / \partial Y^2),
\quad \beta = [\tau \dot{\gamma}^P - (\partial R / \partial \chi) \dot{\chi} ] / (\tau \dot{\gamma}^P).
\end{align}
Denoting an arbitrary rigid body rotation by ${\bf R}_0$, energy $W$ and stress ${\bm \sigma}$ are unaffected by transformations of the form ${\bf F}^E \rightarrow {\bf F}^E {\bf R}_0^{\rm T}$ with ${\bf F}^\xi {\bf F}^P \rightarrow {\bf R}_0 {\bf F}^\xi {\bf F}^P$. However, different forms of ${\bf R}_0$ can affect partitioning of dissipation among inelastic deformation mechanisms: the present choice ${\bf R}_0 = {\bf 1}$, motivated by  physics peculiar to Fig.~\ref{fig1} and metal transformation-plasticity, is not arbitrary \cite{bammann1987,claytonNCM2011}.

\subsection{Reduced-order constitutive model}
Following Molinari and Clifton \cite{molinari1986,molinari1987,molinari1988}, further assumptions
are invoked to enable analytical solutions.
The material is idealized as a non-conductor of both electricity and heat, with $\kappa \rightarrow 0$ so as in \eqref{eq:statics}, $\partial \theta / \partial y$ need not vanish.
Inertial forces are omitted, so other conditions in \eqref{eq:statics} likewise hold.
Invoking \eqref{eq:stressred}, $\sigma_{xx} \approx \sigma_{yy} \approx \sigma_{zz} \approx -p_0 = \text{constant}$, and the first of \eqref{eq:statics} gives $\partial \tau / \partial y \approx 0$.

Heat conduction inhibits localization at low strain rates, and inertia inhibits localization at very high
rates \cite{wright1996}. As explained by Molinari and Clifton \cite{molinari1986,molinari1987,molinari1988}
in the context of complete numerical solutions \cite{shawki1983,wright1985,wright1994},
non-conducting and non-inertial conditions enable accurate predictions of applied strains needed for localization in steels in dynamic torsion experiments, for example those from which
viscoplastic properties are extracted later in Sec.~\ref{sec5}.2 at rates on the order of $10^{3}$--$10^4$/s \cite{gray1994,fellows2001,fellows2001b}. Conduction is also omitted in Ref.~\cite{lieou2019} that models shear bands in steel, like many if not most contemporary computational approaches, usually for numerical efficiency.
In the present analysis, the only regularization mechanism is strain-rate sensitivity \cite{needleman1988}. 

Another assumption \cite{molinari1986,molinari1987,molinari1988,wright2002,shawki1994,anand1987} is
that of rigid viscoplasticity, whereby $\tau/\mu \rightarrow 0$ but $\tau$ remains finite.
In this limit, $\gamma^E \rightarrow 0$. For Fe and steel, $\gamma^E$ at yield is on the order of 0.01, two orders of magnitude smaller than average strain at localization in experiments \cite{fellows2001,fellows2001b} and numerical
results of Sec.~\ref{sec5}.2.
Lastly, in conjunction with rigid-viscoplastic response, the usual assumption $\beta \tau \dot{\gamma}^P \rightarrow \beta_0 \tau \dot{\gamma}$
with $\beta_0 \in (0,1] = \text{constant}$ \cite{molinari1986,molinari1987,molinari1988,shawki1994} is used.
With $J^\xi$ on the order of unity, shear strain, shear strain rate, and the energy balance of \eqref{eq:1stbeta} reduce to
\begin{align}
\label{eq:redmodel}
\gamma \approx \gamma^P + \gamma^\xi \geq 0, \quad \dot{\gamma} \approx \dot{\gamma}^P + \gamma_0^\xi \dot{\xi} \geq 0; \qquad
c_V \dot{\theta} \approx \beta_0 \tau \dot{\gamma} - [\Delta^*\mathbb{G} + (\lambda_T/\theta_T) \theta] \dot{\xi}.
\end{align}
Note that $\beta_0$ implicitly embodies some contributions from $\chi$, $\xi$,  $\dot{\chi}$, and $\dot{\xi}$. 
However, $\beta_0$ does not contain the majority of dissipation or energy storage from
structure transformations such as phase changes or DRX, both of which should be exothermic
for ferrous metals studied later. Rather, the term $-\Delta^* \mathbb{G} \cdot \dot{\xi}$ 
in \eqref{eq:redmodel} accounts for these physics, which should raise the \textit{apparent} Taylor-Quinney factor as measured by temperature rise in experiments \cite{rittel2006,sadjadpour2015}.
If $\gamma^\xi_0 < 0$, shearing from slip must outpace that from transformation. If $\gamma^\xi_0 > 0$, then $\gamma^P < 0$ is not impossible.

In the absence of explicitly resolved elastic shear strain, a viscoplastic flow rule \cite{claytonNCM2011,claytonJDBM2021,nematnasser2004} relates $\dot{\gamma}$ to Kirchhoff-type shear stress $\tau = J^E \sigma_{xy} \geq 0$:
\begin{equation}
\label{eq:flowrule}
\dot{\gamma} = \dot{\gamma_0} \, ( \tau / g)^{1/m}, \quad
g = g(\xi,\chi,\theta; p_0) = g_Y (\xi, p_0) h (\chi) \lambda (\theta); \qquad \beta \tau \dot {\gamma}^P \rightarrow
\beta_0 \tau \dot{\gamma} \geq 0,
\end{equation}
where $\dot{\gamma}_0 = \text{constant} > 0$ and $m = \text{constant} > 0$ are a reference strain rate and rate sensitivity. Flow stress $g > 0$ includes phase- and pressure-dependent yield stress $g_Y > 0$, strain hardening (or softening) function $h > 0$, and thermal softening (or hardening) function $\lambda > 0$. For monotonically increasing $\gamma$, $\chi$ associated with dislocations is defined as follows, with initial values close to unity:
\begin{align}
\label{eq:chi}
\chi(y,t) = \chi_0(Y(y,t)) + \gamma(y,t), \qquad \chi_0 \approx 1.
\end{align}
For arbitrary strain histories outside the current scope, $\gamma$ or $\gamma^P$ are not always appropriate internal state variables \cite{claytonNCM2011}. A mixture rule gives $g_Y$ \cite{claytonJDBM2021,claytonZAMM2024}, and $h$ and $\lambda$ are power-law forms \cite{molinari1986,molinari1987,fressengeas1987}:
\begin{align}
\label{eq:taulaw}
g_Y = g_0 ( 1 - \alpha_0 \xi), \quad \alpha_0 = 1- g_1/g_0 \in (-\infty,1); \quad
h = \chi_0 (1 + \gamma / \gamma_0)^n, \quad \lambda =  ( \theta / \theta_0)^\nu.
\end{align}
Here, $g_0(p_0) > 0$ and $g_1 (p_0) >0$ are strengths of parent and product approximated with the same pressure scaling; $\alpha_0$, $\gamma_0$, $n$, and $\nu$ are dimensionless constants. Inverting \eqref{eq:flowrule}, the flow stress $\tau$ is 
\begin{align}
\label{eq:flowstress}
\tau(\gamma(y,t), \dot{\gamma}(y,t), \xi(y,t),\theta(y,t); \chi_0(y),p_0) = g_0(p_0) \chi_0 ( 1 - \alpha_0 \xi) (1 + \gamma / \gamma_0)^n (\theta / \theta_0)^\nu (\dot{\gamma}/ \dot{\gamma}_0)^m.
\end{align}

A pragmatic assumption for some, but not all, analysis steps of localization in ferrous metals of mild strain-rate sensitivity (i.e., $m$ small relative to unity), deformed in simple shear or torsion, replaces the local strain rate in \eqref{eq:flowstress} with its constant average $\dot{\bar{\gamma}}$  \cite{molinari1986,molinari1987}, as justified in Ref.~\cite{molinari1986}: 
\begin{align}
\label{eq:flowstress2}
\tau(\gamma,\xi,\theta; \dot{\bar{\gamma}}, \chi_0, p_0) \approx g_0(p_0) \chi_0 ( 1 - \alpha_0 \xi) (1 + \gamma / \gamma_0)^n (\theta / \theta_0)^\nu (\dot{\bar{\gamma}}/ \dot{\gamma}_0)^m,
\qquad \dot{\bar{\gamma}} = \upsilon_0 / h_0.
\end{align}
Expression \eqref{eq:flowstress} is favored over \eqref{eq:flowstress2} so is used later in the momentum balance. The latter \eqref{eq:flowstress2} is used sparingly, as necessary, for estimating dissipative contributions to the energy balance \cite{molinari1986,molinari1987} and structure transition kinetics when local strain rates are unknown (e.g., analytical solutions).

A structure transition kinetic model originally used for phase transitions in Fe \cite{boettger1997,claytonCMT2022} and steels \cite{lloyd2022,claytonZAMM2024,claytonTR2024} is adapted to evolve $\xi(y,t)$.
Let $\d \xi = \dot{\xi} \d t$. Then non-negative dissipation is ensured by $-\Delta^* \mathbb{G} \, \d \xi \geq 0$.
Restricting the range of the order parameter to $\xi \in [0,1]$ and considering only forward transformations $\d \xi \geq 0$, 
define the dimensionless driving force $\mathbb{F} = - \Delta^* \mathbb{G}/\beta_F$ and
introduce state-dependent energy barriers $\alpha_F$ and $\beta_F > 0$.
Prescribe $\d \xi = \d \mathbb{F}$ for $\mathbb{F} > \alpha_F / \beta_F$ and $\d \xi = 0$ 
for $\mathbb{F} \leq \alpha_F / \beta_F$, automatically satisfying  $-\Delta^* \mathbb{G} \d \xi \geq 0$
if $\alpha_F \geq 0$\footnote{If $\alpha_F < 0$, 
dissipation from plastic work, microstructure (e.g., $\dot{\chi}$), or conduction must offset
 contributions from $-\Delta^* \mathbb{G} \dot{\xi}$, if negative, to satisfy
\eqref{eq:2ndlaw}. Overlooked previously \cite{claytonZAMM2024,claytonTR2024}, this restriction is verified a posteriori in Sec.~\ref{sec5}.}.
 Integrating this ordinary differential equation (ODE) gives the metastable value $\bar{\xi}(y,t)$, linearizing prior models \cite{boettger1997,claytonCMT2022,claytonZAMM2024}:
 \begin{align}
 \label{eq:xibar}
 \int_0^{\bar{\xi}} { \d \xi} = \int_{\alpha_F / \beta_F}^{-\Delta^* \mathbb{G} / \beta_F} \d \mathbb{F} \, \Rightarrow \, \bar{\xi} = \frac{-\Delta^*\mathbb{G} - \alpha_F }{\beta_F}  \in [0,1] \quad \forall \, -\Delta^* \mathbb{G} \in [\alpha_F, \alpha_F + \beta_F].
 \end{align}
 Denoting a relaxation time constant by $\tau_F \geq 0$, linear kinetics are
 $\tau_F \dot{\xi}(y,t) =  \bar{\xi}(y,t) - \xi(y,t)$. With $\tau_F$ on the order of $10^{-8}$s in Fe and steels \cite{boettger1997,claytonCMT2022,claytonZAMM2024}, for time scales of present interest relaxation should occur very rapidly \cite{claytonTR2024}, that is, $\tau_F \ll 1/ \dot{\bar{\gamma}} \Rightarrow \xi(y,t) \approx \bar{\xi}(y,t) \, \forall -\Delta^* \mathbb{G}(y,t) \in  [\alpha_F, \alpha_F + \beta_F]$. 

Applying $\tau = \tau(\gamma,\xi,\theta,\chi_0)$ of \eqref{eq:flowstress2} and $R = R(\xi,\chi(\gamma,\chi_0))$ of \eqref{eq:chi}, then from \eqref{eq:Gibbs1D} 
 $\Delta^* \mathbb{G} = \Delta^* \mathbb{G}(\gamma,\xi,\theta; \chi_0,\bar{p}_0,H_0)$.
 Letting $\alpha_F = \alpha_F ( p, | {\bf H} |, \tau) \rightarrow \alpha_F({p}_0,H_0,\tau)$ \cite{claytonZAMM2024} and $\beta_F = \beta_F(p_0)$, \eqref{eq:xibar} with $\xi = \bar{\xi}$
 is an implicit equation of the form $\xi = \xi (\gamma,\xi,\theta; \chi_0,\bar{p}_0,H_0)$
 to be solved simultaneously with integration of \eqref{eq:redmodel} and \eqref{eq:flowrule} for $\xi(y,t)$, $\theta(y,t)$, and $\gamma(y,t)$.  Such solutions require numerical methods 
 adopted in Sec.~\ref{sec5}.3 and~\ref{sec5}.4. 
 
 For demonstration purposes, several additional assumptions furnish an explicit equation of the form $\xi = \xi(\gamma; \chi_0,\bar{p}_0,H_0)$ to facilitate strain localization limit analysis \cite{molinari1986,molinari1987,molinari1988}.
 A first-order in $\gamma$ estimate of adiabatic temperature and a first-order in $\xi$ estimate of 
 the interaction force between $\xi$ and $\chi$ are defined as follows, with $r = r(\gamma)$ dimensionless:
 \begin{align}
 \label{eq:thetabar}
 \bar{\theta}(\gamma) = \theta_0 + (\beta_0 g_0/ c_V)\gamma; \qquad 
 \partial R (\xi,\chi(\gamma,\chi_0)) / \partial \xi = \mu (1 - \xi) r(\gamma), \quad r(0) = 0.
 \end{align}
In the second of \eqref{eq:thetabar}, multiplier $(1-\xi)$ 
ensures vanishing force as $\xi \rightarrow 1$.
The form $r(\gamma)$ vanishes at $\gamma =0$ to enable initial thermodynamic equilibrium. A material-dependent form is described in Sec.~\ref{sec5}.1 for ferrous metals.
Note $\bar{\theta}(y,t)$ accounts for increased temperature in a shear band due to plastic work from large $\gamma (y,t)$, but it omits transient contributions from hardening/softening and structure transitions. Set $\theta \approx \bar{\theta}$, $\xi \approx \bar{\xi}$, and use \eqref{eq:thetabar} in \eqref{eq:Gibbs1D}. This gives, with \eqref{eq:flowstress2}, 
 \begin{align}
 \label{eq:xibarsol}
 \beta_F \bar{\xi} & \approx g_0 \chi_0 ( 1 - \alpha_0 \bar{\xi}) (1 + \gamma / \gamma_0)^n [ 1+\beta_0 g_0 \gamma / ( c_V \theta_0) ]^\nu (\dot{\bar{\gamma}}/ \dot{\gamma}_0)^m \gamma^\xi_0   
  \nonumber 
\\ &\quad  + (\lambda_T / \theta_T) ( \beta_0 g_0 \gamma / c_V + \theta_0 - \theta_T) - \mu r (1 - \bar{\xi}) - (\psi_0 + \bar{p}_0 \delta^\xi_0  + \hat{\mu}_0 H_0 M_S + \alpha_F),
 \end{align}
 which can be solved for $\xi = \bar{\xi}$ in the domain for which a valid result in range $(0,1)$ is obtained:
 \begin{align}
 \label{eq:xisoln}
&  \xi (\gamma;\chi_0, \bar{p_0},H_0)  = \begin{cases}
 & 0, \qquad  \qquad  \qquad  \qquad \qquad \qquad \quad [\bar{\xi} \leq 0 ], \\
 & \bar{\xi} = -\Delta^* \mathbb{G}/ \beta_F - \alpha_F / \beta_F
 , \qquad \quad \! [ \bar{\xi} \in (0,1)], \\ 
 & 1, \qquad \qquad  \qquad  \qquad  \qquad \qquad \quad [\bar{\xi} \geq 1];
 \end{cases}
\\
 & 
\bar{\xi} \approx  \frac{ g_0 \chi_0  (1 + \gamma / \gamma_0)^n [1 + \beta_0 g_0 \gamma / ( c_V \theta_0) ]^\nu (\dot{\bar{\gamma}}/ \dot{\gamma}_0)^m \tilde{\gamma}^\xi_0   
 + (\lambda_T / \theta_T) [ \beta_0 g_0 \gamma / c_V - (\theta_T - \theta_0) ] -  \mu r - C_0 }
 {\beta_F  + 
 \alpha_0 g_0 \chi_0  (1 + \gamma / \gamma_0)^n [ 1 + \beta_0 g_0 \gamma / ( c_V \theta_0) ]^\nu (\dot{\bar{\gamma}}/ \dot{\gamma}_0)^m \tilde{\gamma}^\xi_0 - \mu r } ,
   \nonumber \\
   \label{eq:C0}
 & C_0 ({p}_0,H_0)=  \psi_0 + \bar{p}_0 (p_0) \delta^\xi_0 + \hat{\mu}_0 H_0 M_S + \bar{\alpha}_F({p}_0,H_0).
 \end{align}
 The transition barrier $\alpha_F$ depends linearly on $\tau$ in \eqref{eq:alphaF}; $\bar{\alpha}_F$, $\zeta_0$, and $C_0$ are constants for $t \geq 0^+$:
 \begin{align}
\alpha_F (p_0,H_0,\tau) = \bar{\alpha}_F (p_0,H_0) +  \zeta_0(p_0,H_0) \gamma^\xi_0 \tau,
\qquad
\tilde{\gamma}^\xi_0 (p_0,H_0)  = \gamma^\xi_0 [1-\zeta_0 (p_0,H_0) ].
 \label{eq:alphaF}
 \end{align}
   For given material properties, applied strain rate $\dot{\bar{\gamma}}$, pressure $p_0$, and magnetic field $H_0$, transformation $\d \xi  > 0$ initiates at $\gamma = \gamma_1$
  and concludes as $\xi \rightarrow 1$ at $\gamma \rightarrow \gamma_2$, spanning strain domain $\gamma_T$:
  \begin{align}
  \label{eq:gam12}
  \gamma_1 =  \underset{{\gamma} \geq 0 }{\rm argmin} \{ \xi({\gamma}) > 0 \}, \quad \gamma_2 = \underset{{\gamma} \geq \gamma_1 } {\rm argmin}  \{ \xi({\gamma}) \rightarrow 1 \};
  \qquad \gamma_T = \gamma_2 - \gamma_1 \geq 0.
  \end{align}
  Then from \eqref{eq:xibarsol} and \eqref{eq:gam12}, the Gibbs energy difference is approximated as
 \begin{align}
 \label{eq:Gibbsapprox}
  - \Delta^* \mathbb{G} \approx \alpha_F + \beta_F \xi, \quad [\forall \xi \in (0,1) \leftrightarrow \gamma \in (\gamma_1,\gamma_2)].
  \end{align}
 
 \subsection{Reduced governing equations}
In the absence of conduction and inertial forces and the context of the model of Sec.~\ref{sec3}.2,
\eqref{eq:2ndlaw} is obeyed; conservation laws are the momentum balance $\partial \tau / \partial y = 0$ and the energy balance in \eqref{eq:redmodel}. Substitution of \eqref{eq:flowstress2} and \eqref{eq:Gibbsapprox} into the energy balance and dividing by $\lambda (\theta)$--with $\lambda(\bar{\theta})$ for the contribution from $\dot{\xi}$ to the order of approximation in \eqref{eq:xibarsol}--produces the separable first-order ODE
\begin{align}
\label{eq:1stlawsep}
 & c_V \theta_0^\nu  {\theta}^{-\nu} \d {\theta}  \approx \{ \beta_0 g_0 \chi_0 [ 1 - \alpha_0 \xi(\gamma)] (1 + \gamma / \gamma_0)^n (\dot{\bar{\gamma}}/ \dot{\gamma}_0)^m 
 [1 + (\zeta_0 \gamma^\xi_0 / \beta_0) (\d \xi / \d \gamma)] 
  \nonumber \\ & \qquad +  \theta_0^\nu ( \theta_0 + \beta_0 g_0 \gamma / c_V )^{-\nu}
 [\bar{\alpha}_F + \beta_F \xi(\gamma) - (\lambda_T / \theta_T) ( \theta_0 + \beta_0 g_0 \gamma / c_V )] (\d \xi / \d \gamma) \} \d {\gamma},
\end{align}
where $\xi(\gamma)$ is given by \eqref{eq:xisoln} at fixed $\chi_0,p_0,H_0$.
Integrating \eqref{eq:1stlawsep} from initial conditions $\theta(y,0) = \theta_i (y)$ and $\gamma= 0$ produces the temperature history $\theta(y,t) = \theta(\gamma(y,t))$:
\begin{align}
\label{eq:thetasol}
\theta(\gamma)  & \approx  \biggr{\{} \theta_i^{1-\nu} + \frac{1-\nu}{c_V \theta_0^\nu} \int_0^\gamma 
 \beta_0 g_0 \chi_0 [ 1 - \alpha_0 \xi(\hat{\gamma})] (1 + \hat{\gamma} / \gamma_0)^n (\dot{\bar{\gamma}}/ \dot{\gamma}_0)^m
\d {\hat{\gamma}} \nonumber
\\
\nonumber 
&
+ \frac{1-\nu}{c_V \theta_0^\nu } \int_{\gamma_1}^{\min(\gamma,\gamma_2)} 
\zeta_0 \gamma^\xi_0 
g_0 \chi_0 [ 1 - \alpha_0 \xi(\hat{\gamma})] (1 + \hat{\gamma} / \gamma_0)^n (\dot{\bar{\gamma}}/ \dot{\gamma}_0)^m
\frac {\d \xi}{ \d \hat{ \gamma}}
\d {\hat{\gamma}}
\\
& 
+ \frac{1-\nu}{c_V \theta_0^\nu } \int_{\gamma_1}^{\min(\gamma,\gamma_2)} 
\biggr{(} 1+ \frac{\beta_0 g_0 \hat{\gamma} }{ c_V \theta_0} \biggr{)}^{-\nu}
 [\bar{\alpha}_F + \beta_F \xi(\hat{\gamma}) - (\lambda_T / \theta_T) ( \theta_0 + \beta_0 g_0 \hat{\gamma} / c_V )] \frac {\d \xi}{ \d \hat{ \gamma}}
\d \hat{\gamma}
\biggr{\}}^{\textstyle{\frac{1}{1-\nu}}}
.
\end{align}
Substituting $\xi$ and $\d \xi / \d \gamma$ derived from \eqref{eq:xibarsol} gives polynomial integrands of $\hat{\gamma}$ in \eqref{eq:thetasol}, though possibly integrable analytically, are too cumbersome from which to draw transparent conclusions on localization criteria.  
Numerical integration of the energy balance is undertaken with full \eqref{eq:xibarsol} in
Sec.~\ref{sec5}. For purposes of illustration,
$\xi$ is here linearly interpolated between $\gamma_1$ and $\gamma_2$ of \eqref{eq:gam12}:
\begin{align}
\label{eq:xint}
 \xi(\gamma) \approx \frac{\gamma - \gamma_1} {\gamma_T} {\mathsf H}(\gamma - \gamma_1) + 
\frac{\gamma_2 - \gamma} {\gamma_T}  {\mathsf H}(\gamma - \gamma_2), \quad 
 \{ \, \forall \, \gamma_1 \in [0,\infty), \gamma_2 \in (0,\infty), \text{ and } \gamma_T \in (0,\infty) \}.
\end{align}
The left-continuous Heaviside function is ${\mathsf H}(\cdot)$, having derivative of the delta function $\delta(\cdot)$ and
the following integration rule:
\begin{align}
\label{eq:heaviside}
 {\mathsf H}(x) = 
 \begin{cases}
 & 0, \quad \forall \, x \leq 0, \\
 & 1,  \quad \forall \, x > 0;
 \end{cases}
 \quad \frac{\d}{\d x} {\mathsf{H}} (x) = \delta (x); \quad
 \int_0^x \frac{ \d f(y) }{ \d {y} } {\mathsf H} ({y}) \d {y} = {f}(x) {\mathsf H}(x) - f(0).
\end{align}
In \eqref{eq:xint}, $\xi = 0 \, \forall \, \gamma \leq \gamma_1$, $\xi = 1 \, \forall \, \gamma \geq \gamma_2$,
and $\d \xi / \d \gamma \approx 1/\gamma_T \, \forall \, \gamma$ $\in (\gamma_1,\gamma_2)$,
$\d \xi / \d \gamma =0 \, \forall \, \gamma \leq \gamma_1$, 
$\d \xi / \d \gamma =0 \, \forall \, \gamma \geq \gamma_2$.
Limiting cases are defined as
\begin{align}
\label{eq:limcase}
& \underset{ \gamma_1 \rightarrow \infty}{\lim \xi (\gamma)} = 0, \qquad 
 \underset{ \gamma_2 \rightarrow 0}{\lim \xi (\gamma)} = 1, \qquad 
\underset{ \gamma_2 \rightarrow \infty, \, \gamma_1 \in (0,\infty)} {\underset{ \gamma \rightarrow \infty} {\lim \xi (\gamma)}} = {\rm constant} \in (0,1).
\end{align}
The first special case of \eqref{eq:limcase} implies structure transitions never occur as $\gamma \rightarrow \infty$,
the second that structure transitions are always complete regardless of $\gamma$,
and the third that the structure order parameter gets ``stuck'' at a constant intermediate value 
as $\gamma \rightarrow \infty$. The analysis requires $\xi(\gamma)$ be a continuous function of $\gamma$ over  semi-infinite non-negative domain $\gamma \in [0,\infty]$. The pathological case $\gamma_1 = \gamma_2 \in (0,\infty)$, whereby $\xi$ would be discontinuous at $\gamma = \gamma_1 = \gamma_2$, is excluded from the analysis.

Inserting \eqref{eq:xint} in \eqref{eq:thetasol} and integrating by parts with \eqref{eq:heaviside} gives an approximate analytical solution, excluding the special cases in \eqref{eq:limcase}:
\begin{align}
\theta (\gamma) & \approx 
{\pmb{ \biggr{[} }} \theta_i^{1-\nu} + \frac{(1-\nu)  \beta_0 g_0 \chi_0 (\dot{\bar{\gamma}}/ \dot{\gamma}_0)^m   \gamma_0} {(1+n) c_V \theta_0^\nu} [(1 + \gamma/ \gamma_0)^{1+n}-1]
\nonumber
\\
\nonumber
& \quad 
- \frac{(1-\nu)  \alpha_0 \beta_0 g_0 \chi_0  (\dot{\bar{\gamma}}/ \dot{\gamma}_0)^m \gamma_0 } { (1+n) c_V \theta_0^\nu} [(1 + \gamma/ \gamma_0)^{1+n}-(1 + \gamma_2 / \gamma_0)^{1+n}] {\mathsf H}(\gamma - \gamma_2)
\\
\nonumber
& \quad 
+ \frac{(1-\nu)  \zeta_0 g_0 \chi_0  (\dot{\bar{\gamma}}/ \dot{\gamma}_0)^m  \gamma^\xi_0 \gamma_0 } { (1+n) c_V \theta_0^\nu \gamma_T } [(1 + \gamma / \gamma_0)^{1+n}-(1 + \gamma_1 / \gamma_0)^{1+n}] {\mathsf H}(\gamma - \gamma_1)
\\
\nonumber
& \quad 
+ \frac{(1-\nu)  \zeta_0 g_0 \chi_0  (\dot{\bar{\gamma}}/ \dot{\gamma}_0)^m \gamma^\xi_0 \gamma_0 } { (1+n) c_V \theta_0^\nu \gamma_T } [(1 + \gamma_2 / \gamma_0)^{1+n}-(1 + \gamma / \gamma_0)^{1+n}] {\mathsf H}(\gamma - \gamma_2)
\\
\nonumber
& \quad 
+ 
 \frac{(1-\nu)  \alpha_0 \beta_0 g_0 \chi_0  (\dot{\bar{\gamma}}/ \dot{\gamma}_0)^m \gamma_0 \gamma_1 } { (1+n) c_V \theta_0^\nu \gamma_T }
\biggr{(} 1 + \frac{\zeta_0 \gamma^\xi_0}{\beta_0 \gamma_T} \biggr{)}
   [(1 + \gamma/ \gamma_0)^{1+n}-(1 + \gamma_1 / \gamma_0)^{1+n}] {\mathsf H}(\gamma - \gamma_1)
 \\
\nonumber
& \quad 
+ 
 \frac{(1-\nu)  \alpha_0 \beta_0 g_0 \chi_0  (\dot{\bar{\gamma}}/ \dot{\gamma}_0)^m \gamma_0 \gamma_1 } { (1+n) c_V \theta_0^\nu \gamma_T }
 \biggr{(} 1 + \frac{\zeta_0 \gamma^\xi_0}{\beta_0 \gamma_T} \biggr{)}
  [(1 + \gamma_2 / \gamma_0)^{1+n}-(1 + \gamma / \gamma_0)^{1+n}] {\mathsf H}(\gamma - \gamma_2)
%
\end{align}
%
\begin{align}
\label{eq:thetan}
& \quad 
- 
 \frac{(1-\nu)  \alpha_0 \beta_0 g_0 \chi_0  (\dot{\bar{\gamma}}/ \dot{\gamma}_0)^m \gamma_0^2 } { (2+n) c_V \theta_0^\nu \gamma_T } 
 \biggr{(} 1 + \frac{\zeta_0 \gamma^\xi_0}{\beta_0 \gamma_T} \biggr{)}
 [(1 + \gamma/ \gamma_0)^{2+n} - (1 + \gamma_1/ \gamma_0)^{2+n} ]  {\mathsf H}(\gamma - \gamma_1)
 \\
\nonumber
& \quad 
- 
 \frac{(1-\nu)  \alpha_0 \beta_0 g_0 \chi_0  (\dot{\bar{\gamma}}/ \dot{\gamma}_0)^m \gamma_0^2 } { (2+n)  c_V \theta_0^\nu \gamma_T }
 \biggr{(} 1 + \frac{\zeta_0 \gamma^\xi_0}{\beta_0 \gamma_T} \biggr{)}
  [(1 + \gamma_2/ \gamma_0)^{2+n} - (1 + \gamma/ \gamma_0)^{2+n} ]  {\mathsf H}(\gamma - \gamma_2)
\\
\nonumber
& \quad 
+ 
 \frac{(1-\nu)  \alpha_0 \beta_0 g_0 \chi_0  (\dot{\bar{\gamma}}/ \dot{\gamma}_0)^m \gamma_0^2 } { (1+n) c_V \theta_0^\nu \gamma_T }
 \biggr{(} 1 + \frac{\zeta_0 \gamma^\xi_0}{\beta_0 \gamma_T} \biggr{)}
  [(1 + \gamma/ \gamma_0)^{1+n} - (1 + \gamma_1/ \gamma_0)^{1+n} ] {\mathsf H}(\gamma - \gamma_1)
 \\
\nonumber
& \quad 
+ 
 \frac{(1-\nu)  \alpha_0 \beta_0 g_0 \chi_0  (\dot{\bar{\gamma}}/ \dot{\gamma}_0)^m \gamma_0^2 } { (1+n)  c_V \theta_0^\nu \gamma_T } 
 \biggr{(} 1 + \frac{\zeta_0 \gamma^\xi_0}{\beta_0 \gamma_T} \biggr{)}
  [(1 + \gamma_2/ \gamma_0)^{1+n} - (1 + \gamma/ \gamma_0)^{1+n} ]  {\mathsf H}(\gamma - \gamma_2)
\\
\nonumber
& \quad +
\frac{\theta_0^{1-\nu}}{\beta_0 g_0 \gamma_T } \biggr{(} \bar{\alpha}_F - \frac{\beta_F \gamma_1}{ \gamma_T}\biggr{)} 
\biggr{\{} \biggr{[1} + \frac{\beta_0 g_0 \gamma}{c_V \theta_0} \biggr{]}^{1-\nu} -  \biggr{[1} + \frac{\beta_0 g_0 \gamma_1}{c_V \theta_0} \biggr{]}^{1-\nu} \biggr{ \}} {\mathsf H}(\gamma - \gamma_1)
\\
\nonumber
& \quad +
\frac{\theta_0^{1-\nu}}{\beta_0 g_0 \gamma_T } \biggr{(} \bar{\alpha}_F - \frac{\beta_F \gamma_1}{ \gamma_T}\biggr{)} 
\biggr{\{}  \biggr{[1} + \frac{\beta_0 g_0 \gamma_2}{c_V \theta_0} \biggr{]}^{1-\nu} - \biggr{[1} + \frac{\beta_0 g_0 \gamma}{c_V \theta_0} \biggr{]}^{1-\nu} \biggr{ \}} {\mathsf H}(\gamma - \gamma_2)
\nonumber
\\
& 
\nonumber
\quad -
\frac{\theta_0^{2-\nu}c_V \beta_F }{(\beta_0 g_0 \gamma_T)^2 } 
\biggr{\{} \biggr{[1} + \frac{\beta_0 g_0 \gamma}{c_V \theta_0} \biggr{]}^{1-\nu} -  \biggr{[1} + \frac{\beta_0 g_0 \gamma_1}{c_V \theta_0} \biggr{]}^{1-\nu} \biggr{ \}} {\mathsf H}(\gamma - \gamma_1)
\\
\nonumber
& \quad -
\frac{\theta_0^{2-\nu}c_V \beta_F }{(\beta_0 g_0 \gamma_T)^2 } 
\biggr{\{} \biggr{[1} + \frac{\beta_0 g_0 \gamma_2}{c_V \theta_0} \biggr{]}^{1-\nu} -  \biggr{[1} + \frac{\beta_0 g_0 \gamma}{c_V \theta_0} \biggr{]}^{1-\nu} \biggr{ \}} {\mathsf H}(\gamma - \gamma_2)
\\
\nonumber
& \quad +
\frac{(1-\nu)\theta_0^{2-\nu}c_V \beta_F }{(2-\nu)(\beta_0 g_0 \gamma_T)^2 } 
\biggr{\{} \biggr{[1} + \frac{\beta_0 g_0 \gamma}{c_V \theta_0} \biggr{]}^{2-\nu} -  \biggr{[1} + \frac{\beta_0 g_0 \gamma_1}{c_V \theta_0} \biggr{]}^{2-\nu} \biggr{ \}} {\mathsf H}(\gamma - \gamma_1)
\\
\nonumber
& \quad +
\frac{(1-\nu) \theta_0^{2-\nu}c_V \beta_F }{(2-\nu) (\beta_0 g_0 \gamma_T)^2 } 
\biggr{\{} \biggr{[1} + \frac{\beta_0 g_0 \gamma_2}{c_V \theta_0} \biggr{]}^{2-\nu} -  \biggr{[1} + \frac{\beta_0 g_0 \gamma}{c_V \theta_0} \biggr{]}^{2-\nu} \biggr{ \}} {\mathsf H}(\gamma - \gamma_2)
\\
\nonumber
& \quad -
\frac{(1-\nu) \theta_0^{2-\nu} \lambda_T }{(2-\nu) \beta_0 g_0 \gamma_T \theta_T } 
\biggr{\{} \biggr{[1} + \frac{\beta_0 g_0 \gamma}{c_V \theta_0} \biggr{]}^{2-\nu} -  \biggr{[1} + \frac{\beta_0 g_0 \gamma_1}{c_V \theta_0} \biggr{]}^{2-\nu} \biggr{ \}} {\mathsf H}(\gamma - \gamma_1)
\\
& \quad -
\frac{(1-\nu) \theta_0^{2-\nu} \lambda_T }{(2-\nu) \beta_0 g_0 \gamma_T \theta_T } 
\biggr{\{} \biggr{[1} + \frac{\beta_0 g_0 \gamma_2}{c_V \theta_0} \biggr{]}^{2-\nu} -  \biggr{[1} + \frac{\beta_0 g_0 \gamma}{c_V \theta_0} \biggr{]}^{2-\nu} \biggr{ \}} {\mathsf H}(\gamma - \gamma_2)
{\pmb{ \biggr{]}} }^{\textstyle{\frac{1}{1-\nu}}}.
\end{align}

Linear momentum conservation is, with $\theta = \theta(\gamma;\theta_i,\chi_0,{p}_0,H_0)$
and $\xi = \xi(\gamma;\chi_0,{p}_0,H_0)$ in the rate-dependent flow stress expression \eqref{eq:flowstress},
\begin{align}
\label{eq:linmomcons}
&\frac{\partial \tau(\gamma(y,t),\dot{\gamma}(y,t),\theta_i(y),\chi_0(y),\bar{p}_0,H_0)) }{\partial y }= 0 \quad \Rightarrow \quad
\frac{\partial \tau(y(Y,t),t)}{\partial Y} = F_{yY} \frac{\partial \tau}{ \partial y} = 0 \nonumber \\
& \Rightarrow \quad
\tau_A = \tau(Y_A,t) = \tau(Y_B,t) = \tau_B, \qquad [F_{yY}(Y,t) = (J^E)^{1/3}J^\xi(Y,t) > 0],
\end{align}
where $Y_A$, $Y_B$ are coordinates of any two material points the slab with shear stresses $\tau_A$, $\tau_B$ \cite{molinari1986,molinari1987}. In the linear momentum balance, \textit{local transient} rate dependence of $\tau (\dot{\gamma} (y,t), \bullet)$ is enabled, as in the original viscoplastic flow rule \eqref{eq:flowstress}. Approximate form \eqref{eq:flowstress2} has only been used thus far for evaluation of the structure transition model and integration of the balance of energy, the latter following the same assumption invoked in Refs.~\cite{molinari1986,molinari1987}.

\section{Localization analysis \label{sec4}}

\subsection{Localization conditions}
The $L_\infty$ localization definition of Molinari and Clifton \cite{molinari1986,molinari1987,molinari1988}
is implemented. Accordingly, localization of shear strain $\gamma(Y,t)$ occurs at material point $B$ with $Y = Y_B$ and
time $t_c$ when $ \gamma(Y_B,t) / \gamma(Y_A,t) = \gamma_B(t) / \gamma_A(t) \rightarrow \infty$ with increasing time $t \geq t_c$ for every point  $A$  with $Y_A \neq Y_B$.

Integrating the equality $\tau_A^{1/m} = \tau_B^{1/m}$ from \eqref{eq:linmomcons} to any time $t = t_a > 0$,
with integration limits $\gamma_A = \gamma (Y_A,t_a)$ and $\gamma_B = \gamma(Y_B,t_a)$, initial conditions $\chi_{0A} = \chi_0(Y_A)$, $\chi_{0B} = \chi_0(Y_B)$, flow stress \eqref{eq:flowstress},
and functional forms $\xi = \xi(\gamma(Y,t))$ and $\theta = \theta(\gamma(Y,t))$ consistent with Sec.~\ref{sec3}.2 and Sec.~\ref{sec3}.3 gives
\begin{align}
\label{eq:Lc1}
& \int_0^{t_a} [g_0 \chi_{0A} \{ 1-\alpha_0 \xi(\gamma(Y_A,t))\} ]^{1/m}[ 1+ \gamma(Y_A,t)/\gamma_0]^{n/m}[\theta (\gamma(Y_A,t))/\theta_0]^{\nu/m} \dot{\gamma}_0^{-1} \dot{\gamma} (Y_A,t) \d t 
\nonumber
\\ 
&
= \int_0^{t_a} [g_0 \chi_{0B} \{ 1-\alpha_0 \xi(\gamma(Y_B,t))\} ]^{1/m}[ 1+ \gamma(Y_B,t)/\gamma_0]^{n/m}[\theta (\gamma(Y_B,t))/\theta_0]^{\nu/m} \dot{\gamma}_0^{-1} \dot{\gamma} (Y_B,t) \d t 
\nonumber
\\
&
\Rightarrow 
 \int_0^{\gamma_A}\chi_{0A}^{1/m} \{ 1-\alpha_0 \xi(\gamma) \}^{1/m}( 1+ \gamma/\gamma_0)^{n/m}[\theta (\gamma)/\theta_0]^{\nu/m}   \d \gamma  
\nonumber
\\ 
&
\qquad \qquad =  \int_0^{\gamma_B}\chi_{0B}^{1/m} \{ 1-\alpha_0 \xi(\gamma) \}^{1/m}( 1+ \gamma/\gamma_0)^{n/m}[\theta (\gamma)/\theta_0]^{\nu/m}   \d \gamma.  
\end{align}
When $L_\infty$ localization occurs, $t_a \rightarrow t_c$, $\gamma_B \rightarrow \infty$, and $\gamma_A \rightarrow \gamma_{Ac}$, where $\gamma_{Ac} > 0$ is finite:
\begin{align}
\label{eq:Lc2}
 & \int_0^{\gamma_{Ac}}\chi_{0A}^{1/m} \{ 1-\alpha_0 \xi(\gamma) \}^{1/m}( 1+ \gamma/\gamma_0)^{n/m}[\theta (\gamma)/\theta_0]^{\nu/m}   \d \gamma  
\nonumber
\\ 
&
\qquad =  \underset{\gamma_B \rightarrow \infty}{\lim} \int_0^{\gamma_B}\chi_{0B}^{1/m} \{ 1-\alpha_0 \xi(\gamma) \}^{1/m}( 1+ \gamma/\gamma_0)^{n/m}[\theta (\gamma)/\theta_0]^{\nu/m}   \d \gamma
=    \underset{\gamma_B \rightarrow \infty}{\lim} \int_0^{\gamma_B} I(\gamma) \d \gamma.
\end{align}

Since $\gamma_{Ac}$ is finite by definition, the integral on the left, and thus the right side, of \eqref{eq:Lc2} must both be bounded, meaning $L_\infty$ localization occurs if and only if $I(\gamma)$ is integrable as $\gamma \rightarrow \infty$ \cite{molinari1986,molinari1987}.
For the framework of \eqref{eq:flowstress}, \eqref{eq:xint} or \eqref{eq:limcase}, and \eqref{eq:thetan}, $I$ is proportional to a function of power $q$ of a linear function $P(\gamma) = c_1 \gamma + c_2$ as $\gamma \rightarrow \infty$. With $c_1 > 0$ and $|c_2|$ both finite, $c_1 \gamma \gg c_2$ as
$\gamma \rightarrow \infty$, so $P^{\,q} \sim \gamma^{\,q}$ and a corresponding generic localization criterion is then \cite{molinari1986,molinari1987}
\begin{align}
\label{eq:Ifunc}
I_\infty = \underset{\lim \gamma \rightarrow \infty}{I (\gamma) } \sim \underset{\lim \gamma \rightarrow \infty} {[P(\gamma)]^q} \sim \underset{\lim \gamma \rightarrow \infty}{\gamma^{\, q}} \quad \Rightarrow   \quad L_\infty \text{ localization possible} \, \Leftrightarrow q < -1.
\end{align}

First consider the case of conditions in \eqref{eq:xint}: $\gamma_1 \geq 0$, $\gamma_2 < \infty$, $\gamma_2 > \gamma_1$.  As $\gamma \rightarrow \infty$, $\xi = 1$ and all functions of $\gamma$ in \eqref{eq:thetan} except the first two cancel, so
\begin{align}
\label{eq:case1}
& \underset{\gamma \rightarrow \infty} {\lim} [ \chi_0^{1/m} (1 - \alpha_0 \xi)^{n/m} (1 + \gamma / \gamma_0)^{n/m} ] \sim \gamma^{\frac{n}{m}}, \quad
\underset{\gamma \rightarrow \infty} {\lim} \theta(\gamma) \sim \gamma^{\frac{1+n}{1 -\nu}}
\quad \Rightarrow  \quad
I_\infty \sim  \gamma^{\frac{\nu (1+n)}{m(1 -\nu)} + \frac{n}{m}}.
\end {align}
The $L_\infty$ conditions in \eqref{eq:Ifunc} then match those for a similar viscoplastic model in Refs.~\cite{molinari1986,molinari1987}:
\begin{align}
\label{eq:case1b}
 q = ({\nu + n})/[{m (1-\nu)}] < -1 \quad \Rightarrow \quad \nu + n + (1-\nu) m < 0, \qquad [ \forall \, m> 0, \nu < 1]. 
\end{align}

Now consider the limiting special cases in \eqref{eq:limcase}. In the first two, $\xi$ is constant throughout the deformation history, so structure changes make no contribution to $\theta$ or $\tau$, though $\theta$ and $\tau$ are generally affected by the fixed initial value $\xi = \xi_0 = 0$ or $\xi = \xi_0 = 1$.  In these two cases,
relations in \eqref{eq:case1} and \eqref{eq:case1b} clearly apply verbatim. 
For the third limiting case in \eqref{eq:limcase}, structure changes are frozen beyond some finite strain value
$\gamma_f$, where $\xi(\gamma \geq \gamma_f) = \xi_f \in (0,1)$. Again, as $\gamma \rightarrow \infty$,
$\xi$ ceases to contribute to $\theta$ or $\tau$ apart from constant scaling factors and additive constants,
so  \eqref{eq:Ifunc}--\eqref{eq:case1b} continue to hold.
Here in Sec.~\ref{sec4}.1, since both phases have the same values of $m, n$, and $\nu$, the $L_\infty$ possibility conditions are unaffected by structure transitions.  However, structure transitions do affect the applied (i.e., average) critical strain at which localization occurs, as shown in Sec.~\ref{sec4}.2.

\subsection{Stress-strain response}
The integrand of $\int_0^{t_c} \tau^{1/m} \d t$ in \eqref{eq:Lc1} is non-negative and tends to zero only as $t \rightarrow t_c \Leftrightarrow \gamma_B \rightarrow \infty $ when $L_\infty$ localization occurs at point $B$. 
Therefore, among all possible locations $Y$ in the slab, localization will occur at the earliest possible $t_c$, at point $Y = Y_B$ for which the rightmost integral in \eqref{eq:Lc2} is a minimum \cite{molinari1986,molinari1987}.  For the current initial-boundary value problem, different locations $Y$ are
distinguished only by different possible initial conditions $\chi_0(Y)$ and/or $\theta_i(Y)$. 
If $\chi_0$ and $\theta_i$ are uniform, $L_\infty$ localization is impossible since all points $Y$ are indistinguishable so will share the same stress-strain-temperature history (i.e., the trivial homogeneous solution
$\gamma = \gamma(t)$, independent of $Y$, is the only physical solution in that case).
The localization threshold at $Y$ changes with $\theta_i(Y)$; whether it increases or decreases depends on material properties. 
The localization threshold at $Y_B$ drops as $\chi_{0B}$ decreases because $\chi_0 > 0$, $m > 0$,
and to at least a close approximation, $\underset{\gamma_B \rightarrow \infty }{\lim} \int_0^{\gamma_B} I(\gamma) \d \gamma \propto \chi_{0B}^{1/m}$.

Henceforth, the analysis focuses on non-uniform $\chi_0$ and sets $\theta_i = \theta_0 = \text{constant}$.  The $L_\infty$ definition can be extended slightly such that localization can occur simultaneously at multiple points having the same minimum value $\chi_0 = \chi_{0B}$.
Take $\chi_0(Y) = 1 - \delta \chi_0 (Y)$ where $\delta \chi_0 \in [0,1)$ is a smooth perturbation field.  Localization ensues at any point(s) $Y_B = {\rm argmax} \{ \delta \chi_0 \}$ where $\chi_{0B}$ is a minimum.
In the usual case $m \ll 1$, any finite magnitude of the localization threshold integral is very sensitive to the
magnitude of the defect quantified by $\delta \chi_0$.
Each shear band approaches a singular surface at $Y_B$ across which displacement suffers a jump discontinuity as $\gamma \rightarrow \infty$. It is assumed that $\gamma$ and $\dot{\gamma}$ are continuous functions of $Y$ except at such singular point(s) $Y_B$ at $t \geq t_c$.

Average strain $\bar{\gamma}$ in the slab, whose average strain rate is $\dot{\bar{\gamma}}$ since 
$\upsilon_0 = \text{constant}$, can be 
derived from an identity stemming from the generalized theorem of Gauss \cite{hill1972,claytonIJP2003,claytonIJSS2003} with $\hat{y} = (J^E)^{1/3} Y$:
\begin{align}
\label{eq:gammabar}
\bar{\gamma}(t) = \frac{\upsilon_0 t}{ h_0} = \frac{1}{h_0} \int_0^{h_0} \gamma (Y(\hat{y})) \d \hat{y}.
\end{align}
The factor of $(J^E)^{1/3}$ arises since
$h_0$ is the coordinate of the top of the slab \textit{after} application of static pressure $p_0$; see Fig.~\ref{fig1} and \eqref{eq:varphi}--\eqref{eq:FP}.
If $p_0 = 0$, $\hat{y} = Y$. 

In the shear band, from \eqref{eq:Lc1}--\eqref{eq:Ifunc}, $\tau^{1/m} / \dot{\gamma} \rightarrow 0$
at localization point $B$ as $t \rightarrow t_c$. If $\dot{\gamma}$ remains bounded, $\tau \rightarrow 0$
as $\gamma_B \rightarrow \infty$, a condition assumed here and implicit in results of Refs.~\cite{molinari1986,molinari1987}. At any other time $t < t_c$, since ${\gamma}(Y)$ is
continuous, local strain $\gamma_A$ at at least one location $Y_A$ (that can in principle vary with $t$) must match average strain ${\bar{\gamma}} = \upsilon_0 t / h_0$.  If this $Y_A$ is time-independent, then $\dot{\gamma}_A = \dot{\bar{\gamma}} = \upsilon_0/h_0$ identically.  In that case, stress exaclty from \eqref{eq:flowstress} or, if not, approximated from \eqref{eq:flowstress2} at that location,
logically assuming $\delta \chi_0 (Y_A) \approx \delta \bar{\chi}_0 = (1/h_0) \int_0^{h_0} \delta \chi_0 (Y(\hat{y})) \d \hat{y}$ there, is
\begin{align}
\label{eq:flowstress3}
\bar{\tau}(\bar{\gamma},\xi(\bar{\gamma}),\theta (\bar{\gamma})) \approx g_0 (1 - \delta \bar{\chi}_0) ( 1 - \alpha_0 \xi(\bar{\gamma})) (1 + \bar{\gamma} / \gamma_0)^n [\theta(\bar{\gamma}) / \theta_0] ^\nu [\dot{\bar{\gamma}}/ \dot{\gamma}_0]^m, \qquad [t < t_c].
\end{align}
Since $\partial \tau / \partial Y = 0$ by \eqref{eq:linmomcons}, \eqref{eq:flowstress3}
is an equally valid estimate of $\tau$ in the entire $Y$-domain for $t < t_c$.

Localization in the $L_\infty$ sense can occur only across shear band(s) of infinitesimal thickness for $\bar{\gamma}$ to remain bounded in the limit $t \rightarrow t_c$.
The average localization strain $\bar{\gamma}_c$ is computed as follows. First, the point $Y_B$ is identified as any location having $Y_B = {\rm argmax} \{ \delta \chi_0(Y) \} = {\rm argmin} \{ \chi_0(Y) \} $.
A numerical value is calculated from the right side of \eqref{eq:Lc2}, as $\gamma_B \rightarrow \infty$.
This integral converges so long as \eqref{eq:case1} is obeyed; otherwise, it may diverge, indicating $L_\infty$ localization is impossible.
If converged, the left side of \eqref{eq:Lc2} is set equal to this value at every point $Y_A$ having $\chi_{0A} > \chi_{0B}$
and solved implicitly for $\gamma_{Ac}(Y)$ at each such point covering the $Y$-domain.
The critical average strain at localization, $\bar{\gamma}_c$, is then found by integrating $\gamma = \gamma_{Ac}(Y)$ over $Y$ in \eqref{eq:gammabar}, excluding singular point(s) $Y_B$ \cite{molinari1986,molinari1987}.
Extension of \eqref{eq:gammabar} would be required to quantify displacement discontinuities \cite{claytonIJSS2003}.

For admissible, but otherwise arbitrary, material parameters, both sides of \eqref{eq:Lc2} appear to require numerical integration. However, for the first two special cases in \eqref{eq:limcase}, analytical forms of the integrals exist.  For the first condition, 
 $\gamma_1 \rightarrow \infty \Rightarrow  \xi(Y,t) = \xi_0 = 0$; for the second, 
 $\gamma_2 \rightarrow 0 \Rightarrow \xi(Y,t) = \xi_0 = 1$. In each of these two special cases,
 denoting the hypergeometric function by $\mathcal{F}$, the analytical solution for either side of
 \eqref{eq:Lc2} in terms of local strain $\gamma(Y)$ at corresponding material point $Y$ is the following, not derived in prior work in absence of structure transitions \cite{molinari1986,molinari1987,molinari1988}:
 \begin{align}
 \label{eq:HG1}
& \int_0^\gamma I(\hat{\gamma}) \d \hat{\gamma} = \Xi(\gamma) \, \cdot \, \mathcal{F} \bigr{(} -b, (1+c)/a, 1 + (1+c)/a; -(1 + \gamma/\gamma_0)^a / N_0 \bigr{)} ;  
\nonumber \\
  &
 L_0 =[ \chi_0 (1 -\alpha_0 \xi_0) K_0^{\nu/(1-\nu)} ]^{1/m}, \quad
 K_0 =  \frac{(1-\nu)  (1 -\alpha_0 \xi_0) \beta_0 g_0 \chi_0 (\dot{\bar{\gamma}}/ \dot{\gamma}_0)^m   \gamma_0} {(1+n) c_V \theta_0}  
= \frac{1}{1+N_0}; \nonumber 
\\
& a = 1+ n, \qquad b = \nu/ [(1-\nu)m], \qquad c = n/m; \nonumber
\\
& \Xi = \frac{L_0 \gamma_0}{1+c} (1+ \gamma/\gamma_0)^{1+c}
[(1+ \gamma/\gamma_0)^a + N_0 ]^b \bigr{[} (1 +\gamma/\gamma_0)^a/N_0 + 1 \bigr{]}^{-b};
\\
& 
\mathcal{F} (A,B,C;z) = \sum_{k = 0}^\infty \frac{ (A)_k (B)_k}{(C)_k} \frac{z^k}{k!} = 
1 + \frac{A B}{C} \frac{z}{1!} + \frac{A(A+1) B (B+1)}{C(C+1)} \frac{z}{2!} + \cdots .
  \end{align}
Here, $L_0$, $K_0$, $N_0$, $a$, $b$, and $c$ are constants at $Y$ depending on material parameters, phase fraction $\xi_0$, initial defect field $\chi_0(Y)$, and average strain rate $\dot{\bar{\gamma}}$, while $\Xi$ is a function of strain $\gamma(Y,t)$.
Note $  \chi_0^{-1/[(1-\nu)m]} L_0 \propto (1-\alpha_0 \xi_0)^{1/[(1-\nu)m]}$. 
However, as this scaling factor is identical on both sides of \eqref{eq:Lc2}, $t_c$ and $\bar{\gamma}_c$ are unaffected by $  \chi_0^{-1/[(1-\nu)m]} L_0$.
Nonetheless, $\alpha_0$ affects the localization threshold via $N_0$, albeit in a nonlinear manner whose
effects are not transparent in \eqref{eq:HG1}. From
\eqref{eq:taulaw}, $\alpha_0 > 0$ corresponds to a plastically softer transformed material versus the parent, 
$\alpha_0 < 0$ to a stiffer one.


In more general situations where structure changes are transient,
 thermomechanical properties such as $\alpha_F,\beta_F,\lambda_T,\theta_T,\gamma^\xi_0$, and $\psi_0$ 
in the model of Sec.~\ref{sec3}.2 affect $\gamma_1$ and $\gamma_2$ of \eqref{eq:xint} via \eqref{eq:gam12}, $\theta$ via \eqref{eq:thetan}, and $\bar{\tau}$--$\bar{\gamma}$ response via \eqref{eq:gammabar} and \eqref{eq:flowstress3}.
The localization threshold integral on the right of \eqref{eq:Lc2} is also affected by microstructure changes, though whether it
is reduced or increased depends in a complex way on material properties and cannot be deduced by direct inspection.
If $\gamma_2$ is finite, then transformation $\xi: 0 \rightarrow 1$ will always occur at $Y_B$ in the $L_\infty$ limit, as $\gamma_B \rightarrow \infty$.
Transition can lower or raise values of $\bar{\gamma}_c$ and $t_c$.
As the strain in the impending band accommodates more of the average strain, $\gamma$ necessarily drops elsewhere in the domain for the same value of $\bar{\gamma}$.
If transition is $\gamma$-driven, localization may reduce transitions outside the band even if promoted inside.

Applied pressure $p_0$ and magnetic field $H_0$ enter the preceding analysis only through the otherwise constant initial yield strength $g_0 (p_0)$ in the flow rule of \eqref{eq:flowrule} and \eqref{eq:flowstress},
and the otherwise constant functions $\beta_F(p_0)$ and $C_0(p_0,H_0)$ in \eqref{eq:xibarsol}--\eqref{eq:C0} for transformations in Sec.~\ref{sec3}.2.  These loading conditions implicitly affect transition kinetics via $\gamma_1,\gamma_2$, and $\gamma_T$. Since $p_0$ and $H_0$ are constants for $t \geq 0^+$, then any other nominally constant property such as $n$ or $\nu$ can be taken to depend on $p_0$ and/or $H_0$. The derived localization conditions and
stress-strain calculations are of identical form in such cases, now using properties corresponding to the applied $p_0$ and $H_0$.

Some localization mitigation strategies emerge from the foregoing analysis.
If \eqref{eq:flowstress} is accurate with $m,n$, and $\nu$ similar in all phases, processing strategies or environmental
conditions (e.g., $p_0$, $H_0$) that increase the sum $\nu + n + (1-\nu) m$ should inhibit adiabatic shear banding according to \eqref{eq:Ifunc}--\eqref{eq:case1b}.
Reduction of microstructure defects (i.e., minimizing $\delta \chi_0$) would also seem beneficial.
From the second of \eqref{eq:case1}, $\theta_B \rightarrow \infty$ as $\gamma_B \rightarrow \infty$ when $(1+n)/(1-\nu) > 0$.
In the usual case $\nu < 1$ \cite{molinari1987}, any finite melt temperature would be exceeded at $Y_B$ for $n > -1$.  The solid
might also be expected to fracture as $\gamma_B \rightarrow \infty$. Melting and fracture are not analyzed explicitly now, but these mechanisms impose physical constraints on the numerical results discussed in Sec.~\ref{sec5}.4.

\subsection{Phase-dependent strain hardening and thermal softening}

After undergoing structural transition, a transformed microstructure could exhibit different
viscoplastic properties than the original.
The constitutive theory of Sec.~\ref{sec3}.2, localization analysis of Sec.~\ref{sec4}.1, and procedures of Sec.~\ref{sec4}.2 can be extended to situations where strain hardening and thermal softening
functions $h$ and $\lambda$ depend on order parameter $\xi$.  However, some analytical forms derived in Sec.~\ref{sec3}.2, Sec.~\ref{sec3}.3 and Sec.~\ref{sec4}.2 no longer hold, with equations instead requiring numerical iteration.  Integration of the stress equilibrium condition $\tau_A = \tau_B$ inherent in \eqref{eq:Lc1} requires strain rate sensitivity $m$ be constant, independent of structural changes, an assumption maintained throughout herein.
Let ($n_0,\nu_0, \gamma_0^0$) be constants for the parent structure whereby order parameter $\xi = 0$, let ($n_1,\nu_1,\gamma_0^1$) be those for the transformed structure at $\xi = 1$, and take
\begin{align}
\label{eq:nnuxi}
n(\xi) = (1-\xi) n_0 + \xi n_1, \quad \nu(\xi) = (1-\xi)\nu_0 + \xi \nu_1, \quad \gamma_0(\xi) = (1-\xi)\gamma^0_0 + \xi \gamma_0^1.
\end{align}
Instead of \eqref{eq:flowstress}, a more general composite flow stress is
\begin{align}
\label{eq:flowstresscomp}
\tau (\gamma,\dot{\gamma},\theta,\xi; \chi_0, p_0)= g_0(p_0) \chi_0 (1 -\alpha_0 \xi) (1 + \gamma / \gamma_0 (\xi) )^{n (\xi)} (\theta / \theta_0)^{\nu (\xi)} (\dot{\gamma} / \dot{\gamma}_0)^{m}.
\end{align}
Approximation \eqref{eq:flowstress2} could be applied with the same generalization. The transition model of Sec.~\ref{sec3}.2 still holds, but now \eqref{eq:xibarsol} must
be solved iteratively for $\xi(\gamma) \in (0,1)$, with the closed-form solution of $\bar{\xi}$ following \eqref{eq:xisoln} no longer applicable at $\bar{\xi} > 0$. Instead,
 \begin{align}
 \label{eq:xibarsol2}
 \beta_F \bar{\xi} & \approx g_0 \chi_0 ( 1 - \alpha_0 \bar{\xi}) [1 + \gamma / \gamma_0(\bar{\xi}) ]^{n (\bar{\xi})} [ \theta(\gamma) / \theta_0 ] ^{\nu(\bar{\xi})} (\dot{\bar{\gamma}}/ \dot{\gamma}_0)^m \tilde{\gamma}^\xi_0   
  \nonumber 
\\ & \qquad  \qquad \qquad \qquad  \qquad \qquad 
+ (\lambda_T / \theta_T) ( \theta - \theta_T) - \mu (1 - \bar{\xi}) r -C_0.
 \end{align}
Using numerical techniques, the first-order approximation \eqref{eq:thetabar} for $\bar{\theta}(\gamma)$ is also eschewed here and in later calculations. Instead,
the energy balance in the last of \eqref{eq:redmodel} is integrated numerically, specifically in Secs.~\ref{sec5}.3 and~\ref{sec5}.4 using an explicit-implicit predictor-corrector method, for $\theta(\gamma)$:
\begin{align}
\label{eq:theta2}
\theta = \theta_0 + ({\beta_0}/{c_V} )\int_0^\gamma \tau \, \d \hat{\gamma}
+({1}/c_V) \int_{\gamma_1}^{\gamma_2} [\alpha_F + \beta_F \xi - ({\lambda_T}/{\theta_T}) \theta ]
(\d \xi / \d \hat{\gamma}) \d \hat{\gamma}.
\end{align}
If $\nu$ is not constant
as in \eqref{eq:1stlawsep}, \eqref{eq:thetasol}, and \eqref{eq:thetan}, then \eqref{eq:theta2} becomes
mandatory.
Relations for $C_0$ in \eqref{eq:C0}, $\alpha_F$ in \eqref{eq:alphaF}, $\Delta^* \mathbb{G}$ in  \eqref{eq:Gibbsapprox} and ($\gamma_1,\gamma_2,\gamma_T$) in \eqref{eq:gam12} remain valid. 
The linearized transition model in \eqref{eq:xint}--\eqref{eq:limcase} could still be applied verbatim if it is used
for analytical solutions and limit analysis. However, this simple linearized transition model is neither needed nor used for obtaining the numerical solutions reported in Sec.~\ref{sec5}. 

 Note $\tau$ and $\alpha_F$ depend implicitly on $\theta(\hat{\gamma})$ and $\xi(\hat{\gamma})$ on the right side of \eqref{eq:theta2}, and approximation \eqref{eq:flowstress2} is used in \eqref{eq:theta2} for rate dependence of $\tau$ for calculations in which strain-rate field $\dot{\gamma}(Y,t)$ is unknown.
Given \eqref{eq:theta2}, the average stress-strain response is approximated from \eqref{eq:flowstress3} for $t < t_c$. 
This obviates the need for numerical differentiation of $\gamma(Y,t)$ to obtain a local $\dot{\gamma}$ field prone to inaccuracy as the localization threshold is approached.
Integrated equilibrium conditions \eqref{eq:Lc1} and \eqref{eq:Lc2} still hold with
$n$, $\nu$, and $\gamma_0$ dependent on $\xi(\gamma)$. Solution for the localization integral on the right
of \eqref{eq:Lc2} as $\gamma_B \rightarrow \infty$ and critical strain $\gamma_{Ac}(Y)$ 
at each $Y \neq Y_B$ is follows the same procedure of Sec.~\ref{sec4}.2, but numerical integration is more involved. Average strain
at localization $\bar{\gamma}_c$ again follows from inserting $\gamma_{Ac}(Y)$ into \eqref{eq:gammabar}
and excluding singular point(s) $Y_B$ where $\gamma_B \rightarrow \infty$.

Now invoke, for demonstration purposes, \eqref{eq:xint}--\eqref{eq:limcase} for $\xi(\gamma)$. 
First consider \eqref{eq:xint}. Here, since $\xi =1$ as $\gamma \rightarrow \infty$,
limiting behavior at localization is dominated by the product phase.
Since transformation is complete at some $\gamma_2 < \infty$,
the localization analysis can be recast as a new problem with $\xi = 1$
at $\gamma = \gamma_2$ as initial conditions.
Thus, \eqref{eq:case1} and \eqref{eq:case1b} apply with $n, \nu$ of the product phase:
\begin{align}
\label{eq:case1b2}
 \nu_1 + n_1 + (1-\nu_1) m < 0, \qquad [\text{transformation complete at finite }\gamma = \gamma_2]
 \end{align}
For the second of \eqref{eq:limcase}, \eqref{eq:case1b} applies trivially
with  $(n,\nu) \rightarrow (n_1,\nu_1)$, matching \eqref{eq:case1b2}.
On the other hand, for the first special case in \eqref{eq:limcase},
\eqref{eq:case1b} holds trivially with 
\begin{align}
\label{eq:case1b3}
 \nu_0 + n_0 + (1-\nu_0) m < 0, \qquad [\text{transformation never occurs; }\gamma_1 \rightarrow \infty].
\end{align}
For the first and second limiting cases of \eqref{eq:limcase}, analytical solution \eqref{eq:HG1} still applies 
with ($n, \nu$) replaced by $(n_0, \nu_0)$ in the first case and $(n_1,\nu_1)$ in the second. 
In the third case of \eqref{eq:limcase}, $L_\infty$ localization with \eqref{eq:flowstresscomp} seems affected
by all of $(n_0,\nu_0$, $n_1,\nu_1)$, but an inequality analogous to \eqref{eq:case1b2} and \eqref{eq:case1b3} and a solution like \eqref{eq:HG1} are not readily derived.
If phases have different $n$ and $\nu$ as in \eqref{eq:nnuxi} and \eqref{eq:flowstresscomp},
steps that favor transition to phase $j$ with largest $\nu_j+ n_j + (1-\nu_j) m$ seem prudent for shear band mitigation.

Assertions in \eqref{eq:case1b}, \eqref{eq:case1b2}, and \eqref{eq:case1b3} also appear to hold regardless of
whether linear interpolation in \eqref{eq:xint}, of questionably accuracy for certain phase transitions in Fe and steel in Sec.~\ref{sec5}.3, is invoked. If \eqref{eq:xibarsol2}--\eqref{eq:theta2} are used 
instead---as in all numerical solutions of Sec.~\ref{sec5} regardless of whether or not $n$ and $\nu$ depend
on $\xi$---then changes in $\psi$, $\theta$, and $\tau$ from evolution of $\xi$ will cease
at $\gamma = \gamma_f \in [0,\infty)$ when transformation is complete or stuck. Then, with this frozen state as new initial conditions, the localization integrand reduces to $I_\infty \sim \gamma^q$ with $q$ of the same form as in \eqref{eq:case1b}.

\section{Application to ferrous metals \label{sec5}}

The remainder of this work is focused on Fe and a Ni-Cr steel, with
$\xi$ an order parameter measuring the extent of $\alpha \rightarrow \epsilon$
or $\alpha \rightarrow \gamma$ transformation. The former is favored at high pressure, the latter
at high temperature.  Each material is fully BCC in the ambient state: $\xi_0 = 0$ at $\theta_0 = 300$ K,\ $p_0 = 0$, and $H_0 = 0$. The model and analysis in  Secs.~\ref{sec2}--\ref{sec4} 
can be readily adapted to other structural transitions for which the material
remains viscoplastic as $\xi: 0 \rightarrow 1$. Examples include DRX or deformation twinning, where
recrystallized or twinned microstructures are viscoplastic but of different strength than the parent.
Aspects of DRX and deformation twinning for $\alpha \rightarrow \epsilon$ transformations \cite{rittel2006} are embedded implicitly in the Gibbs free energy and kinetic law of the present structural transformation model, but no distinct order parameters are introduced in the present work for DRX (e.g., grain size) or twinning (e.g., twinned volume fractions).

The framework may be extended later to 
solid $\rightarrow$ liquid transitions (i.e., melting \cite{moss1980,moss1981}) and damage like fracture \cite{minnaar1998,claytonPM2015,claytonJMPS2021,mcauliffe2015,mcauliffe2016} or other softening
mechanisms \cite{minnaar1998,claytonActaMech2020,bucchi2023}. In these cases, transformed material is a viscous liquid or otherwise degraded substance with different deformation kinetics, so the treatment would need major adaptations. Cracks, voids, and cavitation are more likely if applied pressure is tensile versus compressive \cite{voyiadjis2012}.

\subsection{Phase transition model for iron and steel}
The $\alpha \rightarrow \epsilon$ and $\alpha \rightarrow \gamma$ transitions are modeled herein.
Reverse transitions $\epsilon \rightarrow \gamma$ and $\gamma \rightarrow \alpha$ are not addressed,
nor are high pressure-temperature $\epsilon \leftrightarrow \gamma$ transformations. 
Phase diagrams for pure Fe in the absence of external magnetic fields are widely available \cite{andrews1973,boettger1997,li2017}.
The $\alpha$-$\gamma$-$\epsilon$ triple point temperature and pressure are
around $\theta_T = 800$ K and $p_T = 11$ GPa. The transformation pressure at $\theta_0 = 300$ K is $p_{T0} = 13$ GPa \cite{mao1967,duvall1977}.  Transformation volumes $\delta^\xi_0$ for $\alpha \rightarrow \epsilon$ and $\alpha \rightarrow \gamma$
are also known \cite{mao1967}.  The zero-pressure $\alpha \rightarrow \epsilon$ phase boundary point is around 1200 K \cite{moss1980,moss1981,li2017}.
Herein, pressure is relative to ambient space: $p = 0$ corresponds to absolute external pressure of 1 atm.
Using this information, at null shear stress and null external magnetic field, $\lambda_T$ and $\psi_0$
for $\alpha \rightarrow \epsilon$ and $\alpha \rightarrow \gamma$ transitions are obtained from equilibrium conditions implied in \eqref{eq:xibar}
\cite{boettger1997,claytonCMT2022,claytonZAMM2024}:
\begin{align}
\label{eq:meta}
 - \Delta^* {\mathbb G} - \alpha_F = 0.
\end{align} 
In the present theory, $\alpha_F = 0$ when $\tau = 0$ and $H_0 = 0$.
Specific forms of \eqref{eq:alphaF} are, with $p = p_0$,
\begin{align}
\label{eq:alphaspec}
\bar{\alpha}_F(p,H_0) = \bar{\alpha}_H \hat{\mu}_0 H_0 \frac{p}{p_{T}}, 
\quad
\zeta_0 (p) = 1 - { \tilde{\gamma}^\xi_0 (p) }/{ \gamma^\xi_0}  = 
1 - \exp \left[ - \nu^\xi_0 \frac{ p}{p_{T0}} {\mathsf H}( p / p_{T0}) \right].
\end{align}
Constants are $\bar{\alpha}_H$ and $\nu^\xi_0$. 
The phase diagram of Fe under a magnetic field of strength $\hat{\mu}_0 H_0 \approx 2$ T is reported
by Curran \cite{curran1971}. At $\theta_T$, the transformation pressure is reduced by this field by around 1 GPa.
The magnetic effect on kinetic barrier $\alpha_F$ is reduced as pressure decreases.  This information furnishes the value of
$\bar{\alpha}_H$. At $p = 0$, the transformation temperature increases by around 2 K per each T of external
field due to the rightmost term in \eqref{eq:Gibbs1D} \cite{murdoch2021,claytonCMT2022}.

\begin{figure}
\begin{center}
 \subfigure[phase diagram]{\includegraphics[width=0.5\textwidth]{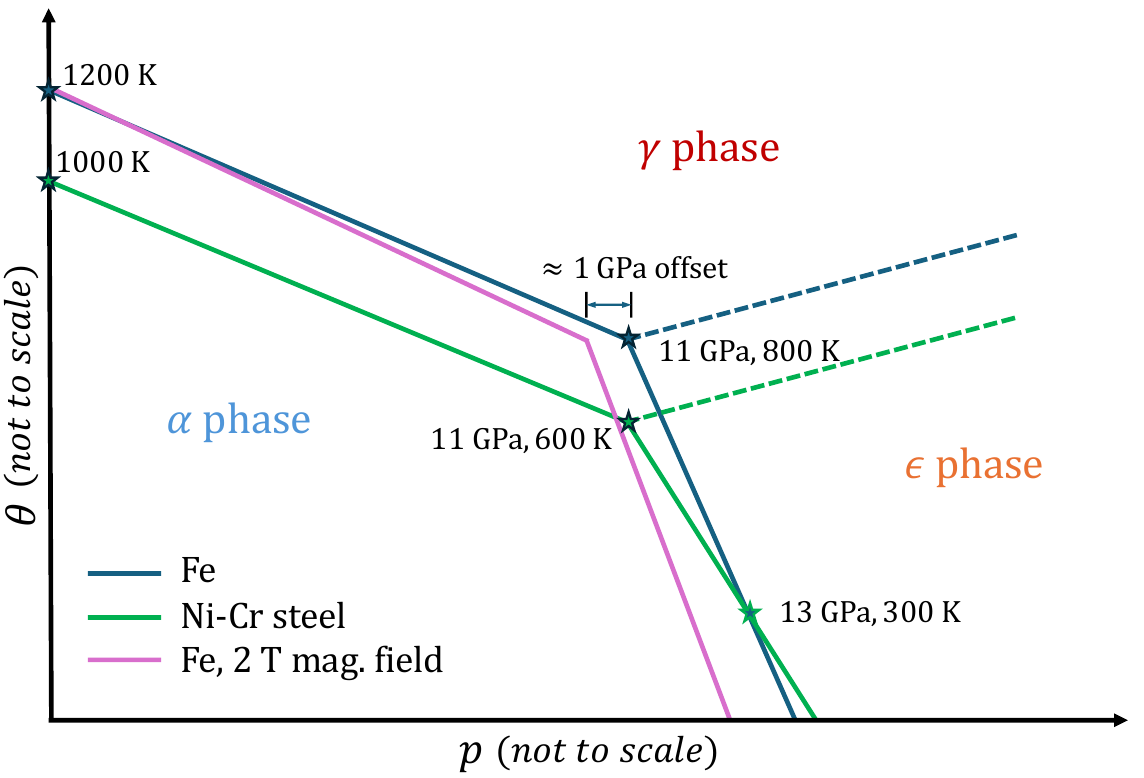} \label{fig2a}} 
 \subfigure[$\alpha \rightarrow \epsilon$ transformation shear]{\includegraphics[width=0.47\textwidth]{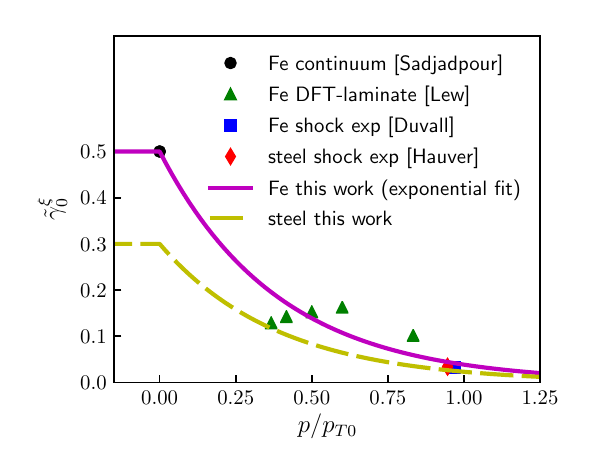}\label{fig2b}} \\
  \end{center}
  \vspace{-0.5cm}
\caption{Phase transformation model:
(a) idealized pressure-temperature ($p$-$\theta$) phase diagrams for Fe and Ni-Cr steel at null shear stress ($\tau = 0$) 
(b) effective $\alpha \rightarrow \epsilon$ transformation shear $\tilde{\gamma}_0^\xi$ for Fe and steel,  external data inferred
from static and shock compression data \cite{mao1967,duvall1977,hauver1976,hauver1979} and other theoretical studies \cite{lew2006,sadjadpour2015}.
Transformation pressure at null shear is $p_{T0}$ (13 GPa, except ref.~\cite{lew2006}), and $\tilde{\gamma}_0^\xi = \gamma_0^\xi$ for $p/p_{T0} \leq 0$. 
}
\label{fig2}       
\end{figure}

The idealized phase diagram corresponding to the model is shown in Fig.~\ref{fig2a}, in the absence of shear stress.
Data points marked by $\star$ are matched exactly. Phase boundaries are depicted as linear in $p$-$\theta$ space,
a reasonable approximation except at high pressures due to nonlinear elastic effects entering
the implicit relationship between $p$ and $\bar{p}_0 = J^E p$. Parameters are listed in Table~\ref{table1}.

The effective transformation shear strain entering the transition kinetic model of \eqref{eq:xibar}--\eqref{eq:gam12}
 is $\tilde{\gamma}^\xi_0$. In austenitic steels, the $\gamma \rightarrow \alpha$ transformation is promoted by shear stress
 \cite{stringfellow1992,turt2005},
 implying $\tilde{\gamma}^\xi_0 \leq 0$ for $\alpha \rightarrow \gamma$ transitions \cite{moss1980,moss1981}.
 In the absence of experimental and theoretical evidence of coupling of pressure to transformation shear strain for $\alpha \leftrightarrow \gamma$
 transitions, $\nu^\xi_0 = 0$ is assumed. On the other hand,
 the $\alpha \rightarrow \epsilon$ transformation shear appears to depend on pressure.
 As $p$ increases, the shear driving force should decrease, since transformation pressures in Fe
 are similar under hydrostatic compression (near-zero shear stress depending on the experimental medium) and shock compression (shear stress on the order of the yield stress) \cite{duvall1977,claytonCMT2022}.  
 The adiabatic temperature rise, $\approx 50$ K at 13 GPa shock pressure, would further reduce the transition pressure under shock conditions, so it does not offset
 any presumed shear driving force. At much lower pressures, shear-dominant experiments suggest, but do not prove,
 a shear-assisted $\alpha \rightarrow \epsilon$ transition in pure Fe. A continuum theory \cite{sadjadpour2015}
 uses a transformation shear strain of 0.5 in the absence of pressure to model these experiments.
 Theoretical calculations with DFT and nonlinear elastic, anisotropic laminate theory \cite{caspersen2004,lew2006}
 show a reduction in transition pressure with increasing shear strain in pure Fe, beginning from a single crystal of the BCC phase.
 
 At constant temperature $\theta = \theta_0$ and null magnetic field, \eqref{eq:meta} produces the following conditions \cite{ma2006},
 recalling $\sigma_{xy} =  \tau / J^E$ and $p =  \bar{p}/J^E$ are Cauchy stresses:
 \begin{align}
 \label{eq:metad} 
 [p(\tau,\theta_0) - p_{T0}] \delta^\xi_0 = \sigma_{xy} \tilde{\gamma}^\xi_0 \qquad 
 \Rightarrow \quad \tilde{\gamma}^\xi_0 =  [p(\tau,\theta_0) - p_{T0}] \delta^\xi_0 /  \sigma_{xy}.
  \end{align}
 Here $p(\tau,\theta_0)$ is transition pressure at shear stress $\tau = J^E \sigma_{xy}$; $p_{T0}$ is transition pressure at $\theta = \theta_0$ and $\tau = \sigma_{xy} = 0$: 13 GPa in experiments \cite{mao1967}. Since $\delta^\xi_0 < 0$ for $\alpha \rightarrow \epsilon$, 
 transition pressure is reduced by $\tau > 0$ for $\tilde{\gamma}^\xi_0 > 0$.
 
 Using the second of \eqref{eq:alphaspec} with \eqref{eq:metad}, $\gamma^\xi_0$ and $\nu^\xi_0$ are fit to the data in Fig.~\ref{fig2b}; the former exactly matches the value from Ref.~\cite{sadjadpour2015} for $p \leq 0$.
 This procedure requires transformation pressure $p$ as a function of $\sigma_{xy}$.
Data points in Fig.~\ref{fig2b} are inferred from shock experiments on Fe and Ni-Cr steel \cite{duvall1977,hauver1979}
using $P = p + \frac{4}{3} \sigma_Y$ \cite{wallace1980,claytonNEIM2019,claytonJMPS2019}, where $P$ is Hugoniot stress and $\sigma_Y$ is approximated as half the static Von Mises yield stress \cite{boettger1997},
so $\sigma_{xy} \approx 2 \sigma_Y / \sqrt{3}$.
In both materials, $P \approx 13$ GPa, while $\sigma_Y \approx 300$ MPa for Fe and $\sigma_Y \approx 500$ MPa for Ni-Cr steel \cite{boettger1997,claytonCMT2022,benck1976,meyer2001}.
The value of $p$ so obtained under shock conditions is corrected for the latent heat to its value at $\theta_0$ using a 50 K adiabatic temperature rise at $P = 13$ GPa \cite{claytonCMT2022}.
Data points inferred from Ref.~\cite{lew2006} convert applied strain component $\epsilon_{xy}$ to shear stress via $ \sigma_{xy} \approx 2 G_0 \epsilon_{xy}$, where $G_0$ is the zero-pressure shear modulus from DFT results in that reference. 
A normalization pressure of $p_{T0} = 30$ GPa is extrapolated for fitting to that data \cite{lew2006} since transformation pressure is elevated in the absence of plasticity and defects \cite{ma2006}.
 
The high-strength steel studied here has thermodynamic properties based on Refs.~\cite{benck1976,hauver1976,hauver1979,moss1980,moss1981}.
The $\alpha$, $\epsilon$, and $\gamma$ phases are BCC, HCP, and FCC, like pure iron.
At ambient conditions $p = 0$ and $\theta_0 = 300$ K, material is entirely in its $\alpha$ phase, mostly bainite with some martensite. Since the shock Hugoniot and inferred transition pressure $p_{T0}$ of this steel and Fe are very similar \cite{hauver1976,hauver1979}, subtle differences in $B_0$, $B'_0$, and $\delta^\xi_0$ due to composition and processing are suitably ignored within the fidelity of the analysis.
Chemistry includes 0.22 \%C, 1.06 \%Ni, and 3.15 \%Ni, among other trace elements, leading to a drop in
the $\alpha$-$\gamma$ phase boundary by around 200 K relative to iron \cite{moss1980,moss1981}. 
Parameters $\lambda_T$ and $\psi_0$ are thus computed using \eqref{eq:meta}; the phase diagram in Fig.~\ref{fig2a} results.
Effects of magnetic field and shear on transition kinetics have not been quantified experimentally.
Thus, the same procedure for Fe is used to calculate $\bar{\alpha}_H$ with a 1 GPa offset from a 2 T field at $p_T = 11$ GPa.
The same value of $\nu^\xi_0$ is assumed for steel as Fe in the absence of data.
The value of $\gamma^\xi_0$ for steel is $\frac{3}{5}$ that of Fe in Fig.~\ref{fig2b} and Table~\ref{table1} to predict similar $\gamma_1$ thresholds in Sec.~\ref{sec5}.3.


\begin{table}
\footnotesize
\caption{Properties or model parameters for polycrystalline Fe and Ni-Cr steel, $\theta_0 = 300$ K}
\label{table1}       
\centering
\begin{tabular}{llrrl}
\hline\noalign{\smallskip}
Parameter [units] & Definition & Value Fe &Value Steel &  Remarks [references] \\
\noalign{\smallskip}\hline\noalign{\smallskip}
$\rho_0$ [g/cm$^3$] & ambient mass density ($\alpha$ phase) & 7.87 & 7.84 & \cite{boettger1997,hauver1979}  \\
${B}_0$ [GPa] & isothermal bulk modulus & 163  & 163 &  \cite{guinan1974,benck1976,hauver1979}   \\
${B}'_0$ [-] & pressure derivative of bulk modulus & 5.29  & 5.29 & inferred for steel \cite{guinan1974,hauver1979}   \\
$\mu$ [GPa] & elastic shear modulus ($\alpha$ phase)  & 83 & 80 &  \cite{gandhi2022,benck1976} \\
$\mu'/\mu$ [1/GPa] & pressure dependence of shear modulus  & 0.023 & 0.024 &  \cite{gandhi2022,guinan1974} \\
$c_V$ [MPa/K] & specific heat at constant volume & 3.54 & 3.48 &  \cite{andrews1973,borvik1999}   \\
$\delta^\xi_0$ [-] &  transition volume change: $\alpha \rightarrow \epsilon$ & -0.0512 & -0.0512 &   from $\alpha,\epsilon$ densities \cite{boettger1997}\\
& transition volume change: $\alpha \rightarrow \gamma$ &  -0.0316 & -0.0316  &   from $\alpha,\gamma$ densities \cite{mao1967}\\
$\gamma^\xi_0$ [-] & transition shear at $p = 0$: $\alpha \rightarrow \epsilon$  & 0.5 & 0.3 &   model fit and Ref.~\cite{sadjadpour2015} \\
& transition shear at $p = 0$: $\alpha \rightarrow \gamma$  & -0.1 & -0.1 &  assumed, Refs.~\cite{stringfellow1992,claytonZAMM2024} \\
$\nu^\xi_0$ [-] & shear-pressure coupling: $\alpha \rightarrow \epsilon$  & 2.56 & 2.56 &   model fit and Ref.~\cite{sadjadpour2015} \\
& shear-pressure coupling: $\alpha \rightarrow \gamma$  & 0.0 & 0.0 &  assumed (no data) \\
$p_{T0}$ [GPa] & transition pressure at $\theta_0$: $\alpha \rightarrow \epsilon$ & 13.0 & 13.0 & \cite{mao1967,duvall1977,hauver1979} \\
$ p_T$ [GPa] & triple point pressure & 11.0 & 11.0 & $\alpha$-$\gamma$-$\epsilon$ triple point  \cite{li2017,moss1980,moss1981}\\
$ \theta_T$ [K] & triple point temperature & 800 & 600 & $\alpha$-$\gamma$-$\epsilon$ triple point  \cite{li2017,moss1980,moss1981}\\
$\theta_M$ [K] & melt temperature at $p=0$ & 1811 & 1783 & \cite{meyer2001,hanim2001} \\
$\psi_0$ [MPa] & free energy offset: $\alpha \rightarrow \epsilon$  & 531 & 531 &    fit to $p$-$\theta$ phase diagram \\
                  & free energy offset: $\alpha \rightarrow \gamma$  & 328 & 328  & fit to $p$-$\theta$ phase diagram \\
 $\lambda_T$ [MPa] & latent heat: $\alpha \rightarrow \epsilon$  & 147 & 183  &    fit to $p$-$\theta$ phase diagram \\
                  & latent heat: $\alpha \rightarrow \gamma$  & 657 &  492 & fit to $p$-$\theta$ phase diagram \\
$\hat{\mu}_0 M_S $ [T]  & saturation magnetization ($\alpha$ phase) & 2.15 &  2.15 &   $ \hat{\mu}_0  H_0  \gtrsim 0.05$ T  \cite{daniel2008} \\
$\bar{\alpha}_H$ [MPa/T] & magnetic transition barrier: $\alpha \rightarrow \epsilon$ & -25.0 & -25.0 & model fit to Ref.~\cite{curran1971} \\
                                        & magnetic transition barrier: $\alpha \rightarrow \gamma$ & -16.1 & -16.1 & model fit to Ref.~\cite{curran1971} \\
 $\beta_{F0}$ [MPa] & controls phase transition width & 90.5 & 90.5 & known for $\alpha \leftrightarrow \gamma$ in Fe \cite{boettger1997,claytonCMT2022,claytonZAMM2024} \\   
 $a_\beta$ [-] & permits $\dot{\xi} > 0$ if $\alpha_0 \gamma_0^\xi < 0$: $\alpha \rightarrow \epsilon$ & 4 & 12 & calibrated; $a_\beta = 1$ for $\alpha \rightarrow \gamma$ \\      
 $ c_r$ [-] & $\chi$-$\xi$ interaction energy: $\alpha \rightarrow \epsilon$
 & 6 & 15 & calibrated; $c_r \rightarrow \infty$ for $\alpha \rightarrow \gamma$ \\                              
 $ {\beta}_0$ [-] & Taylor-Quinney factor & 0.6 & 0.8  & approximated \cite{rittel2006,rittel2017} \\                                        
$ g_0$ [MPa] & initial ambient yield strength ($\alpha$ phase) &167 & 693 & at $p_0= 0$ \cite{benck1976,rittel2006,sadjadpour2015,meyer2001} \\
$ \gamma_0$ [-] & reference plastic strain & 0.01 & 0.01 & convenient for steel \cite{molinari1986,molinari1987}\\
$\dot{ \gamma}_0$ [1/s] & reference plastic strain rate & 1.0 & 1.0 & standard for steel \cite{meyer2001,gray1994}\\
$ \alpha_0$ [-] & flow strength reduction: $\alpha \rightarrow \epsilon$ & -1.55 & -1.55 & $\alpha_0 < 0$: hardening \cite{gandhi2022} \\
& flow strength reduction: $\alpha \rightarrow \gamma$ & 0.50 & 0.50 & TRIP steels \cite{stringfellow1992,tomita1995,claytonJDBM2021} \\
 $n$ [-] & strain hardening exponent & 0.09 & 0.05 & $n > 0$: hardening \cite{sadjadpour2015,meyer2001,gray1994} \\
 $m$ [-] & strain rate sensitivity exponent & 0.089 & 0.065 & $m > 0$: hardening \cite{sadjadpour2015,meyer2001,gray1994} \\
 $\nu$ [-] & thermal softening exponent & -1.6 & -0.33 &  $\nu < 0$: softening \cite{sadjadpour2015,meyer2001,gray1994} \\
\noalign{\smallskip}\hline
\end{tabular}
\end{table}

For phase transitions to be possible according to \eqref{eq:xibarsol} or \eqref{eq:xibarsol2},
the denominator in the fractional expression of \eqref{eq:xisoln} must be positive at $\gamma = \gamma_1$.
If \eqref{eq:xibarsol} is used in conjunction with the more accurate \eqref{eq:theta2} instead of \eqref{eq:thetabar}, then a similar requirement holds.
If $|\mu r|$ is small and $\alpha_0 {\gamma}^\xi_0 < 0$, then $\beta_F$ must
be large enough to ensure phase transitions are admissible over a finite strain interval $\gamma_T$. 
In the absence of shear (i.e., $\dot{\bar{\gamma}} \rightarrow 0$), when transition
is purely pressure-driven, then $\beta_F \rightarrow \beta_{F0}$,
where $\beta_{F0}$ gives the $\alpha \rightarrow \epsilon$ transition width under hydrostatic compression \cite{taylor1991,boettger1997,claytonCMT2022}.  The following function fulfills the stated requirements,
where $a_\beta \geq 1$ is a constant and here $p = p_0$:
\begin{align}
\label{eq:betap}
\beta_F = \beta_{F0} \{ 1 + (a_\beta - 1) [ \{ 1 - (p/p_T) \mathsf{H}(p/p_T) \} \{ 1 - \mathsf{H} ( p/p_T -1) \} ] \}.
\end{align}
The value of $a_\beta$ for $\alpha \rightarrow \epsilon$ transitions in Fe and steel
is chosen such that the denominator of \eqref{eq:C0} is comparable to $\beta_{F0}$
at $\gamma = 0$.  For $\alpha \rightarrow \gamma$ transitions, $a_\beta = 1$ is sufficient,
giving simply $\beta_F = \beta_{F0}$.

In the absence of external pressure or magnetic field, $\alpha \rightarrow \epsilon$
transitions in Fe and Ni-Cr steel do not occur according to model predictions
at $\dot{\bar{\gamma}} = 10^4$/s when dislocation-structure interaction
force $\partial R / \partial \xi \geq 0$ in the second of \eqref{eq:thetabar}.
Driving force from shear stress $\tau$ is insufficient to overcome thermodynamic barriers
from $\lambda_T > 0$ and $\psi_0 > 0$ in \eqref{eq:xibarsol} and \eqref{eq:xibarsol2}.
Since experiments and models \cite{rittel2006,sadjadpour2015} infer $\alpha \rightarrow \epsilon$ transformation is possible under shear-dominant conditions, a non-trivial
 interaction function $r(\gamma) \leq 0$ is merited.
Satisfying requirement $r(0) = 0$ and $\lim_{\gamma \rightarrow \infty} r(\gamma) = \text{constant}$, let
\begin{align}
\label{eq:sech}
& r(\gamma) = {\rm{sech}} [\gamma / (c_r \gamma_0^\xi) ] - 1 \,
\Rightarrow \, R(\chi(\gamma,\chi_0),\xi) = \tilde{R}(\gamma,\chi_0)
+ \mu \{ {\rm{sech}} [\gamma / (c_r \gamma_0^\xi) ] - 1\} ( 1 - \xi/2) \xi ,
\nonumber
\\
& \d r / \d \gamma = \{-1/(c_r \gamma_0^\xi)\}
\{ {\rm sech}[\gamma / (c_r \gamma_0^\xi) ]  {\rm tanh}[\gamma / (c_r \gamma_0^\xi) ] \}, \quad
 (\d r / \d \gamma)_{\gamma = 0} =  (\d r / \d \gamma)_{\gamma \rightarrow \infty} = 0.
\end{align}
Dimensionless constant $c_r$ is chosen such that $\alpha \rightarrow \epsilon$ transitions
begin at strain magnitudes comparable to those inferred from experiments and other models for iron at similar rates \cite{rittel2006,sadjadpour2015}. For $c_r \gamma_0^\xi > 0$ and $\gamma > 0$,
$\d r / \d \gamma < 0$ produces an increase in free energy release rate $-[(\partial R / \partial \chi) \dot{\chi} + (\partial R / \partial \xi) \dot{\xi}]$ entering \eqref{eq:1stlaw} and \eqref{eq:2ndlaw}. 
Contributions proportional to $\dot{\xi}$ are embedded in $\Delta^* \mathbb{G} \cdot \dot{\xi}$. An increase in temperature rate
would result from energy release due to phase changes and exothermic DRX, raising the apparent Taylor-Quinney factor in agreement with experiments \cite{rittel2006,sadjadpour2015}. 

The energy released in conjunction with the $\alpha \rightarrow \epsilon$ transition could be attributed to 
changes in microscopic interfacial energy \cite{turt2005}, dislocation-grain boundary interactions
in rotational DRX \cite{lieou2018,lieou2019}, and deformation twinning \cite{rittel2006,gandhi2022}.
A more detailed model of such physics at the grain- and subgrain-scales is beyond the present scope. For $\alpha \rightarrow \gamma$, adiabatic
temperature rise from $\beta_0 \tau \dot{\gamma}$ is sufficient to enable transitions with
$r=0$. Thus, $c_r \rightarrow \infty$ for $\alpha \rightarrow \gamma$ transitions.

\subsection{Viscoplastic properties for iron and steel}
Recall from \eqref{eq:flowrule}--\eqref{eq:flowstress} that the static flow stress is 
$g (\gamma,\theta,\xi ; p_0,\chi_0) = g_Y (\xi ; p_0) h(\gamma ; \chi_0) \lambda(\theta)$, where
$h$, $\lambda$, and $\chi_0$ are the strain hardening function, thermal softening function, and initial defect parameter.
Flow stress coefficient $g_Y = g_0(1-\alpha_0 \xi)$ entering \eqref{eq:taulaw} is assigned a linear dependence on Cauchy pressure $p  = p_0$ through the pressure derivative of the elastic shear modulus $\mu'$ \cite{claytonNCM2011,gandhi2022}:
\begin{align}
\label{eq:gyp}
g_0(p_0) = [1 + (\mu'/\mu) p_0] g_0(0).
\end{align}

The bottom eight entries in Table~\ref{table1} comprise viscoplastic property values for iron and Ni-Cr steel.
Normalizing parameters $\gamma_0$ and $\dot{\gamma}_0$ are typical values for convenience.
Initial yield $g_0$, strain hardening $n$, thermal softening $\nu$, and rate sensitivity $m$
are matched to fits of experimental data reported in other works on Fe and high-strength steel~\cite{rittel2006,sadjadpour2015,meyer2001,gray1994,benck1976}.

Noting that Taylor-Quinney factor $\beta$ demonstrates dependence on material history and loading rate \cite{claytonJMPS2005,claytonIJSS2005,claytonCOMPB2009,rittel2006},
approximate averages of $\beta_0$ are
estimated from data \cite{rittel2006,rittel2017} on Fe and low-C steels of similar strength to the 
present BCC alloy.
The prescribed value of $\beta_0$ for Fe approximates that measured at high rates $\dot{\gamma} \approx 10^4$/s but lower strains (e.g., $\gamma \lesssim 0.2$), prior
to potential phase transitions or DRX that seem to raise the apparent Taylor-Quinney factor \cite{rittel2006,sadjadpour2015}.

\begin{figure}
\begin{center}
 \subfigure[strain hardening]{\includegraphics[width=0.4\textwidth]{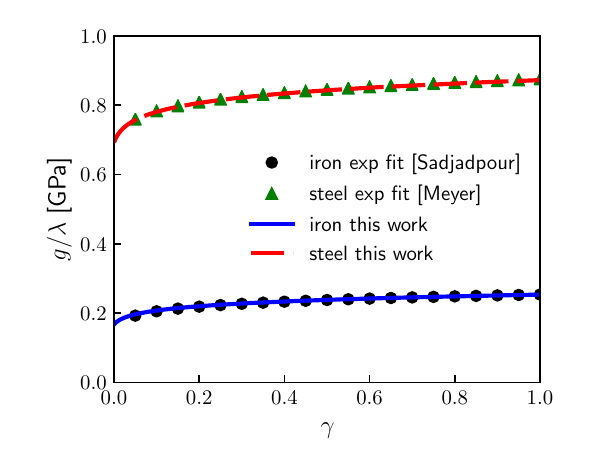} \label{fig3a}} 
 \subfigure[thermal softening]{\includegraphics[width=0.4\textwidth]{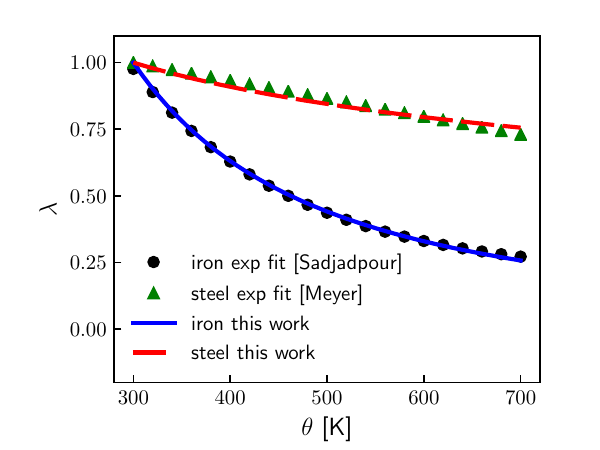}\label{fig3b}}
  \end{center}
  \vspace{-0.5cm}
\caption{Viscoplastic model and experimental data fits \cite{sadjadpour2015,meyer2001} for $p_0 = 0$, $\xi = 0$, $\chi_0 = 1$:
(a) static flow stress $\tau \rightarrow g_Y h / \lambda$ at $\theta_0 = 300$ K and $\dot{\gamma} = \dot{\gamma}_0$, 
(b) thermal softening factor $\lambda = \tau / (g_Y h)$ at $\gamma = 0$ and $\dot{\gamma} = \dot{\gamma}_0$
}
\label{fig3}       
\end{figure}

Calibrated results are shown in Fig.~\ref{fig3}, where experimental data pertain to the $\alpha$ phase.
 Strain hardening and thermal softening are fit well by the present model, for BCC iron and steel.
Rate sensitivity is well fit for steel, and for $\alpha$-Fe at $\dot{\gamma} \lesssim 10^2$/s if $m = 0.001$ is prescribed. Experiments \cite{rittel2006} suggest rate sensitivity increases at higher loading rates.
This behavior cannot be matched by any viscoplastic model like \eqref{eq:flowrule} with
a single value of $m$. Taking $m = 0.089$ as in Table~\ref{table1} fits the flow stress at $10^4$/s.
Data for steel \cite{gray1994,meyer2001} also show an increase in rate sensitivity for rates up to $7 \times 10^3$/s. While a value of $m = 0.004$ is suggested to encompass low-rate data \cite{meyer2001}, setting $m = 0.065$ matches the experimental calibration \cite{meyer2001} to high-rate data \cite{gray1994} exactly at $\dot{\gamma} = 10^4$/s.

Parameter $\alpha_0 = 1 - g_1/g_0$ relates ambient yield strength $g_1$ of the product ($\gamma$ or $\epsilon$)
to $g_0$ of the parent ($\alpha$) phase.
Flow stress data on transformed phases of Fe and this Ni-Cr steel are scarce, if not nonexistent.
Experiments cannot be performed at ambient conditions to obtain $g_1$ since 
$\gamma$ is stable only at very high $\theta$ and reverts to $\alpha$ upon cooling to room temperature, 
and $\epsilon$ is stable only at very high $p$ and reverts to $\alpha$ upon depressurization to atmospheric pressure.
Data on austenitic TRIP steels, whose $\gamma$ and $\alpha$ phases are stable at lower temperatures,  usually show strengthening with the $\gamma \rightarrow \alpha$ martensitic transformation \cite{claytonJDBM2021,stringfellow1992,tomita1995}.
Softening for $\alpha \rightarrow \gamma$ has been speculated elsewhere for this steel \cite{moss1980,moss1981},
with further decreases in flow stress possible due to break-up of ferromagnetic domains.
Representative value $\alpha_0 = 0.5$ is prescribed for $\alpha \rightarrow \gamma$ of both materials.

\begin{figure}
\begin{center}
 \subfigure[strain hardening]{\includegraphics[width=0.4\textwidth]{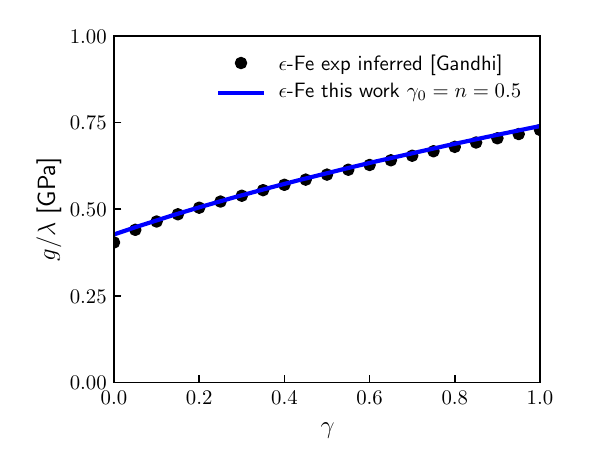} \label{fig4a}} 
 \subfigure[thermal softening]{\includegraphics[width=0.4\textwidth]{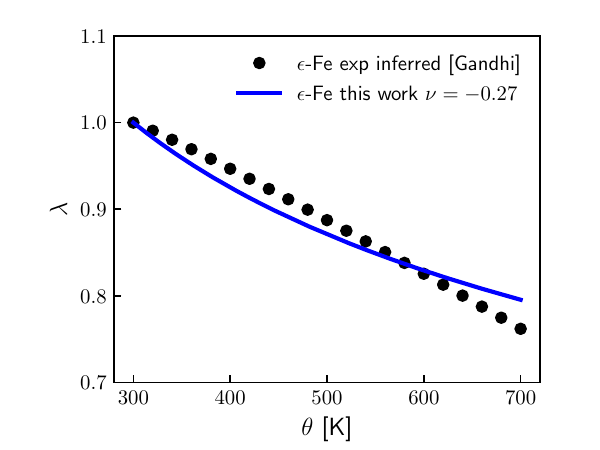}\label{fig4b}}
   \end{center}
  \vspace{-0.5cm}
\caption{Viscoplastic model and inferred fits from pressure-shear impact experiments \cite{gandhi2022} for $\epsilon$-Fe:
(a) flow stress $\tau \rightarrow g_Y h / \lambda$ at $\theta_0 = 300$ K and $\dot{\gamma} = \dot{\gamma}_0$, 
(b) thermal softening $\lambda = \tau / (g_Y h)$ at $\gamma = 0$ and $\dot{\gamma} = \dot{\gamma}_0$
}
\label{fig4}       
\end{figure}

Regarding the $\epsilon$ phase, recent dynamic pressure-shear plate impact experiments and computational modeling probe the shear strength of Fe at $\dot{\gamma} \approx 10^5$/s and $p \in [10,42]$ GPa.
In most experiments, pressures are high enough to completely transform the sample to its $\epsilon$ phase.
Assuming $m$ is the same in each phase and correcting for pressure dependence via \eqref{eq:gyp},
$g_1/g_0 \approx 2.55$ is extracted, leading to $\alpha_0 = -1.55$ (i.e., transformation strengthening).
Since rate sensitivity is omitted in the calculations in Ref.~\cite{gandhi2022} and the experiments
cover only a single strain rate and single starting temperature, parameters $m$, $n$, and $\nu$
cannot be determined uniquely. However, values of $n$ and $\nu$ can be calibrated to match
 inferred strain hardening and thermal softening of $\epsilon$-Fe as modeled in Ref.~\cite{gandhi2022}.
Results are shown in Fig.~\ref{fig4}; $\gamma_0$ is adjusted to 0.5 to fit the high hardening rate.

Localization criteria in \eqref{eq:case1b} are evaluated 
for $\alpha$-Fe (specifically \eqref{eq:case1b3}), $\epsilon$-Fe (specifically \eqref{eq:case1b2}), and
Ni-Cr steel in Table~\ref{table2}. Experimental flow stress information for $\epsilon$-steel is not available;
parameters $n$, $\nu$, and $m$ are assigned as constants to match the experimental fits in Fig.~\ref{fig3}.
Accordingly, $L_\infty$ localization is predicted for $\alpha$-steel.
For $\alpha$-Fe, since the assumption of constant $m$ breaks down at very high strain rates,
minimum and maximum values from Ref.~\cite{sadjadpour2015} are used to establish bounds on the
$L_\infty$ criterion. Even when a very high rate sensitivity of $m = 0.385$ is assumed,
localization is predicted for $\alpha$-Fe because of its pronounced thermal softening.
Both $\nu$ and $m$ are considered uncertain for $\epsilon$-Fe, given the limited domain of initial conditions
probed experimentally \cite{gandhi2022}. Bounds in Table~\ref{table2} consider the same wide range of $m$ as
for $\alpha$-Fe and $\nu$ of $\alpha$-Fe or Fig.~\ref{fig4b}.
Localization is impeded by the high strain hardening exponent $n$, but is still possible when the minimum value of $\nu = -1.6$ is used. Regardless of which value is used for $m$, $L_\infty$ localization is predicted to be impossible in $\epsilon$-Fe when the larger value of $\nu = -0.27$ is used. 

The analysis of Sec.~\ref{sec4}.1 with results in Table~\ref{table2} suggests that the possibility for adiabatic shear localization, but not necessarily the critical average strain $\bar{\gamma}_c$ as shown in numerical results of Sec.~\ref{sec5}.4, should be reduced if the $\alpha \rightarrow \epsilon$ transformation occurs.
This prediction concurs with dynamic experiments on Fe at low and high pressure in which $\alpha \rightarrow \epsilon$ transformation was either inferred or measured \cite{rittel2006,sadjadpour2015,gandhi2022} but no
shear localization occurred. No evidence of $\alpha \rightarrow \epsilon$ transformation was reported
from other torsion experiments on Fe and a similar high-strength BCC steel at modestly lower strain rates in which shear bands did arise \cite{fellows2001,fellows2001b}.
Since $n$, $\nu$, and $m$ are unknown for $\gamma$-Fe and $\gamma$-steel,
the $L_\infty$ criterion cannot be evaluated for these isolated high-temperature phases.
However, dynamic experiments on other steels \cite{cho1990,syn2005,jo2020} indicated $\alpha \rightarrow \gamma$
occurred within adiabatic shear bands, albeit with reversion to other structures upon cooling. Since no evidence exists for a higher hardening rate or less thermal softening in the $\gamma$ phase relative to the $\alpha$ phase, it is concluded here that $L_\infty$ localization should be possible in $\gamma$-Fe and $\gamma$-steel.

\begin{table}
\footnotesize
\caption{Estimated bounds on localization parameters and criteria for $\alpha$- and $\epsilon$-iron and  Ni-Cr steel}
\label{table2}       
\centering
\begin{tabular}{lccccc}
\hline\noalign{\smallskip}
Material & $n$ & $\nu$ & $m$ & $\nu + n + (1-\nu) m$ & $L_\infty$ localization  \\
\noalign{\smallskip}\hline\noalign{\smallskip}
$\alpha$-Fe & 0.09 & -1.6 & $[8 \times 10^{-5},0.385]$ & [-1.51, -0.51] & yes \\
$\epsilon$-Fe & 0.5 & [-1.6, -0.27]  & $[8 \times 10^{-5},0.385]$ &  [-1.10, 0.72] & maybe \\
Ni-Cr steel & 0.05 & -0.33 & $[4 \times 10^{-3},0.065]$ &[-0.27,-0.19] & yes \\
\noalign{\smallskip}\hline
\end{tabular}
\end{table}

\subsection{Homogeneous solutions}
For $\chi_0 = 1$ and $\delta \chi_0 = 0$ everywhere in the slab, 
fields $\gamma(Y,t)$, $\theta(Y,t)$, and $\xi(Y,t)$ become independent
of $Y$. Strain rate is constant everywhere: $\dot{\gamma} = \dot{\bar{\gamma}} = \upsilon_0 / h_0$.
Shear strain is simply $\gamma = \bar{\gamma} = \dot{\bar{\gamma}} t$.

Prescribed now are initial conditions $\xi_0 = 0$; $\theta_0 = 300$ K;
$p_0 = 0, 5$, or 10 GPa; and $\hat{\mu}_0 H_0 = 0, 1,$ or 2 T. The applied
strain rate is $\dot{\gamma} = 10^4$/s, that most suitable for calibrated properties of Table~\ref{table1}.
Responses for Fe and steel are obtained using analytical-numerical methods of Secs.~\ref{sec4}.2 and~\ref{sec4}.3.  Different parameter sets from Table~\ref{table1} approriate to $\alpha \rightarrow \epsilon$ or $\alpha \rightarrow \gamma$ transformations are used. 

Results pertaining to phase transitions---pressure, magnetic field, initiation shear strain, initiation shear stress, initiation temperature, and completion shear strain---are listed in Table~\ref{table3}.
Initiation strain $\gamma_1$ and completion strain $\gamma_2$  are always reduced with increasing pressure.
Magnetic fields also decrement $\gamma_1$ and $\gamma_2$.
These behaviors are physically expected given the phase diagrams in Fig.~\ref{fig2a} and
underlying experimental data.  Temperature $\theta_1$ decreases as $\gamma_1$ decreases
with lower cumulative plastic work, while $\tau_1$ can conceivably increase or decrease with $\gamma_1$ 
depending on the competition between strain hardening and thermal softening.
Pressure $p_0$ tends to increase $\tau$ via \eqref{eq:gyp} before and after
transformation initiation.  

Because the Ni-Cr steel is plastically stiffer than pure Fe, $\tau$ is larger in steel relative to iron at the same $\gamma$, so its adiabatic temperature rise due to plastic working tends to be more rapid.
Thus, the temperature-induced $\alpha \rightarrow \gamma$ transformation commences at lower $\gamma_1$,
also because the $\alpha$-$\gamma$ phase boundary is lower by some 200 K in steel versus Fe in Fig.~\ref{fig2a} \cite{moss1980,moss1981}.
For both Fe and steel, the $\alpha \rightarrow \gamma$ transformation is predicted to occur
at strain $\gamma_1$ of one to two orders of magnitude larger than its $\alpha \rightarrow \epsilon$ counterrpart.
However, model parameters, especially those entering \eqref{eq:betap} and \eqref{eq:sech}, are highly uncertain
for $\alpha \rightarrow \epsilon$ transitions under dominant shear, as experiments have inferred but not proven
that it occurs \cite{rittel2006,sadjadpour2015}. So any conjecture
that $\alpha \rightarrow \epsilon$ is favored over $\alpha \rightarrow \gamma$ in these materials
under dynamic shear from initially ambient conditions is speculative.
Were elastic strains to be included, $\gamma_1$ and $\gamma_2$ could be expected to increase by $\frac{1}{2}$ to $1 \frac{1}{2}$\% given of $\mu$ in Table~\ref{table1} and magnitudes of $\tau$ at yield.
Similarly small increases would be anticipated for results in Sec.~\ref{sec5}.4.

\begin{table}
\footnotesize
\caption{Transition initiation shear strain $\gamma_1$, stress $\tau_1$, and temperature $\theta_1$,
and transition completion strain $\gamma_2$, for iron and high-strength Ni-Cr steel at strain rate $\dot{\bar{\gamma}} = 10^4$/s,
pressure $p_0$, and magnetic field $H_0$}
\label{table3}       
\centering
\begin{tabular}{lccccccc}
\hline\noalign{\smallskip}
Material & Transition & $p_0$ [GPa] & $\hat{\mu}_0 H_0$ [T] & $\gamma_1$ & $\tau_1$ [GPa] & $\theta_1$ [K] & $\gamma_2$  \\
\noalign{\smallskip}\hline\noalign{\smallskip}
Fe & $\alpha \rightarrow \epsilon$ & 0 & 0 & 0.292 & 0.465 & 322 & 7.34 \\
 & & 5 & 0 & 0.245 & 0.510 & 321 & 4.05 \\
 & & 10 & 0 & 0.145 & 0.560 & 314 & 1.91  \\
  & & 10 & 1 & 0.129 & 0.562 & 312 & 1.77 \\
   & & 10 & 2 & 0.110 & 0.559 & 310 & 1.61 \\
 & $\alpha \rightarrow \gamma$ & 0 & 0 & 34.38 & 0.085 & 1210 & 72.31 \\
 & & 5 & 0 & 20.76 & 0.117 & 1027 & 45.16 \\
 & & 10 & 0 & 11.97 & 0.165 & 855 & 27.24 \\
  & & 10 & 1 & 11.43 & 0.169 & 840 & 26.34 \\
   & & 10 & 2 & 10.91 & 0.173 & 824 & 25.47 \\
 Steel & $\alpha \rightarrow \epsilon$ & 0 & 0 & 0.305 & 1.370 & 396 & 4.707  \\
 & & 5 & 0 & 0.296 & 1.524 & 404 & 2.985 \\
 & & 10 & 0 & 0.156 & 1.697 & 360 & 0.574 \\
  & & 10 & 1 & 0.125 & 1.703 & 348 & 0.467 \\
   & & 10 & 2 & 0.086 & 1.699 & 333 & 0.357 \\
 & $\alpha \rightarrow \gamma$ & 0 & 0 & 2.98 & 1.080 & 1131 & 4.63 \\
 & & 5 & 0 & 2.09 & 1.250 & 964 & 3.27 \\
 & & 10 & 0 & 1.41 & 1.440 & 810 & 2.24 \\
  & & 10 & 1 & 1.37 & 1.447 & 795 & 2.19 \\
   & & 10 & 2 & 1.32 & 1.456 & 781 & 2.15 \\
\noalign{\smallskip}\hline
\end{tabular}
\end{table}

\begin{figure}
\begin{center}
 \subfigure[$\epsilon$-phase order parameter]{\includegraphics[width=0.45\textwidth]{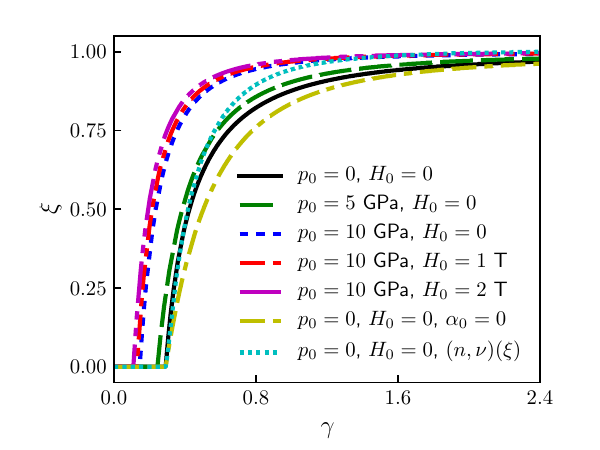} \label{fig5a}} 
 \subfigure[shear stress]{\includegraphics[width=0.45\textwidth]{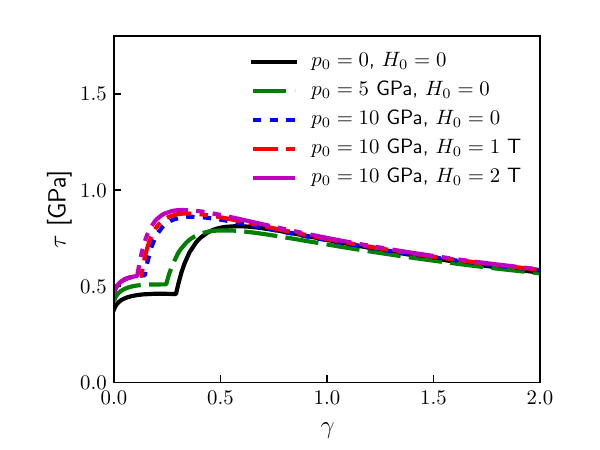}\label{fig5b}} \\
 \subfigure[shear stress]{\includegraphics[width=0.45\textwidth]{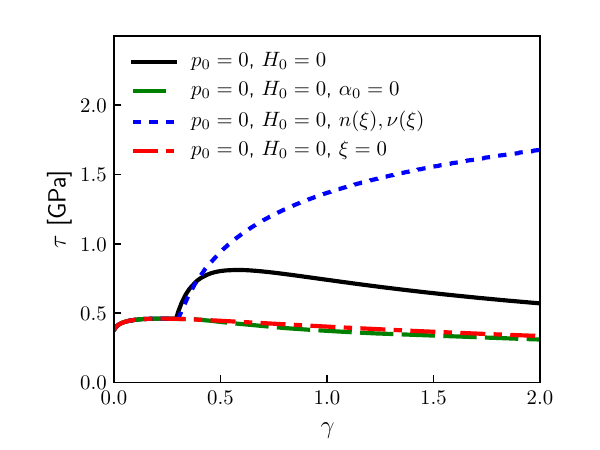}\label{fig5c}} 
  \subfigure[temperature]{\includegraphics[width=0.45\textwidth]{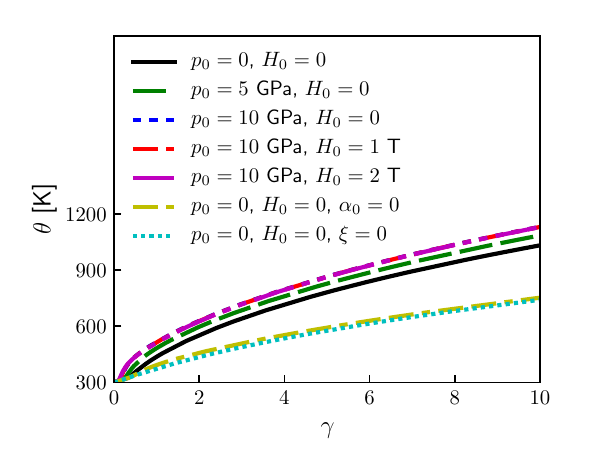}\label{fig5d}} 
  \end{center}
  \vspace{-0.5cm}
\caption{Predictions for iron with $\alpha \rightarrow \epsilon$ transformation, homogeneous straining at $\dot{\gamma} = 10^4$/s,
variable external pressure $p_0$ and magnetic field $H_0$:
(a) order parameter $\xi$ denoting volume fraction of $\epsilon$ phase 
(b) shear stress $\tau$ for nominal material parameters in Table~\ref{table1}
(c) shear stress $\tau$ for special cases of null transformation ($\xi = 0$), null transformation hardening ($\alpha_0 = 0$),
and phase-dependent strain hardening and thermal softening parameters $n(\xi)$ and $\nu(\xi)$
(d) temperature $\theta$ 
}
\label{fig5}       
\end{figure}

\begin{figure}
\begin{center}
 \subfigure[$\gamma$-phase order parameter]{\includegraphics[width=0.45\textwidth]{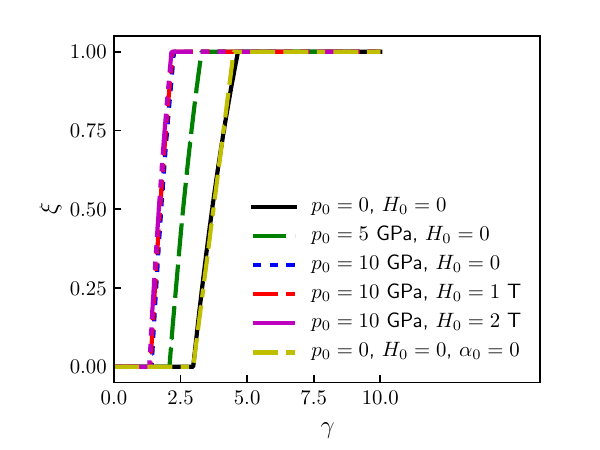} \label{fig6a}} 
 \subfigure[shear stress]{\includegraphics[width=0.45\textwidth]{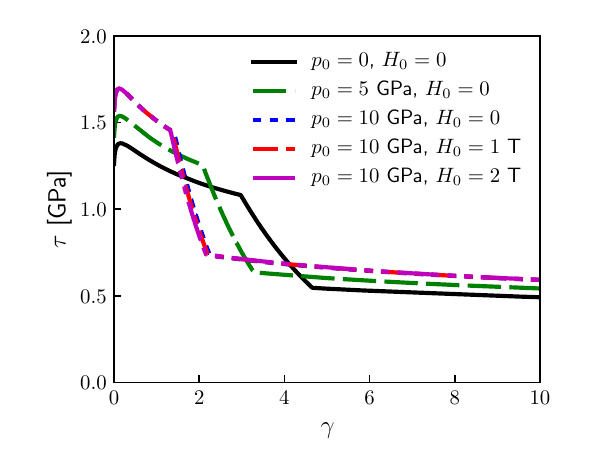}\label{fig6b}} \\
 \subfigure[shear stress]{\includegraphics[width=0.45\textwidth]{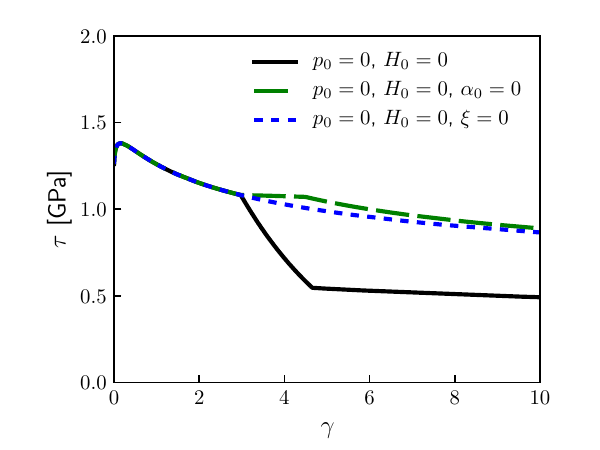}\label{fig6c}} 
  \subfigure[temperature]{\includegraphics[width=0.45\textwidth]{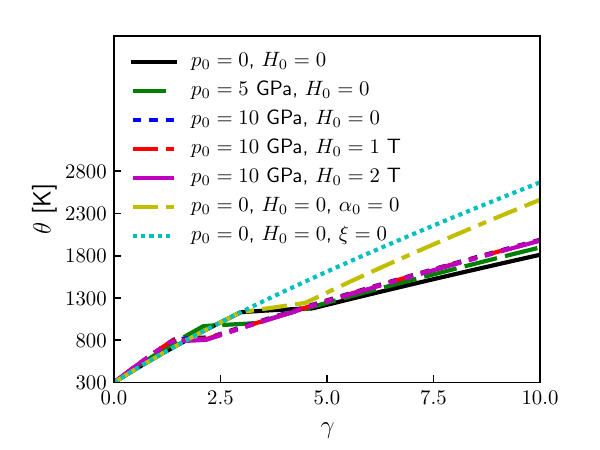}\label{fig6d}} 
  \end{center}
  \vspace{-0.5cm}
\caption{Predictions for Ni-Cr steel with $\alpha \rightarrow \gamma$ transformation, homogeneous straining at $\dot{\gamma} = 10^4$/s,
variable external pressure $p_0$ and magnetic field $H_0$:
(a) order parameter $\xi$ denoting volume fraction of $\gamma$ phase 
(b) shear stress $\tau$ for nominal material parameters in Table~\ref{table1}
(c) shear stress $\tau$ for special cases of null transformation ($\xi = 0$) and null transformation softening ($\alpha_0 = 0$)
(d) temperature $\theta$ 
}
\label{fig6}       
\end{figure}

Predicted strain histories of $\xi$, $\tau$, and $\theta$
are shown in Fig.~\ref{fig5} for iron undergoing the $\alpha \rightarrow \epsilon$ transformation
and in Fig.~\ref{fig6} for Ni-Cr steel undergoing the $\alpha \rightarrow \gamma$ transformation.
Trends in Fig.~\ref{fig5} are similar for Ni-Cr steel undergoing $\alpha \rightarrow \epsilon$ transition (not shown);
trends in Fig.~\ref{fig6} are similar for Fe undergoing $\alpha \rightarrow \gamma$ (not shown).
In all cases, $\xi > 0$ at some threshold $\gamma_1$, beyond which $\xi \rightarrow 1$, increasing monotonically
with $\gamma$.  

For the $\alpha \rightarrow \epsilon$ transformation in Fig.~\ref{fig5b}, since the $\epsilon$ phase
is modeled via $\alpha_0 < 0$ as having higher strength than the $\alpha$ phase \cite{gandhi2022},
$\tau$ increases with increasing $\xi$, but thermal softening reduces $\tau$ at large $\gamma$.
For the special case of $n(\xi)$ and $\nu(\xi)$ shown in Fig.~\ref{fig5a}, whereby
a speculative high strain hardening rate and low thermal softening rate of $\epsilon$-Fe are implemented (Fig.~\ref{fig4}, \cite{gandhi2022}),
$\tau$ continuously increases with $\gamma \geq \gamma_1$, leading to a stable material response and
precluding possible $L_\infty$ localization, in concurrence with Table~\ref{table2} and discussed more in Sec.~\ref{sec5}.4.
For $\gamma \leq 10$, $\theta < 1200$ K for cases in Fig.~\ref{fig5d}, so the $\epsilon \rightarrow \gamma$
transition, excluded a priori from the current modeling, appears unlikely at $p_0 = 0$.
At the highest $p_0$ considered, 10 GPa, $\theta_T = 800$ K is approached or exceeded, so  $\epsilon \rightarrow \gamma$ transformation might be possible,
but it could equally likely be inhibited by shear (i.e., $\gamma^\xi_0 <0$) like the $\alpha \rightarrow \gamma$ transition.

For the $\alpha \rightarrow \gamma$ transformation in Fig.~\ref{fig6b}, 
$\tau$ decreases with increasing $\xi$ since the austenitic phase is presumed softer than the starting
martensitic-bainitic BCC phase \cite{moss1980,moss1981}, as modulated by $\alpha_0 > 0$.
The effect is clear in Fig.~\ref{fig6c}, whereby when $\alpha_0 = 0$ or transition is suppressed,
strain softening is more gradual.
Transition commences at lower $\gamma$ under finite pressure since the slope of the $\alpha$-$\gamma$ phase
boundary is negative in Fig.~\ref{fig2a}. Once transformation initiates, the rate of temperature
increase with strain in Fig.~\ref{fig6d} decreases due to the drop in shear stress and reduced
rate of plastic work.  Transformation widths $\gamma_T = \gamma_2 - \gamma_1$
are within an order of magnitude in Fig.~\ref{fig5a} and Fig.~\ref{fig6a}, namely $0.8 \lesssim \gamma_T \lesssim 8$, if threshold $\xi \gtrsim 0.99$ is
used to define $\gamma_2$ for $\alpha \rightarrow \epsilon$ transitions as has been done for results in Table~\ref{table3}.
The linear interpolation \eqref{eq:xint} appears reasonably realistic for $\alpha \rightarrow \gamma$ 
transitions in Fig.~\ref{fig6a} but less so for $\alpha \rightarrow \epsilon$ transitions in Fig.~\ref{fig5a}. The latter complete relatively slowly versus $\gamma$
 due to $a_\beta > 1$ and  $|r(\gamma)| > 0$ in \eqref{eq:betap} and \eqref{eq:sech}.

\subsection{Localization predictions}
Localization is predicted numerically using methods outlined in Sec.~\ref{sec4}.1--Sec.~\ref{sec4}.3.
Results are calculated for the two material and structural transformation combinations for which
more prior evidence exists and constitutive parameters in Table~\ref{table1} are most certain:
$\alpha \rightarrow \epsilon$ in iron in shear \cite{caspersen2004,rittel2006,lew2006,sadjadpour2015}
and $\alpha \rightarrow \gamma$ in various high-strength steels in shear \cite{moss1980,moss1981,syn2005,jo2020}.
These are the combinations for which homogeneous solutions have been reported in Fig.~\ref{fig5} and Fig.~\ref{fig6}. 
The $\alpha \rightarrow \gamma$ transition in Fe requires extreme shear deformation
according to $\gamma_1 > 10$ for all loading conditions and special cases in Table~\ref{table3}.
The $\alpha \rightarrow \epsilon$ transition has not been reported in any low-pressure
experiments on Ni-Cr steel, though it is theoretically possible according to results
in Table~\ref{table3}, albeit obtained using several parameters borrowed from Fe in the absence of experimental data on this particular steel.

\begin{figure}
\centering
\subfigure [initial defect]{ \includegraphics[width=0.45\textwidth]{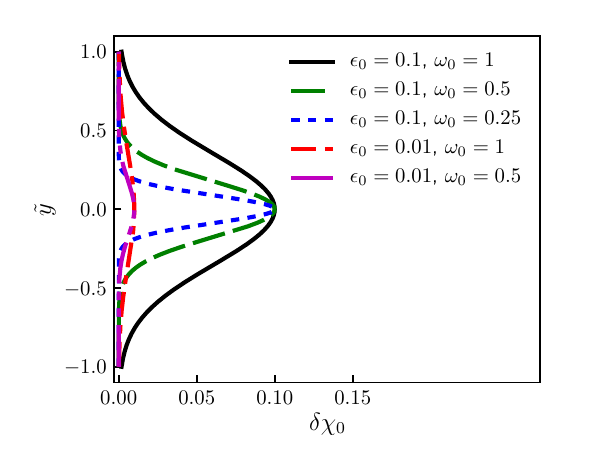} \label{fig7a}}
\subfigure [average strain at localization]{ \includegraphics[width=0.45\textwidth]{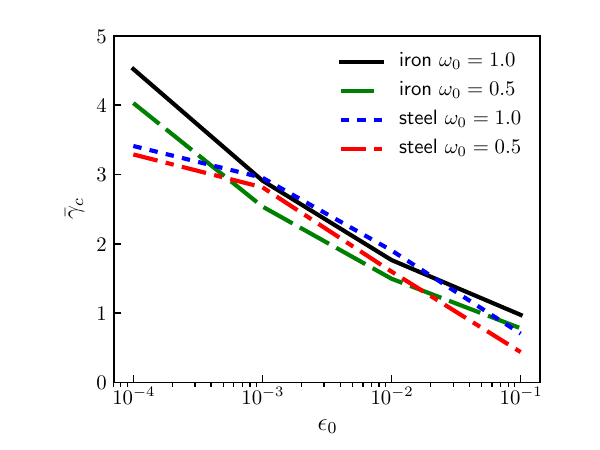} \label{fig7b}}
\caption{Strength defects: (a) initial centered defect distributions (i.e., perturbation fields in initial strength $\chi_0(\tilde{y}) = 1 - \delta \chi_0(\tilde{y})$) for magnitude and width parameters $\epsilon_0$ and $\omega_0$ 
(b) effect of $\epsilon_0$ and $\omega_0$ on critical average localization strain $\bar{\gamma}_c$
for iron and steel at $\dot{\bar{\gamma}} = 10^4$/s, $p_0 =0$, and $H_0 = 0$. Normalized coordinate spanning the slab is $\tilde{y} \in [-1,1]$.
\label{fig7} }      
\end{figure}

As discussed in Sec.~\ref{sec4}.2, initial strength defects quantified by perturbation
distribution $\delta \chi_0(Y)$ are used to instigate localization. Otherwise, if material
properties and initial conditions are perfectly homogeneous, the homogeneous solution
 is the unique solution for the given boundary conditions, even if the stress-strain response is unstable.
 Recall $\hat{y}$ is the initial coordinate at $t = 0^+$ after application of $p_0$ but before shearing deformation $\gamma(t)$.  With the height of the slab in Fig.~\ref{fig1} denoted by $h_0$, define the dimensionless coordinate $\tilde{y}$ with origin at the midpoint of the slab and introduce the
 defect distribution of strength $\epsilon_0$ and relative width $\omega_0$ \cite{langer2017,le2018}:
 \begin{align}
 \label{eq:deltachi}
 & \delta \chi_0 (\tilde{y}(Y)) = \epsilon_0 \exp \biggr{[} - \frac{4 \tilde{y}^2}{\omega_0^2} \biggr{]},
 \qquad \tilde{y}(Y) = \frac{2}{h_0}\biggr{[} \hat{y}(Y) - \frac{h_0}{2} \biggr{]} = 
  \frac{2}{h_0}\biggr{[} (J^E(p_0))^{1/3} Y - \frac{h_0}{2} \biggr{]} \in [-1,1],
 \nonumber 
 \\
 & \delta \bar{\chi}_0 = \frac{1}{2}\int_{-1}^1 \delta \chi_0 \, \d \tilde{y}
 = \frac {\sqrt{\pi}}{4} \epsilon_0 \omega_0 \, {\rm erf} \biggr{[} \frac{2}{\omega_0} \biggr{]} \approx 
 0.4431 \, \epsilon_0 \omega_0, \qquad [0 < \omega_0 \leq 1, \epsilon_0 \geq 0].
 \end{align}
  Value $\omega_0=1$ yields non-negligible $\delta \chi_0$ over the full slab; see
 Fig.~\ref{fig7a} for  $\delta \chi_0$ versus $\epsilon_0$ and $\omega_0$.
 
 According to Sec.~\ref{sec4}.2, the critical average strain at
 the onset of $L_\infty$ localization, $\bar{\gamma}_c$, is calculated via \eqref{eq:gammabar}, where $\gamma (Y) \rightarrow \gamma_c (Y) $ in the integrand is the local strain at any point $Y \neq Y_B$ as $\gamma(Y_B) \rightarrow \infty$.  Per \eqref{eq:deltachi}, localization ensues at the midpoint of the slab, namely at
 $\hat{y} = \frac{1}{2} h_0 \leftrightarrow  \tilde{y} = 0$.  
 To obtain numerical solutions, strain and space domains are discretized into dimensionless increments  $\d {\gamma}$ and $\d \tilde{y}$. Setting $\d \gamma, \d \tilde{y} \lesssim 10^{-4}$ is found small enough that $\bar{\gamma}_c$ is independent of grid size.
 
 Outcomes of the calculation for
 iron and steel at null pressure and magnetic field (i.e., $p_0 = 0$, $H_0 = 0$) with all material properties from Table~\ref{table1} are shown in Fig.~\ref{fig7b} for different choices of $\epsilon_0$ and $\omega_0$.
 In each case, $\bar{\gamma}_c$ decreases with increasing $\epsilon_0$ and decreasing $\omega_0$. The former trend, with near power-law dependence, agrees with results in Refs.~\cite{molinari1986,molinari1987} on two other, presumably non-transforming, steels. 
The latter trend implies that a more localized defect distribution is more detrimental than a diffuse
one, even though $\delta \bar{\chi}_0$ increases with increasing $\omega_0$ at fixed $\epsilon_0$.

 In obtaining results that follow in the remainder of Sec.~\ref{sec5}.4, fixed values $\epsilon_0 = 0.01$ and $\omega_0 = 0.5$ are used.  These choices produce $\bar{\gamma}_c$ on the order of 1.5 in each material, values that are comparable to applied shears (i.e., twice the symmetric strain tensor component) reported in dynamic torsion experiments on iron and a similar high-strength steel \cite{fellows2001,fellows2001b}.
 Subsequent calculations consider the same range of initial pressure $p_0 = 0$, 5, or 10 GPa
 and magnetic field $\hat{\mu}_0 H_0 = 0$, 1, or 2 T as in Sec.~\ref{sec5}.3, as well as an additional 
 condition of a higher 5-T magnetic field. The applied strain
 rate is $\dot{\bar{\gamma}} = \upsilon_0 / h_0 = 10^4$/s.
 
 Several special cases are also considered at null external pressure and null magnetic field.
 Special case $\alpha_0$ gives same initial yield strength in parent and transformed phases,
 where nominally $\alpha_0 < 0$ for $\epsilon$-Fe (i.e., transformation hardening) and
 nominally $\alpha_0 > 0$ for $\gamma$-steel (i.e., transformation softening). 
Special case $\xi = 0$ disables phase transitions entirely.  Finally, special case labeled as $n(\xi), \nu(\xi)$
applies \eqref{eq:nnuxi} for $\alpha \rightarrow \epsilon$ in iron, producing a much higher strain hardening rate and reduced thermal softening rate for $\epsilon$-Fe as in Fig.~\ref{fig4}. Recall these $\epsilon$-Fe strength parameters are non-unique since they are obtained from calibrations \cite{gandhi2022} wherein rate dependence is omitted and starting temperatures are the same in all dynamic high-pressure shear experiments.

Table~\ref{table4} lists average critical strain $\bar{\gamma}_c$, corresponding average temperature $\bar{\theta}_c$, and average transformed volume fraction $\bar{\xi}_c$ for all loading conditions and special cases.
Critical average temperature and order parameter are obtained analogously to \eqref{eq:gammabar}:
\begin{align}
\label{eq:thetaxibar}
\bar{\theta}_c = \frac{1}{h_0} \int_0^{h_0} \theta_c (\hat{y}) \d \hat{y} = \frac{1}{2} \int_{-1}^1 \theta_c (\tilde{y}) \d \tilde{y}, \qquad
\bar{\xi}_c = \frac{1}{h_0} \int_0^{h_0} \xi_c (\hat{y}) \d \hat{y} = \frac{1}{2} \int_{-1}^1 \xi_c (\tilde{y}) \d \tilde{y},
\end{align}
where $\theta_c$ and $\xi_c$ are local values at the onset of shear localization and singular
point $Y_B$ at $\hat{y} = \frac{1}{2} h_0 \leftrightarrow \tilde{y} = 0$ is excluded from the numerical integration \cite{molinari1986,molinari1987}.
Values of stress $\bar{\tau}_c$ are approximated using \eqref{eq:flowstress3} at $\bar{\gamma} \rightarrow \bar{\gamma}_c$, just prior to load collapse at the instant of adiabatic shear localization, $t \rightarrow t_c$.

First consider iron possibly undergoing the $\alpha \rightarrow \epsilon$ transition.
For standard properties (i.e., not special cases), critical average strain $\bar{\gamma}_c$
increases as $p_0$ increases from 0 to 5 GPa then decreases as $p_0$ increases to 10 GPa.
Critical average strain further decreases with increasing magnetic field $H_0$.
Average transformed fraction $\bar{\xi}_c$ increases with $p_0$ and $H_0$. In all standard cases,
$\bar{\xi}_c > 0.92$ implying most of the domain has transformed when the localization threshold is reached.
Critical average temperature $\bar{\theta}_c$ follows a similar trend to $\bar{\gamma}_c$.
Stress $\bar{\tau}_c$ tends to increase as $\bar{\gamma}_c$ decreases since less overall thermal softening has occurred at the onset of localization at lower applied strain.
Dissipation from phase transformation tends to raise the temperature and increase thermal softening, while increases in transformed fraction tend to increase the yield strength for $\alpha_0 < 0$.
Thus, a competition exists among mechanisms related to phase transition that influence $\bar{\tau}_c$.

For the first special case with $\alpha_0 = 0$, $\bar{\gamma}_c$ is significantly reduced relative to the
standard case at null pressure and magnetic field, from 1.497 to 1.226. 
The stress at localization is severely reduced, from 0.644 GPa to 0.354 GPa. Thus, if transformation hardening
is omitted, shear localization is promoted. For the second special case with $\xi(t) = 0$,
$\bar{\gamma}_c$  is slightly increased relative to the nominal case, to 1.544 from 1.497.
Since dissipation from transformation does not arise or contribute in this special case, thermal softening is postponed
and localization occurs later in time.

For the final special case with $n$, $\nu$, and $\gamma_0$ dependent on the $\epsilon$-phase fraction,
$L_\infty$ localization is impossible. This is theoretically anticipated from Table~\ref{table2} and confirmed numerically
by a non-converging localization integral on the right side of \eqref{eq:Lc2} whose rate of increase gets larger as $\gamma$ increases to extreme values. In this case,
$n_1 = 0.5$ and $\nu_1 = -0.27$ are used in the calculation, with $m = 0.089$ regardless of phase. However, the
 temperature $\theta$ at the $Y$-location where $\gamma = \bar{\gamma}$ reaches the melt temperature $\theta_M$
 at $\bar{\gamma} = 5.35$. This value is assigned as the critical strain $\bar{\gamma}_c$ in Table~\ref{table4}.
 At this strain, the material has completely transformed ($\bar{\xi}_c \rightarrow 1$) and
 stress is large, $\bar{\tau}_c > 2$ GPa.
 The result shows that if hardening rate increase and thermal softening decrease with $\alpha \rightarrow \epsilon$ transformation in Fe,
 the strain and time to load collapse are substantially increased, as is the shear stress just prior to collapse.
 
\begin{figure}
\begin{center}
 \subfigure[local strain]{\includegraphics[width=0.45\textwidth]{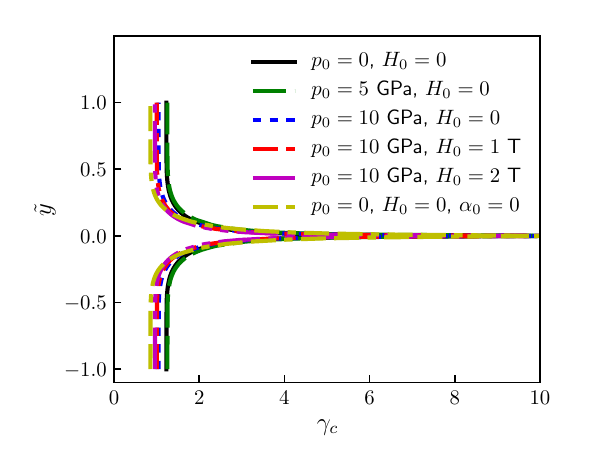} \label{fig8a}} 
 \subfigure[local temperature]{\includegraphics[width=0.45\textwidth]{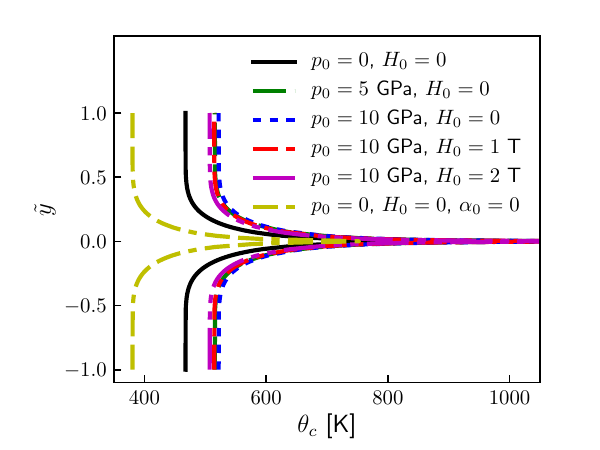}\label{fig8b}} \\
 \subfigure[local $\epsilon$ volume fraction]{\includegraphics[width=0.45\textwidth]{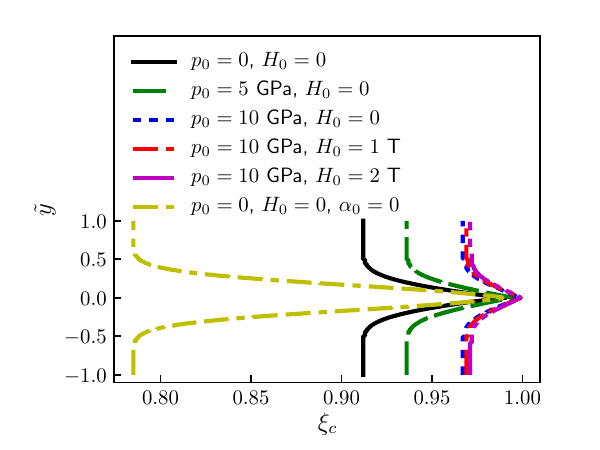}\label{fig8c}} 
  \subfigure[average shear stress]{\includegraphics[width=0.45\textwidth]{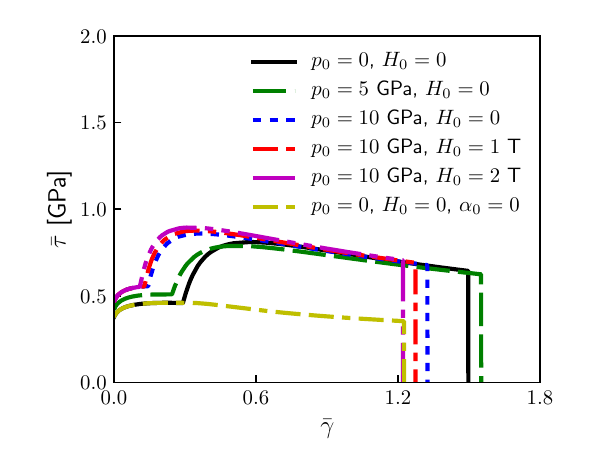}\label{fig8d}} 
  \end{center}
  \vspace{-0.5cm}
\caption{Adiabatic shear localization predictions for iron with $\alpha \rightarrow \epsilon$ transformation,  $\dot{\bar{\gamma}} = 10^4$/s,
 external pressure $p_0$ and magnetic field $H_0$:
(a) local shear strain $\gamma_c$
(b) temperature $\theta_c$ 
(c) fraction of $\epsilon$ phase
(d) average shear stress $\bar{\tau}$ vs.~average shear strain $\bar{\gamma}$.
Normalized coordinate spanning the slab is $\tilde{y} \in [-1,1]$.
Setting $\alpha_0 = 0$ gives yield strength independent of $\xi$; in all other cases, $\epsilon$-Fe is stiffer than $\alpha$-Fe.
}
\label{fig8}       
\end{figure}

 Reported in Figs.~\ref{fig8a}, \ref{fig8b}, and \ref{fig8c} are contours, for iron with possible $\alpha \rightarrow \epsilon$ transformation, of local strain $\gamma_c$, temperature $\theta_c$, and order parameter $\xi_c$ at the onset of adiabatic shear localization.  From \eqref{eq:deltachi}, these are symmetric about $\tilde{y} = 0$ and span the height of the slab: $\hat{y} \in [0,h_0] \leftrightarrow \tilde{y} \in [-1,1]$.
Shown in Fig.~\ref{fig8d} is the corresponding average stress-strain behavior estimated from \eqref{eq:flowstress3}
up to abrupt load collapse at $\bar{\gamma} = \bar{\gamma}_c$ \cite{molinari1986,molinari1987}.
Any deviations in this idealized linear path from the onset of localization to a shear-stress free state cannot be modeled explicitly via the current approach.
From Fig.~\ref{fig8a}, a more diffuse local strain $\bar{\gamma}_c$ tends to increase the overall average critical
strain $\bar{\gamma}_c$ in Table~\ref{table4}, thereby delaying load collapse.
Apparent in Fig.~\ref{fig8b}, localization of $\theta_c$ tends to correlate with $\gamma_c$. In the special
case $\alpha_0 = 0$, the average temperature outside the shear band is around 150 K lower than that of
the nominal case. This is due to the reduction in plastic work in the absence of transformation hardening
when $\alpha_0 = 0$, as well as the reduced dissipated energy from less phase transition.
The latter point is quantified in Fig.~\ref{fig8c}. Transformation also increases with $p_0$ and $H_0$ as anticipated
from results in Sec.~\ref{sec5}.3.  Average shear stress-strain behavior in Fig.~\ref{fig8d} highlights the
earlier transformation hardening with increasing $p_0$ and $H_0$, generally leading to sooner attainment of peak stress, subsequent thermal softening, and earlier stress collapse.

For $\tau \rightarrow 0$ at $\bar{\gamma} > \bar{\gamma}c$, either $\dot{\gamma} \rightarrow 0$ or
$\theta \rightarrow \infty$ with $\nu < 0$ at the localization point $Y_B$, here at $\tilde{y} = 0$.
If the strain rate remains finite, then melting would occur at this singular location in the shear band, if
not also elsewhere. Consider the molten material hypothetically modeled as a viscous Newtonian fluid:
$\tau(\dot{\gamma},\theta) = \eta_0 (\theta/\theta_0)^\nu \dot{\gamma}/\dot{\gamma}_0$, where $\eta_0 > 0$ is a constant.  If $\nu < 0$, $\tau \rightarrow 0$ is possible as $\theta \rightarrow \infty$. However, per this simple model, localization is impossible even upon melting via liberal extension of $L_\infty$ criterion \eqref{eq:case1b} with large $m =1$ and $n= 0$, regardless of $\nu$.

\begin{table}
\footnotesize
\caption{Average critical shear strain $\bar{\gamma}_c$, temperature $\bar{\theta}_c$, transformed volume fraction $\bar{\xi}_c$, and stress $\bar{\tau}_c$ just prior to $t \rightarrow t_c$ at
onset of adiabatic shear localization.
Results for iron and Ni-Cr steel at applied average strain rate $\dot{\bar{\gamma}} = 10^4$/s,
pressure $p_0$, and magnetic field $H_0$. Initial defect distribution is $\epsilon_0 = 0.01$ and $\omega_0 = 0.05$ giving $\delta \bar{\chi}_0 = 2.2 \times 10^{-3}$. Nominal parameters from Table~\ref{table1}; see text for explanation of special cases. }
\label{table4}       
\centering
\begin{tabular}{lcccccccc}
\hline\noalign{\smallskip}
Material & Transition & $p_0$ [GPa] & $\hat{\mu}_0 H_0$ [T] & $\bar{\gamma}_c$ & $\bar{\theta}_c$ [K] &
 $\bar{\xi}_c$ & $\bar{\tau}_c$ [GPa] & Special case  \\
\noalign{\smallskip}\hline\noalign{\smallskip}
Fe & $\alpha \rightarrow \epsilon$ & 0 & 0 & 1.497 & 494 & 0.924 & 0.644 & $\ldots$ \\
 & & 5 & 0 & 1.552 & 544 & 0.946 & 0.625 & $\ldots$  \\
 & & 10 & 0 & 1.325 & 550 & 0.973 & 0.677 & $\ldots$  \\
  & & 10 & 1 & 1.274 & 543 & 0.974 & 0.691 & $\ldots$  \\
  & & 10 & 2 & 1.221 & 535 & 0.976 & 0.705 & $\ldots$  \\
   & & 10 & 5 & 1.038 & 509 & 0.984 & 0.756 & $\ldots$  \\
     & & 0 & 0 & 1.226 & 403 & 0.827 & 0.354 & $\alpha_0 = 0$  \\
          & & 0 & 0 & 1.544 & 406 & 0.000 & 0.362 & $\xi = 0$  \\
            & & 0 & 0 & 5.350$^*$ & 1811$^*$  & 1.000 & 2.030 & $n(\xi), \nu(\xi)$  \\
 Steel &  $\alpha \rightarrow \gamma$ & 0 & 0 & 1.600 & 767 & 0.0105 & 1.187 & $\ldots$  \\
 & & 5 & 0 & 1.279 & 719 &  0.0131 & 1.343  & $\ldots$  \\
 & & 10 & 0 & 0.990 & 661 & 0.0167 & 1.510 & $\ldots$  \\
  & & 10 & 1 & 0.972 & 655 & 0.0174 & 1.513 & $\ldots$ \\
  & & 10 & 2 & 0.954 & 648 & 0.0181 & 1.516 & $\ldots$ \\
   & & 10 & 5& 0.897 & 628 & 0.0204 & 1.527  & $\ldots$  \\
       & & 0 & 0 & 2.668 & 994 & 0.1686 & 1.099 & $\alpha_0 = 0$  \\
             & & 0 & 0 & 2.255 & 930 & 0.0000 & 1.129 & $\xi = 0$  \\
\noalign{\smallskip}\hline 
$^*$melt onset
\end{tabular} 
\end{table}

Now consider the high-strength Ni-Cr steel potentially undergoing the $\alpha \rightarrow \gamma$ phase transition.  
Average quantities at the localization threshold are listed in Table~\ref{table4}.
The special case $n(\xi), \nu(\xi)$ cannot be addressed since strain hardening and thermal softening properties of $\gamma$-phase are unknown for this material.
For the nominal cases in Table~\ref{table4}, critical average strain $\bar{\gamma}_c$ decreases with
increasing pressure $p_0$ and magnetic field $H_0$. Commensurately, temperature $\bar{\theta}_c$ 
decreases due to less plastic dissipation, and $\bar{\tau}_c$ tends to increase since less thermal softening
has ensued. In contrast to pure Fe undergoing the $\alpha \rightarrow \epsilon$ transformation,
for this steel undergoing the $\alpha \rightarrow \gamma$ transition, the average transformed volume fraction
is low at localization: $\bar{\xi}_c \lesssim 0.02$. The local temperature and pressure are not high enough
\textit{outside} the vicinity of the shear band to enable the $\gamma \rightarrow \alpha$ transformation before 
load collapse. 
For the first special case with $\alpha_0 = 0$, transformation softening upon transition to the $\gamma$ phase
is suppressed in the model. The average critical strain increases from 1.600 to 2.668. Temperature $\bar{\theta}_c$
and order parameter $\bar{\xi}_c$ increase simultaneously with the increase in plastic work.
For the second special case with $\xi(t) = 0$, omission of the phase transition increases the
average strain to 2.255 and the temperature reached upon adiabatic shear localization. 

\begin{figure}
\begin{center}
 \subfigure[local strain]{\includegraphics[width=0.45\textwidth]{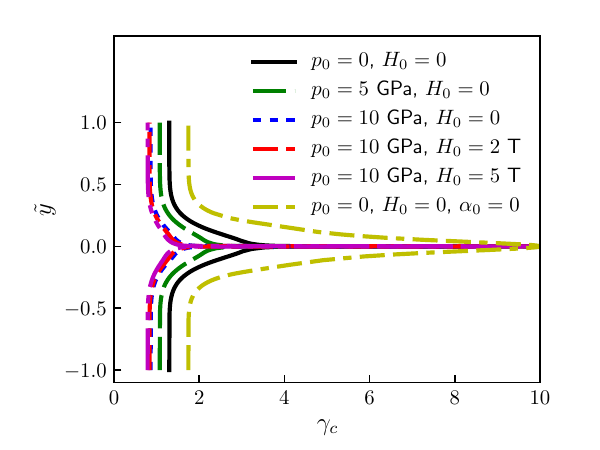} \label{fig9a}} 
 \subfigure[local temperature]{\includegraphics[width=0.45\textwidth]{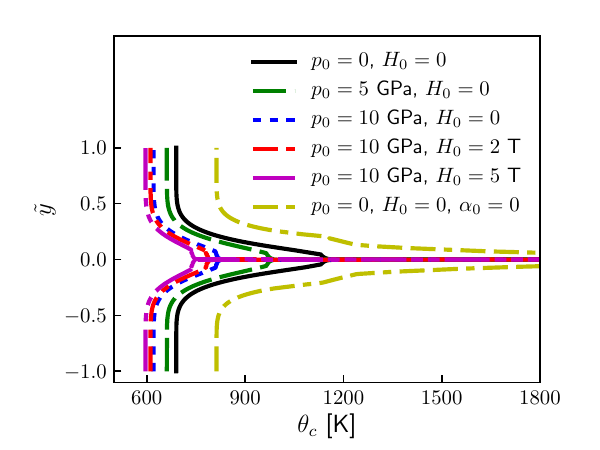}\label{fig9b}} \\
 \subfigure[local $\gamma$ volume fraction]{\includegraphics[width=0.45\textwidth]{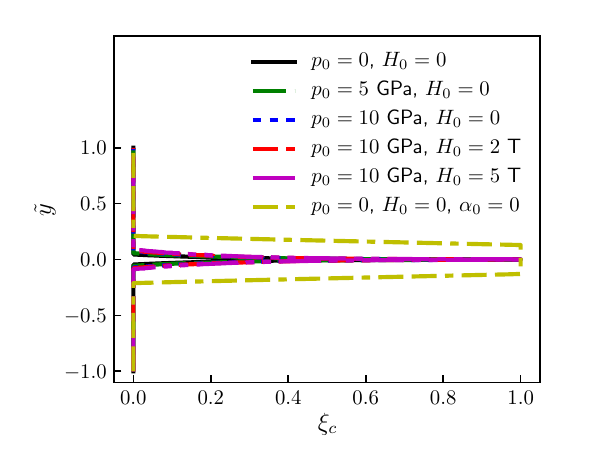}\label{fig9c}} 
  \subfigure[average shear stress]{\includegraphics[width=0.45\textwidth]{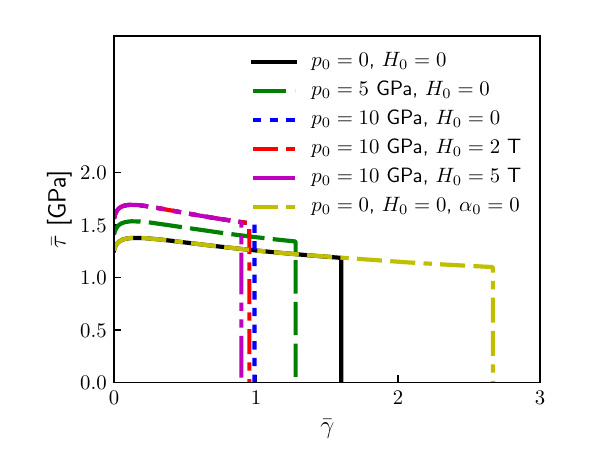}\label{fig9d}} 
  \end{center}
  \vspace{-0.5cm}
\caption{Adiabatic shear localization predictions for Ni-Cr steel with $\alpha \rightarrow \gamma$ transformation,  $\dot{\bar{\gamma}} = 10^4$/s,
 external pressure $p_0$ and magnetic field $H_0$:
(a) local shear strain $\gamma_c$
(b) temperature $\theta_c$ 
(c) fraction of $\gamma$ phase
(d) average shear stress $\bar{\tau}$ vs.~average shear strain $\bar{\gamma}$.
Normalized coordinate spanning the slab is $\tilde{y} \in [-1,1]$.
Setting $\alpha_0 = 0$ gives yield strength independent of $\xi$; in all other cases, $\gamma$ phase is softer than $\alpha$ phase.
}
\label{fig9}       
\end{figure}

Shown in Figs.~\ref{fig9a}, \ref{fig9b}, and \ref{fig9c} are contours of $\gamma_c$, $\theta_c$, and $\xi_c$ at the onset of localization in Ni-Cr steel undergoing possible $\alpha \rightarrow \gamma$ phase transition. 
Average stress-strain behavior up to load collapse at $\bar{\gamma} = \bar{\gamma}_c$ is predicted in Fig.~\ref{fig9d}.
From Fig.~\ref{fig9a}, the more diffuse the local strain $\bar{\gamma}_c$, the larger the average critical
strain $\bar{\gamma}_c$ in Table~\ref{table4}. This trend agrees with that of Fe in Fig.~\ref{fig8a}. However,
for Ni-Cr steel with $\alpha \rightarrow \gamma$ transition, the strain distribution is much more
localized about the singular point at $\tilde{y} = 0$ than predictions for Fe in Fig.~\ref{fig8a}.
The special case $\alpha_0 = 0$ witnesses the most strain diffusion, correlating with the latest onset
of stress collapse.
Localization of $\theta_c$ in Fig.~\ref{fig9b} complements $\gamma_c$ in Fig.~\ref{fig9a}. In the special
case $\alpha_0 = 0$, average $\theta$ far outside the shear band is about 100 K higher than 
the nominal case at null pressure and magnetic field. This is due to the increase in plastic work in the absence of transformation softening
when $\alpha_0 = 0$, along with increased dissipation from the phase transition more prominent
for this special case in Fig.~\ref{fig9c}. 

As witnessed in Fig.~\ref{fig9c}, the region where $\xi_c$ is non-negligible is usually confined to a small zone
about the singular point at $\tilde{y} = 0$. In other words, the $\gamma \rightarrow \alpha$ transformation in this
steel occurs only in the highly strained, high-temperature region in the immediate vicinity of the shear band, meaning localization tends to precede transformation.
Average shear stress-strain behavior in Fig.~\ref{fig9d} 
is similar up to localization strain $\bar{\gamma} = \bar{\gamma}_c$ since the average strain and temperature are insufficient to induce the $\alpha \rightarrow \gamma$ transition in most of the slab, including the interrogated material point $Y$ where $\gamma = \bar{\gamma}$. Modest differences among curves at $\bar{\gamma} < \bar{\gamma}_c$ for different pressures in Fig.~\ref{fig9d} result primarily from pressure dependence of yield strength in \eqref{eq:gyp}.

Contours of $\gamma_c$, $\theta_c$, and $\xi_c$ in Figs.~\ref{fig8} and~\ref{fig9} are depicted in dimensionless 
$\tilde{y}$-space. These solutions enable comparison of the \textit{dimensionless} width of high-strain, high-temperature
zones in the vicinity of the fully formed bands at singular surface $\tilde{y} = 0$. 
From such comparisons, relative effects of different transformations, material properties, and superposed pressures and magnetic fields can be deduced.
The \textit{absolute} width of these zones scales linearly with $h_0$ if the characteristic time $t_0 = h_0/ \upsilon_0$, the defect strength $\epsilon_0$, and the normalized defect profile $\omega_0$ are held constant.
Their absolute width is modulated by the initial conditions $\epsilon_0$ and $\omega_0$ and the geometry $h_0$,
similar to several recent works \cite{langer2016,lieou2018}.
In the absence of heat conduction and inertia, their minimum width tends to zero as $h_0$ decreases.
More sophisticated numerical methods, outside the present scope, are required to account for potential effects of conduction and inertia on band morphology.

In summary, model predictions suggest that the $\alpha \rightarrow \epsilon$ transformation, if it occurs, \textit{precedes}
localization in pure Fe, while the $\alpha \rightarrow \gamma$ transformation, if it occurs, either complements or \textit{follows} adiabatic shear localization in the present Ni-Cr steel. To delay localization and mitigate load collapse commensurate with a fully formed shear band, steps should be taken to reduce $\bar{\gamma}_c$, since
the time at which localization persists is $t_c = \bar{\gamma}_c h_0  / \upsilon_0$.
First consider the case when strain hardening and thermal softening
exponents $n$ and $\nu$ of parent and transformed phases match.
If the transformed phase is of the same initial strength ($\alpha_0 = 0$) or initially softer ($\alpha_0 > 0$) than the parent phase, as assumed for $\alpha \rightarrow \gamma$
transitions, then suppression of the transformation should delay localization.
If the transformed phase is initially harder ($\alpha_0 < 0$), as assumed for $\alpha \rightarrow \epsilon$ transformations, then suppression of the transformation can either delay or promote localization
depending on the initial pressure. 
If the strain hardening rate $n$ increases and/or thermal softening $\nu$ decrease dramatically in conjunction with
a change of phase, as is conceivable but not confirmed for $\alpha \rightarrow \epsilon$ transformations
in Fe, then shear localization should be mitigated by such a phase transformation.
In that case, increasing the pressure and adjusting an external magnetic field to enhance transition kinetics
should delay failure that would now occur from localized melting rather than extreme viscoplastic thermal softening.

\section{Conclusions \label{sec6}}

A theoretical analysis of adiabatic shear localization in viscoplastic solids has been undertaken, newly accounting for structural transitions such as phase transformations. 
A reduced-order model for phase transitions affected by shear stress, pressure, temperature, and magnetic fields has been formulated for Fe and a high-strength steel in polycrystalline form. Effects of constitutive parameters, external pressure and magnetic fields, and initial strength defects on localization criteria and applied shear strains required for localization have been quantified for these two metals.

The $\alpha \rightarrow \epsilon$ transformation in pure Fe, if it occurs, is predicted to precede adiabatic shear localization. The $\alpha \rightarrow \gamma$ transformation in Ni-Cr steel, if it occurs, is predicted accompany localization but not precede it.
Findings agree qualitatively with some experimental observations. First, $\alpha \rightarrow \epsilon$
transformations have been witnessed or inferred in Fe under dynamic shear with or without large superposed pressure,
in the absence of adiabatic shear bands \cite{rittel2006,sadjadpour2015,gandhi2022}.
Second, $\alpha \rightarrow \gamma$ transitions in high-strength steels have
been witnessed or inferred within adiabatic shear bands, but not outside their immediate vicinity \cite{syn2005,cho1990}, in dynamic torsion or compression experiments starting from ambient room temperature and pressure.
If the transformed phase is softer than the parent phase, transitions are found to be detrimental in the sense that shear localization is accelerated.
If the hardening rate is higher and thermal softening is lower in the transformed phase, then a phase transition can mitigate adiabatic shear localization.  Melting appears eventually likely even if the viscoplastic stress-strain response is mechanically stable to very large shear strain.

Experiments probing viscoplastic properties of the materials in their transformed, high-pressure, high-temperature states would lend confidence to those model parameters that are presently uncertain. 
Controlled dynamic experiments involving simultaneous sufficiently high pressure and large enough shear strain, without boundary constraints, for localization would provide further validation.
To enable tractable analytical solutions, this study omits explicit elastic shear strain, inertia, and conduction of heat and electricity. Certain aspects of the phase transition model have been linearized. 
Severity of these assumptions should be quantified in the future by comparison with fully resolved in space-time numerical solutions that capture the presently approximated physics.



\bibliography{refs}

\end{document}